\documentclass[12pt]{article}
 
\newlength{\abstractwidth}
\usepackage{amsmath, nccmath}
\usepackage{amsfonts}
\usepackage{amssymb}
\usepackage{latexsym}

\usepackage{color}
\usepackage{graphicx}
\usepackage{tikz}
\usepackage{dsfont}

\usetikzlibrary{arrows,shapes,positioning}
\usetikzlibrary{decorations.markings}
\usepackage[rightcaption]{sidecap}
\tikzstyle arrowstyle=[scale=1]
\tikzstyle directed=[postaction={decorate,decoration={markings,
    mark=at position .65 with {\arrow[arrowstyle]{stealth}}}}]
\tikzstyle reverse directed=[postaction={decorate,decoration={markings,
    mark=at position .65 with {\arrowreversed[arrowstyle]{stealth};}}}]
\usetikzlibrary{positioning}

\usepackage{hyperref}
\definecolor{darkred}{rgb}{0.8,0.1,0.1}
\hypersetup{colorlinks=true, linkcolor=darkred, citecolor=blue, linktoc=page}

\numberwithin{equation}{section}

\renewcommand{\thefootnote}{\fnsymbol{footnote}}
\renewcommand{\thanks}[1]{\footnote{#1}}
\newcommand{\starttext}{
\setcounter{footnote}{0}
\setcounter{section}{0}
\renewcommand{\thefootnote}{\arabic{footnote}}}
\newcommand{\bea}{\begin{eqnarray}}
\newcommand{\eea}{\end{eqnarray}}
\newcommand{\be}{\begin{eqnarray}}
\newcommand{\ee}{\end{eqnarray}}
\newcommand{\bma}{\begin{matrix}}
\newcommand{\ema}{\cr\end{matrix}}
\newcommand{\<}{\langle}
\renewcommand{\>}{\rangle}

\newtheorem{thm}{Theorem}[section]
\newtheorem{lem}[thm]{Lemma}

\def\cA{{\cal A}}
\def\cB{{\cal B}}
\def\cC{{\cal C}}
\def\cD{{\cal D}}
\def\cE{{\cal E}}
\def\cF{{\cal F}}
\def\cG{{\cal G}}
\def\cH{{\cal H}}
\def\cI{{\cal I}}
\def\cJ{{\cal J}}
\def\cK{{\cal K}}

\def\cM{{\cal M}}
\def\cN{{\cal N}}
\def\cO{{\cal O}}

\def\cQ{{\cal Q}}
\def\cR{{\cal R}}

\def\cV{{\cal V}}
\def\cW{{\cal W}}

\def\cZ{{\cal Z}}

\def\mA{\mathfrak{A}}
\def\mB{\mathfrak{B}}
\def\mC{\mathfrak{C}}

\def\mJ{\mathfrak{J}}

\def\mZ{\mathfrak{Z}}

\def\ZZ{{\mathbb Z}}
\def\RR{{\mathbb R}}

\def\CC{{\mathbb C}}

\def\Re{{\rm Re \,}}
\def\Im{{\rm Im \,}}
\def\tr{{\rm tr}}

\def\det{{\rm det \,}}

\def\half{{1\over 2}}

\def\p{\partial}

\def\Sep{{\Sigma _{ab}}}
\def\RRR{R}

\def\a{\alpha}

\def\eps{\epsilon}
\def\f{\varphi}

\def\ep{\varepsilon}
\def\om{\omega}

\def\pbz{\p _{\bar z}}

\def\pbx{\p _{\bar x}}
\def\pby{\p _{\bar y}}

\def\pbz{\p _{\bar z}}
\def\GA{\cG}
\def\kap{\kappa}
\def\oom{\overline{\om}}
\def\KN{{\rm KN}}

\def\no{\nonumber}
\def\sm{\smallskip}

\def\ap{{\alpha^{\prime}}}
\def\halfap#1{\Big(\mkern-2mu{\ap\over 2}\mkern-2mu\Big)^{\mkern-4mu #1}}
\def\invhalfap#1{\Big(\mkern-1mu{2\over \ap}\mkern-1mu\Big)^{\mkern-4mu #1}}
\def\l{\lambda}

\def\cBtree#1{\cB^{\rm tree}_{\{#1\}}}
\def\cBoneloop#1{\cB^{\rm genus-1}_{\{#1\}}}
\def\cBtwoloop#1{\cB^{\rm genus-2}_{\{#1\}}}
\def\cBnottree#1{\cB^{\rm\, !tree}_{\{#1\}}}

\allowdisplaybreaks

\begin{document}
\starttext
\setcounter{footnote}{0}

\begin{flushright}
2020 August 19 \\
revised 2022 July 24 \\
UUITP-29/20
\end{flushright}

\bigskip
\bigskip

\begin{center}

\vskip -0.2in

{\Large \bf Two-loop superstring five-point amplitudes II }

\vskip 0.1in

{ \bf Low energy expansion and S-duality   }

\vskip 0.3in

{{  \bf Eric D'Hoker$^{(a)}$, Carlos R. Mafra${}^{(b)}$,  Boris Pioline$^{(c)}$, Oliver Schlotterer$^{(d)}$}
\footnote{E-mail: {\tt \small dhoker@physics.ucla.edu}; {\tt \small c.r.mafra@soton.ac.uk}; \\ 
\color{black}{} \hskip 0.7in
{\tt \small pioline@lpthe.jussieu.fr};
{\tt \small oliver.schlotterer@physics.uu.se}.}}

\vskip 0.1in

 ${}^{(a)}$ {\sl Mani L. Bhaumik Institute for Theoretical Physics}\\
 { \sl Department of Physics and Astronomy }\\
{\sl University of California, Los Angeles, CA 90095, USA}\\

\vskip 0.1in

${}^{(b)}$ {\sl STAG Research Centre and Mathematical Sciences,} \\ {\sl  University of Southampton, Highfield, Southampton SO17 1BJ, UK}

\vskip 0.1in

 ${}^{(c)}$ { \sl Laboratoire de Physique Th\'eorique et Hautes Energies}\\
{\sl CNRS and Sorbonne Universit\'e, UMR 7589}\\
{\sl Campus Pierre et Marie Curie, 4 Place Jussieu 75252 Paris, France}

\vskip 0.1in

 ${}^{(d)}$ { \sl Department of Physics and Astronomy,} \\ {\sl Uppsala University, 75108 Uppsala, Sweden}

\vskip 0.7in

{\it Dedicated to the life, science, and art of Professor Jean-Loup Gervais}

\newpage

\begin{abstract}
In an earlier paper, we constructed the genus-two amplitudes for five external massless states in Type II and Heterotic string theory, and showed that the $\alpha'$ expansion of the Type II amplitude  reproduces the corresponding supergravity amplitude to leading order. In this paper, we analyze the effective interactions induced by Type IIB superstrings beyond supergravity, both for $U(1)_R$-preserving amplitudes such as for five gravitons, and for $U(1)_R$-violating amplitudes such as for  one dilaton and four gravitons. At each order in $\alpha'$, the coefficients of the effective interactions are given by integrals over moduli space of genus-two modular graph functions,  generalizing those already encountered for four external massless states. To leading and sub-leading orders, the coefficients of the effective interactions $D^2 \cR^5$ and $D^4\cR^5$ are found to match those of $D^4 \cR^4$ and $D^6 \cR^4$, respectively, as required by non-linear supersymmetry.  To the next order, a $D^6 \cR^5$ effective interaction arises, which is independent of the supersymmetric completion of $D^8 \cR^4$, and already arose at genus one. A novel identity on genus-two modular graph functions, which we prove, ensures that up to order $D^6\cR^5$, the five-point amplitudes require only a single new modular graph function in addition to those needed for the four-point amplitude. We check that the supergravity limit of $U(1)_R$-violating amplitudes is free of UV divergences to this order, consistently with the known structure of divergences in Type IIB supergravity. Our results give strong consistency tests on the full five-point amplitude, and pave the way for understanding S-duality beyond the BPS-protected sector.  
\end{abstract}

\end{center}

\newpage

\setcounter{tocdepth}{2} 
\tableofcontents

\newpage

\baselineskip=16pt
\setcounter{equation}{0}
\setcounter{footnote}{0}

\section{Introduction}
\setcounter{equation}{0}
\label{sec:1}

Scattering amplitudes of massless states are the basic observables in string theory and, in principle, are well-defined at arbitrary order in perturbation theory (for reviews see \cite{D'Hoker:1988ta, D'Hoker:2002gw,Witten:2012ga,Witten:2012bh}). They are UV-finite by construction and, in the $\alpha'$ expansion, reduce to supergravity amplitudes plus an infinite series in  $\alpha' $ of effective interactions \cite{Green:1982sw}.  
 In practice, however, the explicit evaluation of superstring amplitudes rapidly becomes prohibitively complicated beyond genus one. For a long time the state of the art has been the four-point genus-two amplitude which was constructed in the Ramond-Neveu-Schwarz (RNS) formalism  (see  \cite{D'Hoker:2005jc} and references therein), reproduced in the pure spinor (PS) formalism and extended to include external fermions \cite{Berkovits:2005df}.  
 
 \sm
 
 Beyond this, partial results have been obtained in the PS formalism for the five-point two-loop amplitude \cite{Gomez:2015uha}, and the four-point three-loop amplitude \cite{Gomez:2013sla}. 
A major obstacle to explicit evaluations  in the PS formalism (in its non-minimal version)
 is due to  the composite $b$-ghost \cite{Berkovits:2005bt}, which diverges at the origin of the cone of pure spinor zero-modes and requires a large number of Wick contractions.  As a consequence, in both cases the string integrand was determined only up to regular terms (multiplied by the usual Koba-Nielsen factor). These
ambiguities do not affect the leading behavior as $\alpha'\to 0$, which was successfully matched to the UV divergence of the respective supergravity integrands.

\sm

Recently, by combining the non-minimal pure spinor formalism with the chiral splitting formalism initially developed for the RNS formalism \cite{D'Hoker:1988ta, DHoker:1989cxq}, we obtained the full genus-two amplitude for five  arbitrary massless external states in Type II and Heterotic strings~\cite{DHoker:2020prr}. 
This result followed from two key requirements imposed on the amplitude, namely BRST invariance along with invariance under ``homology shifts", which consist of the combined action of taking one vertex point around a homology cycle on the genus-two surface, and shifting the corresponding loop momentum. 
It turns out that these requirements are strong enough to fix the chiral amplitude completely, given the operator product expansion (OPE) singularities between the canonical worldsheet fields. The full amplitude is obtained by assembling the chiral amplitudes for the left- and right-movers (or the  chiral amplitude with the Chan-Paton factors for open strings), and integrating over loop momenta, vertex points, and moduli of the genus-two surface. 

\sm

To leading order in the $\alpha'\to 0$ expansion,   the integral giving the string amplitude was shown to reproduce the kinematic numerators of the two-loop five-point supergravity diagrams, which were computed for four-dimensional ${\cal N}=8$ supergravity in \cite{Carrasco:2011mn} and for ten-dimensional Type II supergravity states in~\cite{Mafra:2015mja}. In a companion paper \cite{DMS3}, the genus-two  amplitude for five NS states will be derived from first principles within the RNS formalism.

\sm

In this paper, we shall use the results of \cite{DHoker:2020prr} as the starting point for a systematic analysis of the low energy expansion of the five-point amplitude beyond leading order. Such an analysis is part of a general endeavor
 to understand the structure of the low energy effective action in superstring theories both in perturbation theory and at the non-perturbative  level.  For Type~IIB superstring theory in 10-dimensional Minkowski space-time, S-duality allows one to make sharp and quantitative predictions of  non-perturbative contributions to certain protected couplings. Specifically, combining  perturbative results at tree-level and genus-one orders for the four-graviton scattering amplitude with requirements of space-time supersymmetry and S-duality invariance \cite{Green:1997tv, Green:1998by, Green:1999pu, Green:2005ba,Green:2014yxa}, the axion-dilaton dependence of the coefficients of the effective interactions of the form $\cR^4, D^4\cR^4$ and $D^6\cR^4$ were determined in terms of non-holomorphic modular functions of $SL(2,\ZZ)$. This has been accomplished not only in ten dimensions but also after compactification on a torus, in terms of certain automorphic functions of the U-duality group (see e.g.  \cite{Green:2010wi,Green:2010kv,Bossard:2020xod} and references therein).  
 \sm
 
The analytic structure of the genus-one four-graviton amplitude was established in \cite{DHoker:1994gnm} based on the moduli-space integrand in \cite{Green:1982sw}. Perturbative contributions to the effective interactions $\cR^4, D^4\cR^4$ and $D^6\cR^4$  were extracted and analyzed at  genus one in \cite{Green:1999pv,Green:2008uj,D'Hoker:2015foa}, and at genus two in \cite{DHoker:2005jhf, D'Hoker:2013eea,DHoker:2014oxd}, the analysis being extended up to order  $D^{8}\cR^4$ in \cite{DHoker:2017pvk,DHoker:2018mys}. The  integrand at a given order is a linear combination of ``modular graph functions" (MGFs), a class of real analytic modular functions which arise by integrating products of Green functions over the vertex points \cite{DHoker:2015wxz,DHoker:2017pvk}. However, while these perturbative contributions are under analytic control,  supersymmetry and S-duality no longer appear to determine the full automorphic forms under the S-duality group beyond $D^6 \cR^4$.  

\sm

For five-graviton scattering, the low energy expansion has so far been considered systematically at tree level \cite{Schlotterer:2012ny} and one loop \cite{Green:2013bza,Basu:2016mmk} only, while a preliminary analysis of the two-loop amplitude  at leading order was performed in \cite{Gomez:2015uha}.  A key result from the one-loop analysis in \cite{Green:2013bza} was that the five-point integrand at any order in $\alpha'$ can be expressed
as a linear combination of MGFs similar to the four-point case. Moreover, 
the very same linear combinations were found to govern the five-point $D^{2k}\cR^5$ and 
four-point $D^{2k+2}\cR^4$ interactions for $k=1,2$. Since the tree-level coefficients
are also identical -- namely $\zeta_5$ in case of $D^{4}\cR^4,D^{2}\cR^5$ and $\zeta_3^2$ in case of $D^{6}\cR^4,D^{4}\cR^5$ -- this suggests that both interactions
are related by non-linear supersymmetry and are multiplied by the same automorphic form.

\sm

For the $D^{2k}\cR^5$ and $D^{2k+2}\cR^4$ effective interactions at $k\geq 3$, by contrast, it was found \cite{Green:2013bza} that new  linear combinations of MGFs occur in the five-point amplitude, which 
indicates the presence of new supersymmetric invariants not present at tree level. The first example of this occurs for $k{=}3$, leading to a five-point effective interaction which we denote by $(D^6\cR^5)'$ to distinguish it from the $D^6\cR^5$ interaction   related by non-linear supersymmetry to $D^8\cR^4$.

\sm

Another key aspect of the one-loop analysis in \cite{Green:2013bza} was the study of amplitudes violating the $U(1)_R$ global symmetry of classical ten-dimensional  Type IIB supergravity:  due to a one-loop anomaly \cite{Gaberdiel:1998ui},  $n$-point string amplitudes may violate  the conservation of $U(1)_R$ charge by up to $\pm 2(n-4)$ units  (see e.g.\ \cite{Green:1999qt, Boels:2012zr, Green:2019rhz}).  At five points, this  violation occurs for  1-dilaton  4-graviton scattering, schematically denoted by $\phi \cR^4$, or 3-gravitons 2-Kalb-Ramond fields\footnote{By a slight abuse of nomenclature, we refer to the  complex combination of RR and NS two-form fields in Type IIB supergravity as the Kalb-Ramond field, and denote its 3-form field strength by $G$. In our conventions the dilaton fluctuation $\phi$ carries 2 units of $U(1)_R$-charge, $G$ carries one unit and $\cR$ is neutral. }, denoted by $G^2 \cR^3$, which are both maximally R-violating amplitudes in the language of \cite{Boels:2012zr}. In this case the automorphic form multiplying these interactions can no longer be invariant under S-duality, but must carry a modular weight so as to cancel the phase variation of the interaction vertex under S-duality. At low orders in
$\alpha'$, the analysis of \cite{Green:2013bza} indicates that the 
automorphic form for $U(1)_R$-violating interactions is related to the
automorphic function for the $U(1)_R$-preserving ones by a raising
operator (or modular derivative), which suggests that both interactions are
part of the same supersymmetric invariant. However, this correspondence 
breaks down for $k=5$, where a $U(1)_R$-violating interaction of the form
$D^{12} G^2 \cR^3$ arises which is not related to any $U(1)_R$-preserving
interaction of type $D^{10}\cR^5$.

\sm

In this paper, we analyze the first few orders in the low energy expansion of the genus-two 5-point amplitude of \cite{DHoker:2020prr}, for various choices of external massless states of Type IIB and IIA superstrings. In general, we find that, at each order, the integrand on genus-two moduli space is a linear combination of genus-two MGFs, a class of real-analytic Siegel modular functions which arise by integrating products of Arakelov Green functions (and partial derivatives thereof) against suitable top forms on multiple copies of the genus-two curve $\Sigma$ \cite{DHoker:2017pvk,DHoker:2018mys}. 
Quite remarkably, we find that the many MGFs occurring at order $D^6\cR^5$ (some of which previously considered in  \cite{Basu:2018bde}) can all be reduced to linear combinations of 5 basic ones $\cZ_1,\dots,\cZ_5$ defined in \eqref{defZs} below, along with the square $\f^2$ of the Kawazumi-Zhang invariant $\f$; the latter occurs in the four-point amplitude at order $D^6\cR^4$ \cite{D'Hoker:2013eea}, and reappears in the five-point amplitude at order $D^4\cR^5$. The graphs for the  relevant genus-two MGFs are presented in Figure~\ref{fig:1}. 

\sm

Moreover, we find that one of these six MGFs can be eliminated by virtue of a novel identity amongst five of them,
\bea
\label{MGIintro}
\cZ_1+\cZ_2 + \cZ_3 + \half \cZ_4 - \f^2=0
\eea
This identity is quite remarkable since it relates different graph topologies, and can be viewed  as a genus-two analogue of the identities between genus-one MGFs proven in 
\cite{DHoker:2015sve,DHoker:2016mwo,DHoker:2016quv,Basu:2016kli}. 
It would be interesting to revisit the analysis of the Laplace equation on genus-two modular graph functions in  \cite{Basu:2018bde} in view of the identity \eqref{MGIintro} and the simpler identities
 \eqref{nuint}.

\sm

In the non-separating degeneration limit, identity (\ref{MGIintro})  implies a novel identity \eqref{MGIell} for genus-one {\sl elliptic} MGFs\footnote{Elliptic MGFs are real-analytic  functions of $(\tau,v)$, which are doubly periodic in $v$ and modular invariant; they can be  obtained from the conventional MGFs of \cite{DHoker:2015wxz} by leaving one vertex position unintegrated, and have also been referred to as {\sl generalized} MGFs in \cite{DHoker:2018mys}.}, which suggests that the identites of \cite{DHoker:2015sve,DHoker:2016mwo,DHoker:2016quv,Basu:2016kli} may admit far reaching generalizations in the elliptic  and Siegel cases. The identity \eqref{MGIintro} is motivated by the analysis of degeneration limits in appendix \ref{sec:degen}, and derived in appendix \ref{sec:MGIproof} by exploiting a novel lemma \eqref{lemma}, which relates derivatives  $\partial_{z_i} \cG(z_i,z_j)$  and $\partial_{z_j} \cG(z_i,z_j)$ 
of the Arakelov Green function at arbitrary genus. Another interesting fact is that the 
MGF $\cZ_5$ involving two derivatives of Green functions tends to zero both in the separating and non-separating degenerations, unlike the others which diverge in both limits, so that it leaves no trace in the supergravity limit.

\sm 

The details of the string integrand on moduli space depend on the order in the expansion and the choice of external massless states of the Type IIB multiplet, as follows.
\begin{itemize}
\item
In the $U(1)_R$-preserving sector, at order $D^{2k}\cR^5$ with $k=1, 2$, we find the same integrand (namely the constant measure $d\mu_2$ on the Siegel upper half plane at order $D^2\cR^5$, and the Kawazumi-Zhang measure $\varphi \, d\mu_2$ at order $D^4\cR^5$) as for the four-point amplitude at order $D^{2k+2}\cR^4$, up to overall normalization. This supports the expectation
that the $D^2 \cR^5$ and $D^4 \cR^5$ interactions belong to the same non-linear supersymmetric invariant as the $D^4\cR^4$ and $D^6 \cR^4$ interactions, respectively, and should appear with the same automorphic coefficient in the low energy effective action, denoted by $\cE_{(1,0)}$ and $\cE_{(0,1)}$ in the standard fashion after \cite{Green:1999pv}.

\item In the $U(1)_R$-preserving sector,  at order  $D^6 \cR^5$, we find 
 two distinct kinematic structures, one identical to the tree-level interaction, and the
 other  identical to the genus-one $(D^6 \cR^5)'$ effective interaction. In the former case, the integrand
is proportional to the same combination $\cZ_1-2\cZ_2+\cZ_3$ of genus-two MGFs 
appearing at  order  $D^8 \cR^4$ in four-graviton scattering, with the correct coefficient relative to the tree-level and genus-one amplitude. This confirms that $D^8\cR^4$ and $D^6\cR^5$ belong to a single supersymmetric invariant, with an automorphic coefficient  $\cE_{(2,0)}$ receiving tree-level up to genus-two  contributions (and presumably higher genera as well). By contrast, the integrand for the genus-two $(D^6 \cR^5)'$ 
involves the new MGFs $\cZ_4, \cZ_5$ and $\varphi^2$ (one of which can be eliminated by virtue of \eqref{MGIintro}). Along with the genus-one amplitude computed in  \cite{Green:2013bza}, this predicts
the first two terms in the weak coupling expansion of a new automorphic coefficient  $\cE_{(2,0)'}$
which presumably also involves contributions of arbitrary genera.

\item 
In the $U(1)_R$-violating sector, at orders $\phi D^4\cR^4$ and $\phi D^6\cR^4$, we find the same integrand as in the
$U(1)_R$-preserving sector, up to a relative coefficient $-3/5$ and $-1/3$, respectively. As we explain in section \ref{sec:4}, this is consistent with linear supersymmetry and S-duality, which relate the ratio of coefficients of the $D^{2k}\cR^5$ and $\phi D^{2k+2}\cR^4$ at different loop orders by the action of a raising operator (or modular covariant derivative operator). At the next order, there are again two different kinematic structures $\phi D^8\cR^4$ and $(\phi D^8\cR^4)'$, as in the one-loop 5-point amplitude \cite{Green:2013bza}. For the first, the integrand is equal to the one for $D^6\cR^5$ up to a relative coefficient $1/7$, consistent with linear supersymmetry. For the second, there is no obvious relation between the $(D^6\cR^5)'$ and $(\phi D^8\cR^4)'$ integrands, except for the fact that they are both linear combinations of the same MGFs $\cZ_i, \varphi^2$ (subject to the relation \eqref{MGIintro}). By requiring that the integrated couplings be related by linear supersymmetry, we predict a relation between the divergent parts of the modular integrals on $\cM_2$, which we check against
the behavior of the integrand in the non-separating degeneration limit.

 \item 
 Extracting the supergravity limit of the 1-dilaton, 4-graviton amplitude in any dimension $D$, we confirm the absence of UV divergences in this sector, in agreement with the known structure of UV divergences in supergravity at two loops \cite{Bern:1998ug}. The consistency of the low energy expansion with supersymmetry and S-duality provides a strong check on the full five-point amplitude constructed in~\cite{DHoker:2020prr}. 

\end{itemize}

Before proceeding further, we make two important comments. First, the notation $D^{2k} \cR^5$ is a  moniker for the Taylor coefficient of order $p^{2k+10}$ in the momentum expansion of the 5-graviton amplitude; 
in general it includes both irreducible contributions from local interactions of the form  $D^{2k} \cR^5$
in the low energy effective action, where $\cR$ is the Riemann tensor and $D$ are covariant derivatives, with indices suitably contracted with the metric tensor, as well as reducible contributions from local interactions of the form $D^{2k+2} \cR^4$ and supergravity vertices.  We do not attempt to disentangle these various contributions at two loops, but rather express the kinematic dependence of the Taylor coefficients at two loops in terms of tensorial quantities appearing at tree level or one loop; the procedure for subtracting reducible diagrams is then identical to the one required at these lower orders (see e.g.\ \cite{Richards:2008jg} at genus one).  The same holds for the notation $\phi D^{2k+2} \cR^4$, which is a moniker for the Taylor coefficient of order $p^{2k+10}$ in the momentum expansion of the 1-dilaton 4-graviton amplitude. Note that the constraints of S-duality on the low energy effective action
translate directly into constrains on the corresponding Taylor coefficients in the amplitudes \cite{Green:2013bza}.

\sm

The second comment is that in certain space-time dimensions $D$ correlated with the order in the $\alpha'$ expansion, these local effective interactions can mix with non-local interactions mediated by massless particles. In such cases a sliding scale must be specified to separate these effects \cite{Green:2010sp,Pioline:2018pso}. This is in particular the case for the $D^8\cR^4$ and $D^6\cR^5$ interactions in $D=10$. Since we are mostly interested in the integrand, we shall mostly ignore these issues in this paper, except at
some places in  sections \ref{secAp} and \ref{sec:4}.

\subsection*{Organization}

The remainder of this paper is organized as follows. In section~\ref{sec:2} we review the necessary results from paper
\cite{DHoker:2020prr} on the structure of the genus-two amplitude for five external massless states, and give simplified
effective rules to extract the contribution from bosonic external states. In section~\ref{sec:3} we decompose the genus-two five-point amplitude into a sum of products of kinematic factors times integrals in the vertex points on the genus-two Riemann surface, perform the $\alpha'$ expansion of these integrals up to orders high enough to access the effective interactions of order $D^6 \cR^5$, and prove the above-mentioned identity between genus-two MGFs. In section~\ref{secAp}, we extract the actual effective interactions up to order $D^6 \cR^4$, and present simplified concrete formulas for the separate cases of Type IIA and Type IIB superstrings. In section~\ref{sec:4} we compare our perturbative results with predictions from S-duality and from the structure of UV divergences in supergravity. An overview of the function theory on Riemann surfaces of genus two is presented in appendix~\ref{sec:func}; the detailed calculations of the $\alpha '$ expansion of the genus-two integrals is given in section~\ref{app:exp}; the analysis of the non-separating, separating, and tropical degenerations of the integrals is given in appendix~\ref{sec:degen}; the identity \eqref{MGIintro} is proved in appendix~\ref{sec:MGIproof} and details on the overall
normalization of the genus-two amplitude are given in appendix~\ref{app:convert}.

\subsection*{Acknowledgments}

The research of ED is supported in part by NSF grant PHY-19-14412. BP and OS are grateful to UCLA and the Mani Bhaumik Institute for kind hospitality and creating a stimulating atmosphere during initial stages of this work. CRM is supported by a University Research Fellowship from the Royal Society. OS is grateful to AEI Potsdam for kind hospitality during final stages of this work.
Moreover, OS is supported by the European Research Council under ERC-STG-804286 UNISCAMP. 

\newpage

\section{Review of the four-  and five-point amplitudes}
\setcounter{equation}{0}
\label{sec:2}

In this section, we review the structure of the genus-two chiral superstring amplitude for five massless states, as well as the physical amplitude in Type II string theory obtained by pairing left and right chiral amplitudes constructed in \cite{DHoker:2020prr}. For comparison we also include the genus-two amplitude for four massless NS states, first computed in the RNS formalism in \cite{D'Hoker:2005jc,DHoker:2005jhf,DHoker:2005dys} (based on the genus-two measure constructed in \cite{DHoker:2001kkt,DHoker:2001qqx,DHoker:2001foj,DHoker:2001jaf} which was re-derived using methods of algebraic geometry in \cite{Witten:2013tpa}), and reproduced in the PS formalism and extended to include external fermions in \cite{Berkovits:2005df,Berkovits:2005ng,Gomez:2010ad}. 
Finally, we shall present a set of effective rules to extract the massless Neveu-Schwarz content of the pure spinor building blocks. These rules will allow us to re-express the results of \cite{DHoker:2020prr}, and of section \ref{secAp} of this paper, in terms of the familiar $t_8$ and $\epsilon_{10}$ tensors and thereby facilitate the comparison with the RNS genus-two computation in \cite{DMS3}.

\subsection{Chiral Splitting}

The construction of the full integrand  in \cite{DHoker:2020prr} hinges on chiral splitting \cite{D'Hoker:1988ta, DHoker:1989cxq}, which allows us to decompose the integrand of the amplitude at fixed loop momentum into the product of chiral and anti-chiral amplitudes, associated to the left- and right-movers, respectively, \footnote{Throughout we denote $\delta(k) = ( 2 \pi)^{10} \delta ^{(10)} (\sum _i k_i)$ where $k_i$ are the momenta of the external states. }
\bea
\label{pairing}
\cA^{\rm genus-2}_{(N)} = \delta(k)\, \cN_{(N)} \,
\int _{\cM_2} |d^3 \Omega|^2   \int _{\Sigma^N} 
\int _{\RR^{20}} dp  \, 
\cF_{(N)} (z_i,k_i,p^I) \, \overline{ \tilde \cF_{(N)} (z_i, - \bar k_i, -p^I) } 
\eea
Here, $\cM_2$ is a fundamental domain in the rank 2 Siegel upper-half space, which may be parametrized locally by the period matrix $\Omega_{IJ}$ and $d^3 \Omega=d\Omega_{11}d\Omega_{12}d\Omega_{22}$ is the holomorphic top form on $\cM_2$. The loop momenta for genus two are $p^I=(p^1,p^2)$ with $p^1, p^2 \in \RR^{10}$ and the volume form for the integration over loop momenta is $dp = d^{10}p^1 \, d^{10}p^2$.  The chiral and anti-chiral amplitudes may be further decomposed as follows,
\bea
\label{FK}
\cF_{(N)} & = & \langle\cK_{(N)}  \rangle_0 \, \cI_{(N)} \ ,
\hskip 0.6in
\tilde \cF_{(N)}= \langle \tilde \cK_{(N)}  \rangle_0 \, \cI_{(N)}
\eea
where $\langle\cK_{(N)}\rangle_0$ and $\langle \tilde \cK_{(N)}\rangle_0$ are the left- and right-moving chiral correlators,
which will be discussed in detail in subsection \ref{sec:22}, and $ \cI_{(N)}$ is the chiral Koba-Nielsen factor. Finally, the
prefactor $\cN_{(N)}$ is a normalization factor, which will include the dependence on the dilaton vacuum expectation value, and which we shall fix in section \ref{secAp}.

\sm

The chiral Koba-Nielsen factor depends on the positions of the vertex operators $z_i$, the external momenta $k_i$ and the loop
momenta $p^I$ and is given by the following universal formula, independently of the particular string theory under
consideration,\footnote{Our conventions will follow those of appendix B in \cite{DHoker:2020prr} and are summarized in appendix \ref{sec:func} of this paper. In particular, we adopt the Einstein summation conventions for repeated indices $I,J,\ldots=1,2$ and often  abbreviate the point $z_i$, as an argument of a function, simply by $i$, for example in $\Delta(i,j)=\Delta(z_i,z_j)$ below.}
\bea
\cI_{(N)}(z_i,k_i,p^I) = \exp \left \{  i \pi  \Omega _{IJ} p ^I  \cdot p ^J 
+   \sum_{i=1}^N  2 \pi i p^I \cdot k_i  \int ^{z_i} _{z_0} \om _I  - \sum_{i<j}^N s_{ij} \ln E(z_i,z_j)\right \}
\label{KNN}
\eea
where $\om_I$ are holomorphic Abelian differentials, $\Omega_{IJ}$ are the components of the period matrix, and $E$ is the prime form. The dimensionless kinematic variables $s_{ij}$ are defined by, 
\bea 
s_{ij}=- \frac{ \alpha' }{2}k_i\cdot k_j
\eea
The  chiral Koba-Nielsen factor $\cI_{(N)}$, as well as the full chiral amplitude $\cF_{(N)}$,  enjoy two fundamental properties \cite{D'Hoker:1988ta, DHoker:1989cxq}: they are locally holomorphic in $z_i$ and $\Omega_{IJ}$ and are invariant under combined shifts of the points $z_i$ by homology cycles $\mA_J,\mB_J$, multiplication by a phase, and a shift in loop momenta, given as follows for $\cI_{(N)}$,
\bea
\label{Bmon1}
\cI_{(N)}(z_i,k_i,p^I) 
& =  & e^{- 2 \pi i p_J \cdot k_j} \, \cI_{(N)} (z_i + \delta_{ij} \mA_J , k_i, p^I) 
\no \\
\cI_{(N)}(z_i,k_i,p^I)  
& = &
 \cI_{(N)} (z_i+ \delta_{ij} \mB_J ,  k_i, p^I -  \delta^I_J\,  k_j ) 
\eea
We refer to these combined transformations as {\sl homology shifts}. The complex conjugate of the anti-chiral amplitude $\tilde \cF_{(N)}$ satisfies the above homology shift invariance with inverse phase factor. 
As a result, the integral over loop momenta of the product of chiral and anti-chiral amplitudes is  single-valued in each $z_i$ and produces a well-defined integral over~$\Sigma^N$.

\subsection{The chiral correlator}
\label{sec:22}

The chiral correlator $ \< \cK_{(N)} \>_0 $ depends on the same data as $\cI_{(N)}$, along with the left-moving polarization
vectors $\ep^m_i$ and spinors $\chi^\alpha_i$ describing the external states of the ten-dimensional super-Yang--Mills (SYM)
multiplet. The function $\cK_{(N)}$ further depends on the zero modes of the spinor fields $\theta ^\a,\lambda ^\a$ (subject to the pure spinor constraint $\lambda \gamma ^m \lambda=0$) and may be thought of as a {\sl superfield}. The bracket $\< \cdot
\>_0$ picks up  the coefficient of $(\l\gamma^m\theta)(\l\gamma^n\theta)(\l\gamma^p\theta)(\theta\gamma_{mnp}\theta)$ from $\cK_{(N)}$ in the cohomology of the left-moving BRST
charge \cite{Berkovits:2000fe,Berkovits:2006ik}. It will often be convenient to manipulate the full superfield $\cK_{(N)}$
rather than its component $\< \cK_{(N)}\>_0$ and, by a slight abuse of notation, we shall refer to both as chiral correlators. 

\sm

The chiral correlator $\cK_{(N)}$ is a locally holomorphic $(1,0)$ form in each vertex point~$z_i$, and is invariant under homology shifts but, in contrast to $\cI_{(N)}$ and $\cF_{(N)}$, without phase factors, 
\bea
\label{Bmon2}
\cK_{(N)}(z_i,k_i,p^I)  & =&   \cK_{(N)} (z_i + \delta_{ij} \mA_J , k_i,p^I)  
 \no \\
 \cK_{(N)}(z_i,k_i,p^I)  & = &
 \cK_{(N)} (z_i+ \delta_{ij} \mB_J ,  k_i, p^I -  \delta^I_J\,  k_j ) 
\eea
The anti-chiral correlator  $\langle \tilde \cK_{(N)} \rangle_0$ is expressed analogously in terms of the right-moving 
polarization vectors $\tilde \ep^m_i$,  and right-moving spinors $\tilde \chi^\alpha_i$ for the Type II strings or the right-moving gauge data for Heterotic strings. The corresponding superfield $\tilde \cK_{(N)}$ additionally depends on 
the zero modes of the right-moving spinor fields  $\tilde\theta^\alpha, \tilde\lambda^\alpha$. As usual, the left- and right-moving Weyl spinors $\theta^\alpha, \lambda^\alpha$ and $\tilde\theta^\alpha, \tilde\lambda^\alpha$ have the same chirality for Type IIB strings, or opposite chirality for Type IIA strings.

\sm

The chiral correlator  $\cK_{(4)}$ is independent of loop momenta, and given by \cite{Berkovits:2005df},
\be
\label{defK4}
\cK_{(4)} = T_{1,2|3,4} \, \Delta(4,1)\, \Delta(2,3) +  T_{1,4|2,3} \, \Delta(1,2)\, \Delta(3,4)
\ee
where $\Delta(x,y)=-\Delta(y,x)$ is the standard bi-holomorphic one-form  (see appendix~\ref{sec:func}), 
and the superfield $T_{1,2|3,4}$ is a function of the momenta $k^m_i$, polarization vectors $\ep^m_i$,  spinors $\chi^\alpha_i$, and the zero modes of $\theta^\alpha$ and $\lambda^\alpha$. The anti-chiral correlator 
$\tilde\cK_{(4)}$ is given by the same formula, with $T_{1,2|3,4}$ replaced by $\tilde T_{1,2|3,4}$
which depends on $k^m_i$, $\tilde \ep^m_i$, $\tilde \chi^\alpha_i$, $\tilde \theta^\alpha$ and $\tilde \lambda^\alpha$.

\sm

The chiral correlator $ \cK_{(5)}$ and its counterpart $ \tilde \cK_{(5)}$ for Type II strings were shown in  \cite{DHoker:2020prr} to be linear in the loop momenta $p^I$, and were decomposed as follows,
\bea
\label{K5tW}
\cK_{(5)} & = & \cW + 2 \pi i  \, \hat p^I _m \cV^m _I 
\no \\
\tilde \cK_{(5)} & = & \tilde \cW + 2 \pi i \, \hat p^I_m \tilde \cV_I^m
\eea
where $\hat p^I$ is the shifted loop momentum defined by,
\bea 
\hat p^I = p^I + Y^{IJ} \sum _{i=1}^5 k_i  \, \Im \int ^{z_i} _{z_0} \om_J 
\eea
with $Y^{IJ}$ the inverse of the imaginary part $Y_{IJ}= \Im \Omega _{IJ}$ of the period matrix $\Omega$.

\sm

Several equivalent representations of the chiral correlator $\cK_{(5)}$ were given in sections 5 and 6 of \cite{DHoker:2020prr}, each one manifesting different properties of the integrand in (\ref{pairing}). The representation in terms of superspace building blocks $T^m_{1,2,3|4,5}$ and $S_{1;2|3|4,5}$, to be reviewed below, is given by,\footnote{For reasons to become clear in section~\ref{secAp}, we have restored a factor of ${\ap\over2}$ in order to match with the conventions of
\cite{Gomez:2015uha}, see e.g. (5.40) of that reference.}
\bea
\label{defcv}
\cV^m_I & =&  T^m_{1,2,3|4,5} \, \om_I(2) \Delta (3,4)\Delta (5,1)  + \hbox{ cycl}(1,2,3,4,5)
\no \\
\cW &=&  \halfap{}{\cal Q}_{12} + (1,2|1,2,3,4,5)
\eea
The notation $+\hbox{ cycl}(1,2,3,4,5)$ stands for the addition of all cyclic permutations, while $+(i,j|1,2,3,4,5)$ stands for the addition  of all ordered choices of $i$ and $j$ from the set $\{1,2,3,4,5\}$ for a total of ${5\choose 2}=10$ terms. The function $\cQ_{12}$ is given by,
 \bea
 {\cal Q}_{12} &= & - \partial_1 \cG(1,2) \big[ S_{1;2|3|4,5} \Delta(2,4) \Delta(3,5) 
 + S_{1;2|4|3,5} \Delta(2,3) \Delta(4,5) \big]
 \no \\ && -  \partial_2 \cG(2,1) \big[ S_{2;1|3|4,5} \Delta(1,4) \Delta(3,5) 
 + S_{2;1|4|3,5} \Delta(1,3) \Delta(4,5) \big] 
 \label{altdefcw.2}
\eea
where $\cG(i,j)=\cG(z_i,z_j)$ is the Arakelov Green function (see appendix \ref{sec:arakapp} or \cite[\S 2.4]{DHoker:2017pvk}).

\sm

While the expression \eqref{altdefcw.2} is compact, it does not optimally expose the singularities
of the correlator at coincident vertex positions $z_1 \to z_2$. This is achieved by 
the alternative representation,
\bea
\label{defcw.2}
\cQ_{12} &= &
-\p_1\cG(1,2) \big[  T_{12,3|4,5} \Delta(2,4)\Delta(3,5) + T_{12,4|3,5} \Delta(2,3)\Delta(4,5)\big] 
 \no \\
&&
- S_{2;1|3|4,5}\big[\p_1\cG(1,2) \Delta(2,4)\Delta(3,5) + \p_2\cG(2,1)\Delta(1,4)\Delta(3,5)\big]
\no \\
&& 
- S_{2;1|4|3,5}\big[\p_1\cG(1,2) \Delta(2,3)\Delta(4,5) +
\p_2\cG(2,1)\Delta(1,3)\Delta(4,5)\big] 
\eea
where the  singularity as $z_1\to z_2$ is contained entirely in the first line, while the second and third lines are
manifestly regular due to the cancellation of the poles from $\p_1\cG(1,2)$ and $\p_2\cG(1,2)$. 
In particular,  \eqref{defcw.2} makes it manifest that the residues of kinematic poles
in the integrated amplitude will only feature permutations of $|T_{12,3|4,5}|^2$.

\sm

When discussing the difference between Type IIA and Type IIB amplitudes in sections~\ref{sec.IIB} and~\ref{sec.IIA}, a third representation of the correlator will become convenient, given in terms of 
\bea
\widehat \cV^m_I & =&  C^m_{1,2,3|4,5} \, \om_I(2) \Delta (3,4)\Delta (5,1)  + \hbox{ cycl}(1,2,3,4,5)
\label{defcvCm}
\no\\
\widehat\cW &=&  \halfap{} \widehat{\cal Q}_{12} + (1,2|1,2,3,4,5)
\eea
with
\bea
\widehat {\cal Q}_{12} &= & - s_{12}\partial_1 \cG(1,2) \big[C_{1;2|3|4,5} \Delta(2,4) \Delta(3,5)
 + C_{1;2|4|3,5} \Delta(2,3) \Delta(4,5) \big]
 \no \\
 && -  s_{12}\partial_2 \cG(2,1) \big[ C_{2;1|3|4,5} \Delta(1,4) \Delta(3,5) 
 + C_{2;1|4|3,5} \Delta(1,3) \Delta(4,5) \big]
 \label{hatcw.2}
\eea
Here,  the superfields $C^m_{1,2,3|4,5}$ and $C_{1;2|3|4,5}$ are non-local, but manifestly BRST-closed, building blocks to be described below.  The correlators of \eqref{defcvCm} can be shown to be equivalent to \eqref{defcv} after 
substituting the relations to be given below in \eqref{BRST.2} and discarding total derivatives\footnote{The correlators of \eqref{defcv} and \eqref{defcvCm} may be formally related by the substitution rule
$T^m_{1,2,3|4,5} \rightarrow C^m_{1,2,3|4,5}$  and $S_{1;2|3|4,5} \rightarrow s_{12} C_{1;2|3|4,5}$. This rule mimics similar manipulations observed at one loop \cite{Mafra:2014gsa}.}.
Note that, in both of these representations, $\cQ_{12}$ and $\widehat \cQ_{12}$ are totally symmetric in the omitted
labels $3,4,5$ due to the symmetries of the building blocks as well as
$\Delta(2,3)\Delta(4,5) + {\rm cyc}(3,4,5) = 0$. 

\sm

Similar expressions are valid for the right-moving parts $ \tilde \cW$ and $\tilde \cV$,
with $T^m_{1,2,3|4,5}$ and $S_{1;2|3|4,5}$ replaced by their counterparts
$\tilde T^m_{1,2,3|4,5}$ and $\tilde S_{1;2|3|4,5}$ depending
on the zero modes of $\tilde\theta^\alpha$ and $\tilde\lambda^\alpha$ (with the usual chirality flip for Type IIA).

\subsection{Scalar and vector superspace building blocks}
\label{sec:BBs}

To complete the definition of the integrands, it remains to specify the superspace constituents
referred to above as ``building blocks''. These are kinematic expressions  in pure spinor superspace,
constructed 
using the multiparticle formalism of the standard superfields of ten-dimensional SYM \cite{Mafra:2014oia}.

\subsubsection{Local building blocks}
\label{sec:Ts}

The four-point scalar block $T_{1,2|3,4}$ was constructed in \cite{Berkovits:2005df, Gomez:2010ad} and satisfies, 
\bea
\label{symmT}
Q T_{1,2|3,4} &=& 0 
\no \\
T_{1,2|3,4} &=& T_{2,1|3,4}=T_{3,4|1,2}
\no \\
T_{1,2|3,4} & = &-T_{1,3|4,2}-T_{1,4|2,3} 
\eea
where $Q= \lambda^\alpha D_\alpha$ is the BRST operator of the pure spinor formalism \cite{Berkovits:2000fe} with,
\bea
D_\alpha = {\partial \over \partial \theta^\alpha} + {1\over 2}(\gamma^m \theta)_\alpha  
{\partial \over \partial x^m}
\eea
The derivative with respect to $x^m$ acts on the plane-wave factor $e^{ik\cdot x}$ of each superfield to produce a factor
of $i k_m$.  The properties (\ref{symmT}) along with the antisymmetry of $\Delta(i,j)$ ensure the invariance of \eqref{defK4}
under permutations of the 4 external states. 

\sm

The five-point vector block $T^m_{1,2,3|4,5}$ was constructed in \cite{Mafra:2015mja} so as to satisfy,
\begin{align}
\label{mTQBRST}
Q T^m _{1,2,3|4,5}  &= i k^m_1 V_1 \, T_{2,3|4,5} + i k_2^m V_2 \, T_{3,1|4,5} + i k^m _3 V_3 \, T_{1,2|4,5}
\end{align}
as well as the following symmetry relations,
\begin{align}
\label{mTsym}
T^m _{1,2,3|4,5} &= T^m_{3,4,5|1,2} +  T^m_{2,4,5|1,3} +  T^m_{1,4,5|2,3}\\
T^m _{1,2,3|4,5} &= T^m _{1,3,2|4,5} = T^m _{2,1,3|4,5} = T^m _{1,2,3|5,4}\no
\end{align}
where $V_i$ are the BRST-closed one-particle unintegrated vertex operators. The relations \eqref{mTsym}
ensure that $\cV^m_I$ in \eqref{defcv}  is invariant under permutations of the five external legs.

In addition, a scalar superfield $T_{12,3|4,5}$ was constructed in \cite{Mafra:2015mja} using two-particle superfields obeying,
\begin{equation}
\label{QTij}
QT_{12,3|4,5} = s_{12}(V_1T_{2,3|4,5}-V_2T_{1,3|4,5})
\end{equation}
as well as $T_{12,3|4,5}=T_{12,3|5,4}$ and the ``Jacobi'' symmetry,
\begin{equation}
\label{Tijsym}
T_{12,3|4,5} + T_{12,4|5,3} + T_{12,5|3,4} = 0
\end{equation}
Finally, the five-point scalar blocks in \eqref{altdefcw.2} are given by \cite{DHoker:2020prr},
\bea
S_{1;2|3|4,5} =
{1\over 2} \Big (  i (k_1^m{+}k_2^m{-} k_3^m) T^m_{1,2,3|4,5} + T_{12,3|4,5}+T_{13,2|4,5}+T_{23,1|4,5} \Big ) 
\label{multi.5}
\eea
and satisfy,
\begin{equation}
Q S_{1;2|3|4,5}  =  s_{12} V_1 T_{2,3|4,5}\,,\quad
S_{1;2|3|4,5} = S_{1;2|3|5,4}\,,\quad
T_{12,3|4,5}  = S_{1;2|3|4,5}-S_{2;1|3|4,5}
\end{equation}
Furthermore, we have the following relations between permutations of (\ref{multi.5}),
\bea
\label{multi.10}
S_{1;2|3|4,5}+ S_{1;2|4|5,3}+ S_{1;2|5|3,4} &\cong& 0
\no \\
S_{1;2|3|4,5}+S_{1;3|2|4,5}+S_{1;4|5|2,3}+S_{1;5|4|2,3}  &\cong &0
 \eea
where $\cong$ denotes an equality in the BRST cohomology. Importantly, the bosonic components of the
vector building blocks $T^m_{1,2,3|4,5}$ are proportional to $k^6 \ep^5$ while those of the scalar blocks $T_{12,3|4,5}$ and $S_{1;2|3|4,5}$ are proportional to $k^7\ep^5$, where $\ep$ represents the SYM polarization vector. As a consequence, gravitational components of $T^m_{1,2,3|4,5}\tilde T^m_{1,2,3|4,5}$ and $T_{12,3|4,5} \tilde T_{12,3|4,5} /k_1\cdot k_2$ have the mass dimension of $D^2 \cR^5$.

\subsubsection{Non-local building blocks}
\label{sec:Cs}

Besides the above building blocks, which are polynomials in external momenta, it will be useful to introduce 
the non-local combinations introduced in section 5.4 of \cite{DHoker:2020prr},
\begin{align}
\label{BRST.2} 
 C_{1;3|4|2,5} &= {1\over 4} \Big( \frac{ 3S_{1;3|4|2,5} }{s_{13}} -\frac{ S_{1;4|3|2,5}  }{s_{14}}
 - \frac{S_{1;2|5|3,4} }{s_{12}}  - \frac{ S_{1;5|2|3,4}  }{s_{15}}\Big)
\no \\
C^m_{5,1,2|3,4} &=  T^m_{5,1,2|3,4} - {i\over 4} k_1^m \Big( \frac{ S_{1;2|5|3,4} }{s_{12}}
+ \frac{ S_{1;5|2|3,4}   }{s_{15}}+ \frac{ S_{1;3|4|2,5}   }{s_{13}}
+ \frac{ S_{1;4|3|2,5}  }{s_{14}}\Big)
\no \\
& \hskip 0.75in
- {i\over 4} k_2^m \Big( \frac{S_{2;1|5|3,4} }{s_{12}}  +\frac{ S_{2;5|1|3,4}  }{s_{25}}
+ \frac{S_{2;3|4|1,5} }{s_{23}} + \frac{S_{2;4|3|1,5}  }{s_{24}} \Big) 
\no \\
& \hskip 0.75in
 - {i\over 4} k_5^m \Big( \frac{S_{5;1|2|3,4}  }{s_{15}}+ \frac{S_{5;2|1|3,4}  }{s_{25}}
 + \frac{S_{5;3|4|1,2}  }{s_{35}}+ \frac{S_{5;4|3|1,2} }{s_{45}} \Big)
\end{align}
that are manifestly BRST invariant
\bea
\label{BRST.3}
QC^m_{5,1,2|3,4} & = & 0
\no \\
QC_{1;3|4|2,5}& = & 0
\eea
In addition, they satisfy the following relations \cite{DHoker:2020prr},
\begin{align}
\label{BRST.4} 
ik_2^m C^m_{5,1,2|3,4} &= s_{12} C_{1;2|5|3,4}  +  s_{25} C_{5;2|1|3,4} 
\no \\
ik_3^m C^m_{5,1,2|3,4} &\cong s_{13} C_{1;3|4|2,5}  +
 s_{23} C_{2;3|4|1,5}  +   s_{35} C_{5;3|4|1,2}  
  \no \\
0&\cong s_{12} C_{2;1|5|3,4}  +  s_{25} C_{2;5|1|3,4}  +  s_{23} C_{2;3|4|1,5}  +  s_{24} C_{2;4|3|1,5} 
\no \\
0&\cong C_{2;1|5|3,4}+C_{2;1|4|5,3}+C_{2;1|3|4,5}
\no \\
0 &= C_{2;1|5|3,4} - C_{2;1|5|4,3}
\end{align}
Importantly, the  invariants $C^m_{1,2,3|4,5}$ and $C_{1;2|3|4,5}$, which we call ``two-loop BRST invariants'', can be rewritten in terms of similar BRST invariants $C^m_{1|2,3,4,5}$ and $C_{1|23,4,5}$ \cite{Mafra:2014oia, Mafra:2014gsa} (the ``one-loop BRST invariants'') which occur\footnote{Note that in \cite{Green:2013bza} the object
called $C^m_{1,2,3,4,5}\tilde C^m_{1,2,3,4,5}$ is a shorthand for the leading-order contributions from the correlator
and should not be confused with the holomorphic square of $C^m_{1|2,3,4,5}$.} in the integrand
of the one-loop five-point amplitude \cite{Green:2013bza,Mafra:2016nwr}.
Using the components $\langle C^m_{1|2,3,4,5} \rangle_0$ and $\langle C_{1|23,4,5} \rangle_0$ available for download 
from \cite{PSSsite}, one finds \cite{DHoker:2020prr},
\bea
C^m_{1,2,3|4,5} &\cong &
 - 16 s_{45} C^m_{1|2,3,4,5}
+ 8(k_4^m - k_5^m) s_{45} C_{1|45,2,3}\label{BRST.11} 
\nonumber \\
&&+4 k_2^m \big( s_{45} (C_{1|24,3,5} + C_{1|25,3,4}) + (
s_{13}+s_{23}) C_{1|23,4,5} \big)
\no \\
&&+ 4k_3^m\big( s_{45} (C_{1|34,2,5} + C_{1|35,2,4})  -
(s_{12}+s_{23}) C_{1|23,4,5}) \big)
\no\\
&&- 4(k_1^m + k_2^m + k_3^m)\big( s_{24} C_{1|24,3,5} + s_{25}
C_{1|25,3,4} + (2\leftrightarrow 3) \big)
\eea
and
\begin{align}
C_{1;2|3|4,5} &\cong 4 \big(
s_{24} C_{1|24,3,5} + s_{25} C_{1|25,3,4}
+s_{34} C_{1|34,2,5} + s_{35} C_{1|35,2,4}
 - 2 s_{23} C_{1|23,4,5}   \big)
   \label{BRST.12}
\end{align}
In turn, the components of the  one-loop BRST invariants can be expressed as 
combinations of color-ordered tree amplitudes\footnote{We have $k_j \rightarrow ik_j$ 
and different conventions for $s_{ij}$ in comparison to the definitions in \cite{Mafra:2014gsa}.}
\cite{Mafra:2014oia, Mafra:2014gsa},
\begin{align}
\langle C_{1|23,4,5} \rangle_0 &= s_{45} \big[ s_{34} A_{\rm YM}(1,2,3,4,5) 
- s_{24} A_{\rm YM}(1,3,2,4,5)  \big]
\notag \\
0&= \langle ik_2^m C^m_{1|2,3,4,5} +\big[ s_{23} C_{1|23,4,5} +(3\leftrightarrow 4,5) \big] \rangle_0
\label{BRST.13}
\end{align}
These relations will become useful in section \ref{sec:4} 
when comparing our two-loop results with one-loop
and tree-level amplitudes.

\subsection{Effective rules for bosonic components}
\label{sec:eff}

The bosonic components\footnote{With the techniques of \cite{Mafra:2010pn} to perform the zero-mode integrals over $\lambda^\alpha,\theta^\alpha$, one can obtain direct access to the polarization
dependence of the five-point amplitudes in string and field-theory  for any combination of external bosons and fermions.} of the building blocks $T^m_{1,2,3|4,5}$ and $T_{12,3|4,5}$ in pure spinor
superspace are available for download from the website \cite{PSSsite}.
However, the expressions from  \cite{PSSsite} involve unpleasant rational factors such as   $\frac{13}{7}$ within individual $\langle T^m_{1,2,3|4,5} \rangle_0$ or $\langle T_{12,3|4,5} \rangle_0$, which drop out from BRST invariants. These factors come from an implicit choice of contact terms, which is far from being canonical nor optimal. 

\sm

In order to streamline the expressions for the bosonic components of the local building blocks
and facilitate the comparison with the RNS computation \cite{DMS3},
we shall now give an alternative description of the correlators in \cite{DHoker:2020prr}. The key quantities
are the {\sl effective components} $T^{m, {\rm eff}}_{1,2,3|4,5},T^{\rm eff}_{12,3|4,5}$ and 
$S^{\rm eff}_{1;2|3|4,5}$ defined by\footnote{We are grateful to Alex Edison and Fei Teng 
for discussions that
led to correcting the coefficients of $\epsilon_{10}^m$ in (\ref{Teff.5}) and $t_8$ in (\ref{typeIIA.5})
by factors of 8 in the current v3. Follow-up changes are limited to
section \ref{sec.IIA} and footnote \ref{foophihfour}, i.e.\ the
main results including S-duality checks are unaffected.}
\bea
\label{Teff.5} 
T^{m, {\rm eff}}_{1,2,3|4,5} &=& 8 (k_4\cdot k_5) \big[ \ep^m_1 t_8(f_2,f_3,f_4,f_5)  + (1\leftrightarrow 2,3,4,5) \big] 
\notag \\
&&+ 4i \big[  k_1^m (\RRR_{1;2|3,4,5} + \RRR_{1;3|2,4,5})+ (1\leftrightarrow 2,3)\big] 
\notag \\
&&+ 8 ik_4^m \RRR_{4;5|1,2,3} + 8 ik_5^m \RRR_{5;4|1,2,3} - 
 \frac{1}{2}  (k_4\cdot k_5) \epsilon_{10}^m(\ep_1,f_2,f_3,f_4,f_5)
\notag \\
T_{12,3|4,5}^{\rm eff} &=& (8k_4\cdot k_5 - 4 k_1\cdot k_2) (\RRR_{1;2|3,4,5} - \RRR_{2;1|3,4,5}) + 4k_1 \cdot k_2( \RRR_{2;3|1,4,5} - \RRR_{1;3|2,4,5}) 
\no \\
S^{\rm eff}_{1;2|3|4,5} 
&= & (8k_4\cdot k_5 - 4 k_1\cdot k_2) \RRR_{1;2|3,4,5} - 4k_1\cdot k_2 \RRR_{1;3|2,4,5}
\notag \\
&&+8 (k_3\cdot k_4 \RRR_{4;5|1,2,3} - k_4\cdot k_5 \RRR_{4;3|1,2,5})
\notag \\
&&+8 (k_3\cdot k_5 \RRR_{5;4|1,2,3} - k_4\cdot k_5 \RRR_{5;3|1,2,4})
\eea
which are composed of
\begin{align}
\RRR_{1;2|3,4,5} &= i(\ep_1\cdot k_2) t_8(f_2,f_3,f_4,f_5) - \frac{i}{2} t_8([f_1,f_2],f_3,f_4,f_5)
\notag \\
\label{famous}
t_8(  f_2,  f_3,  f_4,   f_5) &=  {\rm tr}(f_2 f_3 f_4 f_5) - \frac{1}{4}  {\rm tr}(f_2 f_3) {\rm tr}(f_4 f_5)
+ {\rm cyc}(3,4,5)
\end{align}
with Lorentz traces ${\rm tr}(\ldots)$, linearized field strength $f^{mn}_j = \ep^m_j k^n_j - \ep_j^n k^m_j$ 
and its commutators
$[f_1,f_2]^{mn}=  f_1^{mp}f^{pn}_2 - f_2^{mp}f^{pn}_1$. As will be explained below, the
bosonic components of the two-loop five-point amplitude are unchanged when
performing the replacement
\bea
 T^{m}_{1,2,3|4,5}    \rightarrow
T^{m, {\rm eff}}_{1,2,3|4,5} 
\, , \ \ \ \ \ \
 T_{12,3|4,5}   \rightarrow 
T^{\rm eff}_{12,3|4,5} 
\, , \ \ \ \ \ \
  S_{1;2|3|4,5}  \rightarrow 
S^{\rm eff}_{1;2|3|4,5} 
\label{rplce}
\eea
in all terms of the correlator (\ref{K5tW}) and dropping the zero-mode brackets $\langle \ldots \rangle_0$
in the chiral amplitude (\ref{FK}).

\subsubsection{Symmetries and relations of the effective components}

The effective replacement rules (\ref{rplce}) are well-defined at the level of  $\langle {\cal K}_{(5)} \rangle_0$ since all of $T^{m, {\rm eff}}_{1,2,3|4,5} ,T^{\rm eff}_{12,3|4,5} ,S^{\rm eff}_{1;2|3|4,5}$ given by (\ref{Teff.5})
inherit the symmetry relations of the superfields $T^{m}_{1,2,3|4,5} ,T_{12,3|4,5} ,$ $S_{1;2|3|4,5}$
in the BRST cohomology. This is a consequence of the symmetry  of $t_8$,
\bea
\RRR_{1;2|3,4,5} = \RRR_{1;2|4,3,5} = \RRR_{1;2|3,5,4} 
\eea
as well as momentum conservation, transversality of $\ep_i$, and the relation  
$\tr (f_1f_2f_3f_4f_5) = -\tr (f_1f_5f_4f_3f_2)$ used in (\ref{famous}),
\bea
\RRR_{1;2|3,4,5} + \RRR_{1;3|2,4,5} + \RRR_{1;4|2,3,5} + \RRR_{1;5|2,3,4} = 0
\label{Teff.2}
\eea
as well as the identity,
\begin{align}
ik_1^m   \big[ \ep^m_1 t_8(f_2,f_3,f_4,f_5)  + (1\leftrightarrow 2,3,4,5) \big] 
= \RRR_{2;1|3,4,5} + \RRR_{3;1|2,4,5}+\RRR_{4;1|2,3,5}+\RRR_{5;1|2,3,4}
\label{Teff.8} 
\end{align}
where the commutators $[f_i,f_j]$ all drop out from the right-hand side. These basic properties
imply that the effective components in (\ref{Teff.5}) obey
\begin{align}
T^{\rm eff}_{12,3|4,5} &= T^{\rm eff}_{12,3|5,4} &0 &= T^{\rm eff}_{12,3|4,5} + T^{\rm eff}_{12,4|5,3}+ T^{\rm eff}_{12,5|3,4} \notag \\
S^{\rm eff}_{1;2|3|4,5} &= S^{\rm eff}_{1;2|3|5,4} &0 &= S^{\rm eff}_{1;2|3|4,5} + S^{\rm eff}_{1;2|4|5,3}+ S^{\rm eff}_{1;2|5|3,4} 
\label{Teff.11}
\end{align}
as well as
\begin{align}
0 &=S^{\rm eff}_{1;2|3|4,5}+S^{\rm eff}_{1;3|2|4,5}+S^{\rm eff}_{1;4|5|2,3}+S^{\rm eff}_{1;5|4|2,3}
\notag \\
T^{m,{\rm eff}}_{1,2,3|4,5} &= T^{m,{\rm eff}}_{3,4,5|1,2} +  T^{m,{\rm eff}}_{2,4,5|1,3} +  T^{m,{\rm eff}}_{1,4,5|2,3} \label{Teff.10} \\
T^{m,{\rm eff}}_{1,2,3|4,5} &= T^{m,{\rm eff}} _{1,3,2|4,5} = T^{m,{\rm eff}} _{2,1,3|4,5} = T^{m,{\rm eff}} _{1,2,3|5,4} 
\notag
\end{align}
and are related by
\begin{align}
S^{\rm eff}_{1;2|3|4,5} &= \frac{1}{2}\big[ i(k_{12}^m -k_3^m) T^{m,{\rm eff}}_{1,2,3|4,5} 
+T^{\rm eff}_{12,3|4,5}+T^{\rm eff}_{13,2|4,5} +T^{\rm eff}_{23,1|4,5}  \big]
\notag
\\
T^{\rm eff}_{12,3|4,5} &=  S^{\rm eff}_{1;2|3|4,5} - S^{\rm eff}_{2;1|3|4,5}  \notag\\
ik_1^m T^{m,{\rm eff}}_{1,2,3|4,5} &= S^{\rm eff}_{2;1|3|4,5}+S^{\rm eff}_{3;1|2|4,5}
\label{Teff.9} \\
ik_5^m  T^{m,{\rm eff}}_{1,2,3|4,5}  &= S^{\rm eff}_{1;5|4|2,3}+S^{\rm eff}_{2;5|4|1,3}+S^{\rm eff}_{3;5|4|1,2}
\notag\\
ik_3^m (T^{m,{\rm eff}}_{1,2,3|4,5}+T^{m,{\rm eff}}_{3,4,5|1,2}) &= 
T^{\rm eff}_{13,2|4,5} +T^{\rm eff}_{23,1|4,5} - T^{\rm eff}_{34,5|1,2} -T^{\rm eff}_{35,4|1,2} 
\notag
\end{align}
Hence, any relation among the superfields in the BRST cohomology --
see e.g.\ (\ref{mTsym}) to (\ref{multi.10}) -- is preserved by the transition
(\ref{rplce}) to effective bosonic components.

\sm

In fact, we have checked that the bosonic components of any BRST-invariant quantity composed from the building blocks reviewed above can be obtained by using their ``effective'' versions,
\begin{equation}
\label{effrsubs}
(S_{a;b|c|d,e} ,T_{ab,c|d,e},T^m_{a,b,c|d,e} )\rightarrow (S^{\rm eff}_{a;b|c|d,e},T^{\rm eff}_{ab,c|d,e},T^{m,{\rm eff}}_{a,b,c|d,e} )
\end{equation}
This includes all representations of the genus-two correlator \eqref{defcv} since they obviously are BRST invariant.

\sm

\subsubsection{Effective BRST invariants and correlators}

The effective bosonic components (\ref{Teff.5}) not only preserve the relations
of their superspace prototypes but also the two-loop BRST invariants (\ref{BRST.2}):
One can check from the results on the website \cite{PSSsite} that,
\begin{align}
-2880 \langle C_{1;3|4|2,5} \rangle_0  \, \big|_{\rm bos}&= {1\over 4} \Big( \frac{ 3S^{\rm eff}_{1;3|4|2,5} }{s_{13}} -\frac{ S^{\rm eff}_{1;4|3|2,5}  }{s_{14}}
- \frac{S^{\rm eff}_{1;2|5|3,4} }{s_{12}}  - \frac{ S^{\rm eff}_{1;5|2|3,4}  }{s_{15}}\Big)
\notag\\
-2880 \langle C^m_{5,1,2|3,4} \rangle_0  \, \big|_{\rm bos}&=  T^{m,{\rm eff}}_{5,1,2|3,4} - {i\over 4} k_1^m \Big( \frac{ S^{\rm eff}_{1;2|5|3,4} }{s_{12}}
+ \frac{ S^{\rm eff}_{1;5|2|3,4}   }{s_{15}}+ \frac{ S^{\rm eff}_{1;3|4|2,5}   }{s_{13}}
+ \frac{ S^{\rm eff}_{1;4|3|2,5}  }{s_{14}}\Big)\label{BRSTeff}  \\
& \hskip 0.75in
- {i\over 4} k_2^m \Big( \frac{S^{\rm eff}_{2;1|5|3,4} }{s_{12}}  +\frac{ S^{\rm eff}_{2;5|1|3,4}  }{s_{25}}
+ \frac{S^{\rm eff}_{2;3|4|1,5} }{s_{23}} + \frac{S^{\rm eff}_{2;4|3|1,5}  }{s_{24}} \Big) \no \\
& \hskip 0.75in
- {i\over 4} k_5^m \Big( \frac{S^{\rm eff}_{5;1|2|3,4}  }{s_{15}}+ \frac{S^{\rm eff}_{5;2|1|3,4}  }{s_{25}}
+ \frac{S^{\rm eff}_{5;3|4|1,2}  }{s_{35}}+ \frac{S^{\rm eff}_{5;4|3|1,2} }{s_{45}} \Big)\no
\end{align}
Since the chiral amplitude is expressible in terms of the BRST invariants (\ref{BRST.2}) \cite{DHoker:2020prr},
its bosonic components can be equivalently expressed in terms of the effective components.
One can then pass back to a local representation by repeating the integration-by-parts manipulations
 in section 5.4 of the reference with effective components in the place of the superfields:
The bosonic components of the string amplitude are unchanged
when $\langle {\cal K}_{(5)} \rangle_0$ in the chiral amplitude (\ref{FK}) is replaced by
\begin{align}
{\cal K}_{(5)}^{\rm eff} = \omega_I(2) \Delta(3,4) \Delta(5,1) {\cal K}_{1,2,3|4,5}^{I,{\rm eff}} + {\rm cycl}(1,2,3,4,5) \label{Teff.14} 
\end{align}
where
\bea
\label{Teff.15} 
{\cal K}_{1,2,3|4,5}^{I,{\rm eff}} &= &
2\pi p_m^I T^{m,{\rm eff}}_{1,2,3|4,5} - g^I_{2,3} T^{\rm eff}_{23,1|4,5} 
- g^I_{2,1} T^{\rm eff}_{21,3|4,5}  - g^I_{3,1} T^{\rm eff}_{31,2|4,5}  
\notag \\
&& - g_{2,4}^I S^{\rm eff}_{2;4|5|1,3}  - g_{3,4}^I S^{\rm eff}_{3;4|5|2,1}  - g_{1,4}^I S^{\rm eff}_{1;4|5|2,3} 
\no \\
&&- g_{2,5}^I S^{\rm eff}_{2;5|4|3,1}  - g_{3,5}^I S^{\rm eff}_{3;5|4|2,1}  - g_{1,5}^I S^{\rm eff}_{1;5|4|2,3} 
\eea
and (for some odd spin structure $\nu$ whose choice is immaterial for (\ref{Teff.14}))
\begin{align}
g_{i,j}^I = \frac{\partial }{\partial \zeta_I} \ln \vartheta[\nu](\zeta|\Omega) \, , \ \ \ \ \ \ \zeta_I = \int^{z_i}_{z_j} \omega_I
\end{align}
Since the effective components inherit all the relations of the superfields, one can also adapt 
the representation (\ref{K5tW}) in terms of $\hat p^I$ and the Arakelov Green function to the effective components,
\bea
\label{K5tWeff}
\cK^{\rm eff}_{(5)} &= &
2 \pi i  \hat p^I _m \big[  T^{m,{\rm eff}}_{1,2,3|4,5} \, \om_I(2) \Delta (3,4)\Delta (5,1)  {+}{\rm cycl}(1,2,3,4,5) \big]
\no \\ &&
+ \halfap{}\big[ {\cal Q}^{\rm eff}_{12} {+} (1,2|1,2,3,4,5) \big]
\no \\
 {\cal Q}^{\rm eff}_{12} &=&  - \partial_1 \cG(1,2) \big[ S^{\rm eff}_{1;2|3|4,5} \Delta(2,4) \Delta(3,5) 
+ S^{\rm eff}_{1;2|4|3,5} \Delta(2,3) \Delta(4,5) \big] 
\no \\ && 
-  \partial_2 \cG(2,1) \big[ S^{\rm eff}_{2;1|3|4,5} \Delta(1,4) \Delta(3,5) 
 + S^{\rm eff}_{2;1|4|3,5} \Delta(1,3) \Delta(4,5) \big] 
\eea

\subsection{Assembling and expanding}
\label{sec:assemble}

After performing the Gaussian integral over loop momenta, the amplitude (\ref{pairing}) becomes,
\bea
\cA_{(N)}^{\rm genus-2} = \delta(k)
\cN_{(N)}\,  \int _{\cM_2}
d\mu_2\,    \cB_{(N)}(k_i|\Omega)
\label{LRcontractA}
\eea
where $d\mu_2$ is the $Sp(4,\RR)$ invariant measure on the Siegel upper-half plane, normalized as in \cite{Siegel,DHoker:2014oxd},
\be
\label{defmu2}
d\mu_2= \frac{|d^3\Omega|^2}{(\det Y)^3}
\hskip 1in
  {\rm Vol}_2 = \int_{\cM_2} d\mu_2  =
{2^2\pi^3\over 3^3\,5}
\ee
The integrand $\cB_{(N)}(k_i|\Omega)$ (which also depends on the polarizations of the external particles, which we do not exhibit here) is given by an integral over $\Sigma^N$ and over the zero modes of 
$\theta^\alpha,\tilde\theta^\alpha,\lambda^\alpha,\tilde\lambda^\alpha$. For $N=4$, the Gaussian integral
over $p^I$  leads to (see (\ref{defK4}) for ${\cal K}_{(4)}$)
\bea
\cB_{(4)}(k_i| \Omega) =
\int _{\Sigma^4}\frac{\KN_{(4)}}{(\det Y)^2}\,  \langle \cK_{(4)} \tilde \cK_{(4)} \rangle_0
\label{defB4A}
\eea
where $\KN_{(N)}$ is the full Koba-Nielsen factor (as opposed to the chiral one ${\cal I}_{(N)}$ in (\ref{KNN})),
\be
\label{defKN}
\KN_{(N)}(k_i|\Omega) = \prod_{1\leq i<j\leq N} e^{s_{ij} \cG(z_i,z_j | \Omega)} 
\ee
and $\cG(z_i,z_j|\Omega)$ is the Arakelov Green function  (see appendix \ref{sec:func}).
For $N=5$, the integral over loop momenta contains additional terms arising from integrating a 
bilinear term in loop momenta between left and right movers,
\be
\cB_{(5)}(k_i| \Omega) =  {1\over i\pi}
\int _{\Sigma^5} \frac{\KN_{(5)}}{(\det Y)^2}\, \left \<  \cW \, \tilde \cW 
- \halfap{}\pi Y^{IJ} \, \cV^m _I \,\tilde \cV^m _J \right \> _0
\label{defB5}
\eea
For brevity, we shall denote the two terms in the angled bracket by $\langle |\cW|^2-\pi |\cV^m_I|^2\rangle_0$. Upon integration over $\Sigma^5$, (\ref{defB5}) is unchanged when all of $\cW, \cV^m_I$ and the corresponding right-movers are  replaced by their manifestly BRST invariant counterparts $\widehat \cW, \widehat \cV^m_I$  in (\ref{defcvCm}) and (\ref{altdefcw.2}). While the manifestly local or BRST invariant superspace  representations of (\ref{defB5}) apply to any combination of external bosons and fermions, their  NSNS components can be alternatively rewritten by replacing the various building blocks within $\cW$ and $\cV^m_I$ by their ``effective'' versions.
This form will be useful in the discussion of Type IIA
components in section \ref{sec.IIA}.

\sm

By construction, both \eqref{defB4A} and \eqref{defB5} are invariant under modular transformations of $\Omega$, and can
therefore be meaningfully integrated against the invariant measure in \eqref{LRcontractA} over the moduli space $\cM_2$, realized as a fundamental domain of the action of $Sp(4,\mathbb{Z})$ on the Siegel upper half-space. For Type II strings compactified on a torus $T^d$, the measure in \eqref{LRcontractA} is multiplied by the Siegel-Narain theta series
$\Gamma_{d,d,2}$ \cite{AlvarezGaume:1986es}, which is modular invariant by itself.

\sm

The main goal of this paper will be to analyze the low energy expansion of the five-point integrand \eqref{defB5} in powers of
the kinematical invariants $s_{ij}$. The dependence on external momenta arises explicitly through the Koba-Nielsen factor
\eqref{defKN}, and through permutations of the building blocks $|T^m_{1,2,3|4,5}|^2$ of dimension $D^2\cR^5$ and
$|T_{12,3|4,5}|^2, |S_{1;2|3|4,5}|^2$ of dimension $D^4\cR^5$. While the integrals multiplying $|T^m_{1,2,3|4,5}|^2$ and
$|S_{1;2|3|4,5}|^2$ in the representation \eqref{defcw.2} are convergent and
analytic as $s_{ij}\to 0$, the integrals multiplying $|T_{ij,k|l,m}|^2$ have short-distance singularities and give rise to
factors of $1/s_{ij}$. Therefore, only permutations of $|T^m_{1,2,3|4,5}|^2$ and $|T_{12,3|4,5}|^2/s_{12}$
contribute at the lowest order $D^2\cR^5$ in the low energy effective action
\cite{Gomez:2015uha}.  This is the same order in the derivative expansion as the effective interaction $D^4\cR^4$ appearing in
the four-point genus-two amplitude at leading order, and indeed the couplings $D^4\cR^4$ and $D^2\cR^5$ are expected to belong to
a single supersymmetric invariant under non-linear supersymmetry. As we shall see in section \ref{sec:3}, this is still the
case for the genus-two $D^4\cR^5$ and $D^6\cR^4$ interactions, but the next order features a new five-point interaction $(D^6\cR^5)'$ which is no longer related to the corresponding four-point $D^8\cR^4$ interaction.

\newpage

\section{The $\ap$ expansion of genus-two integrals}
\setcounter{equation}{0}
\label{sec:3}

In this section, we shall decompose the integral for the amplitude with five external massless states, given in 
\eqref{defB5} for Type II strings,  into a sum of basic integrals over $\Sigma^5$, in terms of which the full amplitude may be obtained by including suitable permutations of the external states. The low energy expansion of these integrals will be expressed in terms of genus-two MGFs, thereby generalizing a similar analysis carried out for the  genus-two amplitude with four massless external states in
\cite{D'Hoker:2013eea,DHoker:2014oxd,DHoker:2017pvk,DHoker:2018mys}.
These results will be used in section \ref{secAp} to analyze the  low energy expansion of the genus-two four-point and five-point amplitudes.

\subsection{Genus-two integrals occurring in Type II amplitudes}
\label{sec:31}

In order to analyze the $\ap$ expansion of the genus-two four-point and five-point amplitudes in Type II string theory,
it will be useful to list the scalar integrals over four and five copies of the surface $\Sigma$ that occur along with the kinematic factors.
\begin{itemize}
\item The $I$-integrals occur in the four-point amplitude, 
\bea
I_1 
&= \int_{\Sigma^4}
{ {\rm KN}_{(4)} \over (\det Y)^2} \, \Delta(1,2)\, \Delta(3,4)\, \overline{  \Delta(1,2) } \,\overline{  \Delta(3,4) }
\notag \\
I_2 
&= \int_{\Sigma^4} 
{ {\rm KN}_{(4)} \over (\det Y)^2} \, \Delta(1,2)\, \Delta(3,4)\, \overline{  \Delta(2,3) } \, \overline{  \Delta(4,1) }
\label{defIint}
\eea
\item The $J$-integrals arise from the 
contributions $|{\cal V}^m_I|^2$ due to integrating a bilinear term in loop momenta
(with cyclic identification $\omega_I(j{+}5) = \omega_I(j)$), 
\bea
\label{defJintA}
J_{r,s} & = & {i \over 2} \int _{\Sigma ^5} { \KN_{(5)} \over (\det Y)^2} \om_I(r) \, \Delta (r{+}1,r{+}2)\, \Delta (r{+}3, r{+}4)
\no \\ && \hskip 0.6in \times 
Y^{IJ} \oom_J(s) \, \overline{ \Delta (s{+}1,s{+}2) } \, \overline{\Delta (s{+}3,s{+}4) }
\eea
They may all be obtained by cyclic permutations from one of the three basic integrals,
\begin{align}
J_{1,1}&= {i\over2} \int_{\Sigma^5} { {\rm KN}_{(5)} \over (\det Y)^2}\, \omega_I(1) Y^{IJ} \oom_J(1)\,
\, \Delta(2,3)\, \Delta(4,5)\, \overline{  \Delta(2,3) } \,\overline{  \Delta(4,5) }\no \\
J_{1,2}&= {i\over 2} \int_{\Sigma^5} { {\rm KN}_{(5)} \over (\det Y)^2}\,\omega_I(1) Y^{IJ} \oom_J(2)\,
\, \Delta(2,3)\, \Delta(4,5)\, \overline{  \Delta(3,4) } \,\overline{  \Delta(5,1) } \label{defJint} \\
J_{1,3}&= {i\over 2} \int_{\Sigma^5} { {\rm KN}_{(5)} \over (\det Y)^2}\, \omega_I(1) Y^{IJ} \oom_J(3)\,
\, \Delta(2,3)\, \Delta(4,5)\, \overline{  \Delta(4,5) } \,\overline{  \Delta(1,2) } \no
\end{align}
and their complex conjugates.
\item The $F$-integrals involve combinations of  $\cG(1,2) \cG(1,2)$,
\bea
\label{defFint}
F_1
& = & { 1 \over i \pi} \int _{\Sigma ^5} { \KN_{(5)} \over (\det Y)^2}
\p_1 \cG(1,2) \bar \p_1 \cG(1,2)
\, \Delta(2,3)\, \Delta(4,5)\, \overline{  \Delta(2,3) } \,\overline{  \Delta(4,5) }
\\
F_2
& = & { 1 \over i \pi} \int _{\Sigma ^5} { \KN_{(5)} \over (\det Y)^2}
\p_1 \cG(1,2) \bar \p_1 \cG(1,2) 
\, \Delta(2,3)\, \Delta(4,5)\, \overline{  \Delta(2,4) } \,\overline{  \Delta(3,5) }
\no \\
F_3
& = & { 1 \over i \pi} \int _{\Sigma ^5} { \KN_{(5)} \over (\det Y)^2}
\p_1 \cG(1,2)\bar \p_2 \cG(1,2) 
\, \Delta(2,3)\, \Delta(4,5)\, \overline{  \Delta(1,3) } \,\overline{  \Delta(4,5) }
\no \\
F_4
& = & { 1 \over i \pi} \int _{\Sigma ^5} { \KN_{(5)} \over (\det Y)^2}
\p_1 \cG(1,2) \bar \p_2 \cG(1,2) 
\, \Delta(2,3)\, \Delta(4,5)\, \overline{  \Delta(1,4) } \,\overline{  \Delta(3,5) }
\qquad \no 
\eea
\item The $G$-integrals involve combinations of $\cG(1,2) \cG(1,3)$,
\bea
\label{defGint}
G_1
& = & { 1 \over i \pi} \int _{\Sigma ^5} { \KN_{(5)} \over (\det Y)^2} \, 
\p_1 \cG(1,2) \bar \p_1 \cG(1,3)
\, \Delta(2,4)\, \Delta(3,5)\, \overline{  \Delta(2,4) } \,\overline{  \Delta(3,5) }
\\
G_2
& = & { 1 \over i \pi} \int _{\Sigma ^5} { \KN_{(5)} \over (\det Y)^2} \, 
\p_1 \cG(1,2) \bar \p_1 \cG(1,3) 
\, \Delta(2,4)\, \Delta(3,5)\, \overline{  \Delta(2,5) } \,\overline{  \Delta(3,4) }
\no \\
G_3
& = & { 1 \over i \pi} \int _{\Sigma ^5} { \KN_{(5)} \over (\det Y)^2} \, 
\p_1 \cG(1,2) \bar \p_3 \cG(1,3) 
\, \Delta(2,4)\, \Delta(3,5)\, \overline{  \Delta(2,4) } \,\overline{  \Delta(1,5) }
\no \\
G_4
& = & { 1 \over i \pi} \int _{\Sigma ^5} { \KN_{(5)} \over (\det Y)^2} \, 
\p_1 \cG(1,2) \bar \p_3 \cG(1,3) 
\, \Delta(2,4)\, \Delta(3,5)\, \overline{  \Delta(1,4) } \,\overline{  \Delta(2,5) }
\no \\
G_5
& = & { 1 \over i \pi} \int _{\Sigma ^5} { \KN_{(5)} \over (\det Y)^2} \, 
\p_2 \cG(1,2) \bar \p_3 \cG(1,3)
\, \Delta(1,4)\, \Delta(3,5)\, \overline{  \Delta(1,4) } \,\overline{  \Delta(2,5) }
\no \\
G_6
& = & { 1 \over i \pi} \int _{\Sigma ^5} { \KN_{(5)} \over (\det Y)^2} \, 
\p_2 \cG(1,2) \bar \p_3 \cG(1,3)
\, \Delta(1,4)\, \Delta(3,5)\, \overline{  \Delta(1,5) } \,\overline{  \Delta(2,4) }
\no
\eea
\item The $H$-integrals  involve combinations of $\cG(1,2) \cG(3,4)$,
\bea
\label{defHint}
H_1&=& {1\over i\pi} \int_{\Sigma^5} {  \KN_{(5)} \over (\det Y)^2}\,\partial_1 \cG(1,2)\, \overline{ \partial_3 \cG(3,4)}\,
\, \Delta(2,4)\, \Delta(3,5)\, \overline{  \Delta(2,4) } \,\overline{  \Delta(1,5) }
\\
H_2&=& {1\over i\pi} \int_{\Sigma^5} {  \KN_{(5)} \over (\det Y)^2}\,\partial_1 \cG(1,2)\, \overline{ \partial_3 \cG(3,4)}\,
\, \Delta(2,3)\, \Delta(4,5)\, \overline{  \Delta(1,4) } \,\overline{  \Delta(2,5) } \no
\\
H_3&=& {1\over i\pi} \int_{\Sigma^5} {  \KN_{(5)} \over (\det Y)^2}\,\partial_1 \cG(1,2)\, \overline{ \partial_3 \cG(3,4)}\,
\, \Delta(2,3)\, \Delta(4,5)\, \overline{  \Delta(1,5) } \,\overline{  \Delta(2,4) } \no
\\
H_4&=& {1\over i\pi} \int_{\Sigma^5} { \KN_{(5)} \over (\det Y)^2}\,\partial_1 \cG(1,2)\, \overline{ \partial_3 \cG(3,4)}\,
\, \Delta(2,4)\, \Delta(3,5)\, \overline{  \Delta(1,4) } \,\overline{  \Delta(2,5) } \no
\eea
\end{itemize}
All integrals required in the genus-two amplitude with five massless external states may be expressed in terms of the above
integrals and permutations of their vertex labels\footnote{Note that it is convenient to explicitly add the complex conjugates $\bar G_3$ and $\bar G_4$ to the list above in order to quickly identify all the integrals in the genus-two correlator \eqref{defB5}.}.

\subsection{Extracting the singular part of the $F$-integrals at $s_{ij}=0$}
\label{sec:intexp}

All the integrals given in subsection \ref{sec:31} are absolutely convergent for $|s_{ij}|\ll 1$ and admit convergent Taylor
series expansions at $s_{ij}=0$, with the notable exception of the $F$-integrals \eqref{defFint} which have simple poles at
$s_{ij}=0$. In this subsection, we present the analysis needed to extract this singularity for the integral $F_1$ and defer the
cases of the integrals $F_2, F_3, F_4$ to appendix \ref{sec:expF}. The singularity of $F_1$ is due to the non-integrable
singularity at $z_1=z_2$ of the following factor of the integrand of $F_1$,
\be
\p_1 \cG(1,2) \bar \p_1 \cG(1,2) \sim \frac{1}{|z_1-z_2|^2}
\ee
As a result, the integral $F_1$ has a simple pole at $s_{12}=0$. The simple pole in  $F_1$  may be exposed by using the following identity of the integrand, 
\bea
\label{KNbyparts}
 \KN_{(5)} \, \p_1 \cG(1,2) \bar \p_1 \cG(1,2)
& = & 
- {  \KN_{(5)}  \over s_{12}} \Big (  \sum_{k=3}^5 s_{1k} \p_1 \cG(1,2) \, \bar \p_1 \cG(1,k)  + \p_1 \bar \p_1 \cG(1,2)  \Big )
\no \\ && 
+ { 1 \over s_{12}} \bar \p_1 \Big (  \KN_{(5)}  \p_1 \cG(1,2) \Big )
\eea
Since the combination inside the parentheses on the second line is a $(1,0)$ form its Dolbeault differential  $\bar \p_1$ may be recast as a total differential, $\bar \p_1 ( \KN_{(5)} \p_1 \cG(1,2)  ) = d_1 (  \KN_{(5)} \p_1 \cG(1,2)  ) $
whose integral over the closed compact surface $\Sigma$ vanishes. As a result, the term on the second line does not contribute to $F_1$.

\sm

The contribution to the second term in the parentheses on the first line of (\ref{KNbyparts}) is given by (\ref{AraG}). The $\delta(1,2)$ term vanishes provided we assume the following domain for $s_{12}$,
\bea
\label{Res12}
\Re (s_{12}) <0
\eea 
or alternatively vanishes by the ``cancelled propagator" argument in old string theory lingo. The remaining contributions to the integral $F_1$ are therefore given by the first term of the first line in (\ref{KNbyparts}) and by the $\kappa$-term in (\ref{AraG}) for the second term in the parentheses, and we obtain the following formula,
\bea
\label{F1dec}
F_1& = & 
- { 1 \over i \pi }  \sum_{k=3}^5 {s_{1k} \over s_{12}} \int _{\Sigma ^5} { \KN_{(5)} \over (\det Y)^2}
\p_1 \cG(1,2) \bar \p_1 \cG(1,k) |\Delta (2,3)|^2 | \Delta (4,5)|^2 
\no \\ &&
+ { 2 \over s_{12} } \int _{\Sigma ^5} { \KN_{(5)} \over (\det Y)^2}
\kappa(1)  |\Delta (2,3)|^2 | \Delta (4,5)|^2 
\eea
The integrals in \eqref{F1dec} are now absolutely convergent for $|s_{ij}|\ll 1$ and admit a convergent Taylor series expansion at $s_{ij}=0$. The coefficient of $1/s_{12}$ is recognized as the integral $J_{1,1}$, and one can similarly express the first line of (\ref{F1dec}) in terms of permutations of the integrals $G_1,G_2$ defined in (\ref{defGint}), see (\ref{FFFF}).

\subsection{Genus-two modular graph functions up to order $D^6 \cR^5$}

Our aim in this section will be to find the first few terms in the $\alpha'$ expansions for the above integrals, so as to extract the coefficient of the effective interactions up to order $D^6 \cR^5$ in the low energy effective action. In addition to $\Delta(x,y)$, the following combination of holomorphic  $(1,0)$-forms $\om_I$ and their complex conjugates will be ubiquitous in our analysis, and are given as follows,  in components,
\bea
\label{nu0}
\nu (x,y) = { i \over 2} Y^{IJ} \om_I(x)  \oom_J(y) = - \overline{\nu(y,x)}
\eea
On the diagonal $y = x$ they reduce to twice the canonical form $\kappa$ defined by, 
\bea
\kappa(x)  = \frac{1}{2} \nu(x,x) ={ i \over 4} Y^{IJ} \om_I(x)  \oom_J(x) \, , \ \ \ \ \ \ \int_{\Sigma} \kappa = 1
\eea 
Various details of the subsequent computations 
are relegated to appendix~\ref{app:exp}.

\sm

\begin{figure}[h]
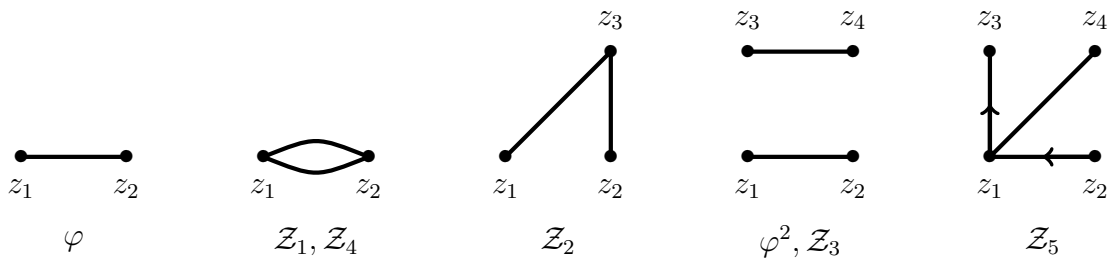

\begin{center}
\tikzpicture[scale=1.4]
\scope[xshift=1.4cm,yshift=0cm]
\draw (-3,0) node{$\bullet$};
\draw (-2,0) node{$\bullet$};
\draw (-3,-0.3) node{$z_1$};
\draw (-2,-0.3) node{$z_2$};
\draw (-2.5, -0.8) node{$\f$};
\draw[ultra thick] (-3,0) -- (-2,0);
\endscope
\scope[xshift=0.7cm,yshift=0cm]
\draw (0,0) node{$\bullet$};
\draw (1,0) node{$\bullet$};
\draw (0,-0.3) node{$z_1$};
\draw (1,-0.3) node{$z_2$};
\draw (0.5, -0.8) node{$\cZ_1, \cZ_4$};
\draw[ultra thick] (0,0) .. controls (0.5, -0.2) .. (1,0);
\draw[ultra thick] (0,0) .. controls (0.5, 0.2) .. (1,0);
\endscope
\scope[xshift=0cm,yshift=0cm]
\draw (3,0) node{$\bullet$};

\draw (4,0) node{$\bullet$};
\draw (4,1) node{$\bullet$};
\draw (3,-0.3) node{$z_1$};
\draw (4,-0.3) node{$z_2$};

\draw (4,1.3) node{$z_3$};
\draw (3.5, -0.8) node{$\cZ_2$};
\draw[ultra thick] (3,0) -- (4,1);
\draw[ultra thick] (4,0) --  (4,1);
\endscope
\scope[xshift=-0.7cm,yshift=0cm]
\draw (6,0) node{$\bullet$};
\draw (6,1) node{$\bullet$};
\draw (7,0) node{$\bullet$};
\draw (7,1) node{$\bullet$};
\draw (6,-0.3) node{$z_1$};
\draw (7,-0.3) node{$z_2$};
\draw (6,1.3) node{$z_3$};
\draw (7,1.3) node{$z_4$};
\draw (6.5, -0.8) node{$\f^2, \cZ_3$};
\draw[ultra thick] (6,0) -- (7,0);
\draw[ultra thick] (6,1) --  (7,1);
\endscope
\scope[xshift=-1.4cm,yshift=0cm]
\draw (9,0) node{$\bullet$};
\draw (9,1) node{$\bullet$};
\draw (10,0) node{$\bullet$};
\draw (10,1) node{$\bullet$};
\draw (9,-0.3) node{$z_1$};
\draw (10,-0.3) node{$z_2$};
\draw (9,1.3) node{$z_3$};
\draw (10,1.3) node{$z_4$};
\draw (9.5, -0.8) node{$\cZ_5$};
\draw[ultra thick] (9,0) -- (10,0);
\draw[ultra thick] (9,0) --  (9,1);
\draw[ultra thick,<-] (9.5,0) -- (10,0);
\draw[ultra thick,->] (9,0) --  (9,0.5);
\draw[ultra thick] (9,0) --  (10,1);
\endscope
\endtikzpicture
\caption{Graphs representing the genus-two modular graph functions $\f$, $\cZ_i$ and $\f^2$ where each line represents the Green function $\cG$ but the structure of the Abelian differentials is not exhibited. The arrows on two of the lines in  $\cZ_5$ indicate the derivatives $\p_1$ and $\bar\p_1$ on $\cG$.} \label{fig:1}
\end{center}
\end{figure}

Up to order $D^6 \cR^5$, we find that the coefficients can all be expressed in terms of the Kawazumi-Zhang invariant $\varphi$ given by any one of the following equivalent expressions (see \cite{D'Hoker:2013eea,Pioline:2015qha} and references therein),
 \begin{align}
\f &=  \int _{\Sigma ^2} { | \Delta (1,2) |^2 \over 4 (\det Y)}  \, \cG(1,2)
\notag \\
&= - { 1 \over 4} Y^{IL} Y^{JK} \int_{\Sigma^2} \om_I(1) \, \oom_J(1) \, \om_K(2) \, \oom_L(2) \, \cG(1,2) \label{KZ2} \\
&=  \int _{\Sigma^2} \nu (1,2) \nu (2,1) \, \cG(1,2)
\notag
\end{align}
and the following convergent integrals on direct products of $\Sigma$, 
\begin{align}
\label{defZs}  
\cZ_1 &  =8 \int _{\Sigma ^2} \kappa(1)\kappa(2)  \cG(1,2)^2 
\notag \\
\cZ_2 &= - \int_{\Sigma^3} { | \Delta(1,2) |^2 \over \det Y}  \kappa(3)  \cG(1,3) \cG(2,3) 
\notag \\
\cZ_3 &=  \int_{\Sigma^4} { | \Delta(1,3) \Delta(2,4) |^2 \over 8 (\det Y)^2}   \cG(1,2) \cG(3,4)  
\no \\
 \cZ_4 &=  -4 \int_{\Sigma^2}  \nu(1,2)  \nu(2,1)  \cG(1,2)^2
 \notag \\
 \cZ_5&=   \frac{16 i}{\pi} \int_{\Sigma^4} \cG(1,4)\, \partial_1 \cG(1,2)\, \bar \partial_1 \cG(1,3)\, 
\nu(2,4) \nu(4,3) \nu(3,2)  
 \end{align}
As is the case for the KZ invariant \eqref{KZ2}, the integrals  \eqref{defZs} are modular invariant functions of the period matrix $\Omega$, and real analytic  away from the separating and non-separating divisors. They belong to the class of genus-two MGFs  introduced in \cite{DHoker:2017pvk}, generalizing the genus-one MGFs of \cite{DHoker:2015wxz}. The relevant graphs keep track of the products of Arakelov Green functions to be integrated, and are displayed in figure~\ref{fig:1}. 

\sm

The integrals $\cZ_1,\cZ_2,\cZ_3$ have appeared previously in the study of the 4-point amplitude \cite{DHoker:2017pvk,DHoker:2018mys}, where their asymptotic behavior near the separating and non-separating divisors in the moduli space $\cM_2$ was investigated in great detail. The integral $\cZ_4$ was introduced in the course of the analysis of the action of the Laplace-Beltrami operator in \cite{Basu:2018bde},  along with several other integrals 
which also occur here in the evaluation of the five-point amplitude, and which we evaluate in terms of the ones above in appendix \ref{sec:A7}. The integral  $\cZ_5$ is novel, and reminiscent of the modular graph forms introduced in \cite{DHoker:2016mwo}, although it is genuinely modular invariant. 
The asymptotics of $\cZ_4$ and $\cZ_5$ near the separating and non-separating divisors is derived in appendix \ref{sec:degen} using similar methods as in  \cite{DHoker:2017pvk,DHoker:2018mys}. 
Genus-two amplitudes for the Heterotic string are expected to involve higher weight generalizations of these MGFs, in parallel with the modular graph forms appearing in Heterotic genus-one amplitudes \cite{Gerken:2018jrq}.

\subsection{Novel modular graph function identities}

The study of the degenerations of the integrals $\cZ_i$ in appendix \ref{sec:degen} suggests that these integrals are not linearly independent, but rather satisfy a remarkable identity, 
\bea
\label{MGI}
\cZ_1+\cZ_2 + \cZ_3 + \half \cZ_4 - \f^2=0
\eea
which we shall prove in appendix \ref{sec:MGIproof}. 
From the point of view of two-dimensional quantum field theory on the genus-two surface, the  identity \eqref{MGI} is quite striking since it relates a combination of one-loop graphs $\cZ_1, \cZ_4$ to a combination of tree-level graphs $\f^2, \cZ_2, \cZ_3$. By contrast,  the alternative expressions for $\f^2, \cZ_2, \cZ_3$ given in appendix \ref{sec:A7}  exclusively relate tree-level graphs to one another and are the result of elementary relations, such as (\ref{omdel}) and (\ref{Delnu}), between Abelian differentials. Thus, the identity \eqref{MGI} is more akin to the identities between genus-one MGFs proven in \cite{DHoker:2015sve,DHoker:2016mwo,DHoker:2016quv,Basu:2016kli} and exposed by
their representations in terms of iterated Eisenstein integrals \cite{Gerken:2019cxz, Gerken:2020yii}.  The proof of \eqref{MGI}, detailed in appendix  \ref{sec:MGIproof}, makes crucial use of a lemma \eqref{lemma} valid at any genus $h$, which allows us, effectively, to convert a derivative $\partial_i \cG(z_i,z_j)$ into a derivative   $-\partial_j \cG(z_i,z_j)$, despite the lack of translational invariance when $h\geq 2$. We anticipate that this property will become  important in future studies of relations between
genus-two MGFs.\footnote{Indeed, the conversion of derivatives has been used to generalize  
(\ref{MGI}) to arbitrary genus and to derive higher-weight identities \cite{DHoker:2020uid} 
since the first version of this work.}

\sm

As a consequence of the genus-two identity \eqref{MGI} in the minimal non-separating degeneration limit $t\to\infty$ (with $t = \frac{\det \Im \Omega}{\Im \Omega_{11}}$ \cite{DHoker:2017pvk}), we also obtain a new identity at genus one, 
\bea
\label{MGIell}
\Delta_\tau \big (F_2^2-2F_4 \big ) = 6 F_2^2 - 4 F_4
\eea
where $F_k(v|\tau)$ is the {\sl elliptic} MGF on a torus $\Sigma _1$ of modulus $\tau$ with $v \in \Sigma_1$,  defined by,
\bea
\label{defFk}
F_k(v\vert \tau) & = & { 1 \over k!} \int _{\Sigma _1} \kappa_1(x) f(x)^k 
\hskip 1in
\kappa_1(x)  = \frac{i}{2\tau_2} \, dx\, d\bar x
\eea
Here  $f(x)= g(x-p_b) - g(x-p_a)$ where $g$ is the Green function $g$ on $\Sigma _1$ and $p_a,p_b$ are
two punctures on $\Sigma_1$ with $v=p_b-p_a$ (see appendix \ref{sec:degen}).
The Laplacian on $\tau$, defined by $\Delta_\tau= 4 \tau_2^2 \p_\tau \p_{\bar\tau}$, acts on 
elliptic functions of $v=u_1+u_2\tau$  by keeping the real coordinates $u_1,u_2$  fixed. 
Identity (\ref{MGIell}) is again reminiscent of the identities proven  in \cite{DHoker:2015sve,DHoker:2016mwo,DHoker:2016quv,Basu:2016kli}.\footnote{In an earlier version of this work, it was left 
as an open problem to derive \eqref{MGIell} directly, without recourse to its genus-two ancestor.
Since then, a direct proof has been given in \cite{Basu:2020pey} using genus-one methods.}

\subsection{Expansion in $\alpha'$ of the basic genus-two integrals} \label{sec:inexp}

In this subsection, we shall list the results of the expansions of the integrals $J_{1,i}$ to order $s_{ij}^2$ and of $F_j, G_j, H_j$ to order $s_{ij}$.  Their derivations are relegated to appendix \ref{app:exp}. We also include the expansion for the integrals $I_1, I_2$ governing the four-point amplitude  \cite{DHoker:2018mys}.
\begin{itemize}
\item For the four-point integrals in \eqref{defIint}, 
\bea
\label{4ptexp}
I_1
&= &
64 - 64 \, s_{12} \, \varphi  + (   24 s_{12}^2 - 16  s_{13} s_{23})  (\cZ_1 - 2 \cZ_2 + \cZ_3) 
\no \\ && 
+ s_{12}^2 (48 \cZ_2 +  8 \cZ_4 -  16 \cZ_3 + 16 \varphi^2)  + {\cal O}(s_{ij}^3)
\no \\
I_2
&=&  32 + 64  \varphi \, s_{13} + 8 (  s_{12}^2 +s_{23}^2 )  (\cZ_1 - 2 \cZ_2 + \cZ_3) 
 \no \\ &&
 -  s_{12}s_{23} (48 \cZ_2 + 8 \cZ_4 -  16 \cZ_3 + 16 \varphi^2)  + {\cal O}(s_{ij}^3)
\eea

\item For the five-point  $J$-integrals in (\ref{defJint}), 
\bea
\label{expJs}
J_{1,1} &=& 128 - 64 (s_{23} + s_{45})  \varphi + 16 s_{23} s_{45} (-2 \cZ_1 + \cZ_3 + 2 \varphi^2) + 
 8 (s_{23}^2 + s_{45}^2) (\cZ_4 + 5 \cZ_1)   \notag \\
&&+ 
 16 ( s_{12} s_{15} + 2 s_{34}^2 - 2 s_{12} s_{34} - 2 s_{15} s_{34} + s_{23} s_{34} + 
    s_{34} s_{45}) (\cZ_1 - 2 \cZ_2 + \cZ_3)  \notag \\
 &&+  32 (s_{12}^2 + s_{15}^2 - s_{15} s_{23} - s_{45} s_{12}) (\cZ_1 -  \cZ_2) + 16 (s_{12} s_{23} + s_{45} s_{15}) (\cZ_1 - \cZ_3)+{\cal O}(s_{ij}^3)
\notag \\
J_{1,2} &=& 32 + 64 s_{35}  \varphi + 8 (s_{12}^2 + s_{34}^2 + s_{45}^2) (\cZ_1- 2 \cZ_2 + \cZ_3)
+  8 s_{15} s_{23} ( 2 \cZ_2 + 3 \cZ_3 + \cZ_4) 
\no \\ &&
 -8  s_{12} (s_{34} + s_{45}) ( \cZ_1 - 4 \cZ_2 +3 \cZ_3)  
 +  8 s_{12} ( s_{15} + s_{23}) (\cZ_1 +  \cZ_2 + \cZ_3 + \tfrac{1}{2} \cZ_4 - \varphi^2)
\no \\ &&
-  4 s_{34} s_{45} (  10 \cZ_2 - 10 \cZ_3 + \cZ_4 + 10 \varphi^2)
 - 4 (s_{23} s_{34} + s_{45} s_{15}) (   2 \cZ_2 + 6 \cZ_3 + \cZ_4 - 6 \varphi^2) 
 \notag \\
 && 
 +  8 (s_{15}^2 + s_{23}^2 - s_{23} s_{45} - s_{34} s_{15}) (\cZ_1 - \cZ_3) 
 +{\cal O}(s_{ij}^3)
\notag \\
J_{1,3} &=& -64 + 64 (s_{45} - s_{13}) \varphi  - 16 (s_{12}^2 + s_{23}^2) (\cZ_1  -  \cZ_2) 
+  8 s_{45}^2 (-2 \cZ_1 + \cZ_3 + 2 \varphi^2)
 \notag \\
&&+  16 (s_{15} s_{23} + s_{12} s_{34} - s_{34}^2 - s_{15}^2) (\cZ_1  -  2 \cZ_2 + \cZ_3) + 
 8 s_{12} s_{23} (  2 \cZ_2 + 3 \cZ_3 + \cZ_4)
 \notag \\
 && +  8 (2 s_{15} s_{34} - s_{12} s_{15} - s_{23} s_{34}) (\cZ_1 -  4 \cZ_2 + 3 \cZ_3)  + 
 8 (s_{15} + s_{34}) s_{45} (\cZ_3 - \cZ_1)
\notag \\
&&+ 8 (s_{12} + s_{23}) s_{45} (\cZ_1 -  2 \cZ_2 - 2 \cZ_3 - \cZ_4 - 4 \varphi^2) +{\cal O}(s_{ij}^3)
\eea

\item For the $F$-integrals (\ref{defFint}), with kinematic poles exposed via (\ref{F1dec}), 
\bea
F_1 &=& \frac{128}{s_{12}} + \frac{64 \varphi (s_{12} - 2 s_{45})}{s_{12}}  
- \frac{ 32 s_{34} s_{35}}{s_{12}} (\cZ_1-  2 \cZ_2 + \cZ_3) 
+  \frac{16 s_{45}^2}{s_{12}} (3 \cZ_1 + \cZ_3 + \cZ_4 + 2 \varphi^2) 
\no \\ &&
+  8 s_{23} (-\cZ_1 -  2 \cZ_2 - \cZ_4)  + 32 (\cZ_1 -  \cZ_2) s_{12}
 \notag \\ &&  
- 8 (5 \cZ_1 -  6 \cZ_2 + 2 \cZ_3 + \cZ_4 + 4 \varphi^2) s_{45}+{\cal O}(s_{ij}^2)\notag
\\
F_2 &=& \frac{64}{s_{12}} +\frac{ 32 \varphi (4 s_{34} - s_{12})}{s_{12}} + \frac{16 s_{34}^2}{
  s_{12}} (\cZ_1 -  2 \cZ_2 + \cZ_3) - \frac{16 s_{35} s_{45}}{
  s_{12}} (2 \cZ_1 +  2 \cZ_2 + \cZ_4 + 2 \varphi^2) 
  \notag \\
&& + 16 \cZ_1 s_{12} + 
 8 (-\cZ_1 - 2 \cZ_2 - \cZ_4) s_{15} - 16 (\cZ_1 + \cZ_3) s_{34}+{\cal O}(s_{ij}^2)\notag
\\
F_3 &=&  -\frac{128}{s_{12}} + 
 \frac{128 \varphi (s_{45} - s_{12})}{s_{12}} + \frac{32 s_{34} s_{35}}{s_{12}} (\cZ_1 -  2 \cZ_2 + \cZ_3)
  - \frac{  16 s_{45}^2}{s_{12} }(3 \cZ_1 + \cZ_3 + \cZ_4 + 2 \varphi^2) \notag \\
&& - 32 s_{12} (\cZ_1 - \cZ_2) + 
 16 s_{45} (3 \cZ_1 -  3 \cZ_2 + \cZ_3 + \cZ_4 + 4 \varphi^2)+{\cal O}(s_{ij}^2)\notag
\\
F_4 &=& -\frac{64}{s_{12}} - 
 \frac{128 \varphi s_{34}}{s_{12}} - \frac{16 s_{34}^2}{s_{12} }(\cZ_1 - 2 \cZ_2 + \cZ_3) + \frac{16 s_{35} s_{45}}{
  s_{12}} (2 \cZ_1  + 2 \cZ_2 + \cZ_4 + 2 \varphi^2)  \notag \\
&&+ 8 s_{12} ( - \cZ_2 - 2 \cZ_1 + 2 \varphi^2) + 
 16 s_{34} (\cZ_1 +  \cZ_2 + \cZ_3 - 2 \varphi^2)+{\cal O}(s_{ij}^2)
 \label{expFs}
\eea

\item For the $G$-integrals  in (\ref{defGint}), 
\bea
G_1 &=& -16 s_{23} \cZ_1  -  16 (s_{25} + s_{34}) \cZ_2  - 16 s_{45}  \cZ_3+{\cal O}(s_{ij}^2)
\notag \\
G_2 &=& -32 \varphi + (s_{14} - s_{15}) \cZ_5  + 4 (s_{12} + s_{13}) ( \cZ_1 + \cZ_4) +    8 s_{23}  \cZ_4  
    \notag \\
&&  -  8 \cZ_2 (s_{24} + s_{25} + s_{34} + s_{35}) + 
   16 s_{45} (- \cZ_3 +  \varphi^2 )+{\cal O}(s_{ij}^2)
   \notag \\
G_3 &=& 16 s_{23} ( \cZ_1  +  \cZ_2) + 16 s_{34} (  \cZ_2 +  \cZ_3)+{\cal O}(s_{ij}^2)\notag
\\
G_4 &=& 8 s_{23} ( \cZ_2  - \cZ_4) + 8 s_{35} (  \cZ_2 + 2 \varphi^2) + 8 s_{34} ( \cZ_2 + 2 \cZ_3 - 2 \varphi^2)+{\cal O}(s_{ij}^2)
\notag \\
G_5 &=&  - 16 s_{23} (\cZ_1 +  \cZ_2)+{\cal O}(s_{ij}^2)
\notag \\
G_6 &= &8 s_{23} ( \cZ_4 - \cZ_2)+{\cal O}(s_{ij}^2)
\label{expGs}
\eea
\item For the $H$- integrals in (\ref{defHint}), 
\bea
H_1 &=& - 16 s_{13} ( \cZ_2 + \cZ_3)+{\cal O}(s_{ij}^2)\notag
\\
H_2 &=& - 8 s_{13} (\cZ_2 + 2 \cZ_3 - 2 \varphi^2) +{\cal O}(s_{ij}^2)\notag
\\
H_3 &=& - 8 s_{13} ( \cZ_2 + 2 \cZ_3 - 2 \varphi^2 ) +{\cal O}(s_{ij}^2)\notag
\\
H_4 &=& - 8 s_{13} (\cZ_2 + 2 \cZ_3 - 2 \varphi^2 ) +{\cal O}(s_{ij}^2)
\label{expHs}
\eea
\end{itemize}

\subsection{Decomposing the five-point correlator}
\label{sec:decB5}

The five-point integrand \eqref{defB5} is expressible via permutations and complex conjugation of the
integrals discussed above and will be decomposed into four sectors
 according to the appearance and arguments of the Arakelov Green functions $\cG$,
\bea
\cB_{(5)} = \cB_{(5)}^J + \cB_{(5)}^F + \cB_{(5)}^G + \cB_{(5)}^H 
 \label{splitB.1}
\eea
where the superscripts indicate the type of integral involved in the decomposition.
The first part comprises the contractions of the vector blocks  between the left- and right-movers, resulting from integrating
out the loop momenta,\footnote{Throughout, complex conjugation on an integral will leave the kinematic variables $s_{ij}$ unchanged.}
\begin{align}
\cB_{(5)}^J &= i \int_{\Sigma^5}  {  {\rm KN}_{(5)} \over  (\det Y)^2} \, Y^{IJ}  \cV^m_I \tilde \cV^m_J 
\notag \\
&= 2 \Big\{
 J_{1,1} T^m_{5,1,2|3,4} \tilde T^m_{5,1,2|3,4}+ J_{1,2} T^m_{5,1,2|3,4} \tilde T^m_{1,2,3|4,5} 
+ \overline{J_{1,2}} T^m_{1,2,3|4,5} \tilde T^m_{5,1,2|3,4}    \label{splitB.2} \\
&\ \ \ \ + J_{1,3} T^m_{5,1,2|3,4} \tilde T^m_{2,3,4|5,1} 
+ \overline{J_{1,3}} T^m_{2,3,4|5,1} \tilde T^m_{5,1,2|3,4}   +  {\rm cycl}(1,2,3,4,5) \Big\} 
\notag 
\end{align}
The remaining three parts, $\cB_{(5)}^F, \,  \cB_{(5)}^G$ and $ \cB_{(5)}^H $ are organized by the number of labels shared between the scalar building blocks ${\cal Q}_{ab}$ and $\tilde {\cal Q}_{cd}$ defined in (\ref{altdefcw.2}) and the positions of the  derivatives on the Arakelov Green functions. It will be convenient to express these contributions as sums over permutations of more elementary building blocks, 
\begin{align}
\cB_{(5)}^F &= \cB_{12}^{F} + (1,2|1,2,3,4,5) \notag \\
\cB_{(5)}^G &= \big[ \cB_{1;23}^{G} + \cB_{1;32}^{G} + (2,3|2,3,4,5) \big] + (1\leftrightarrow 2,3,4,5)
\label{splitB.5}  \\
\cB_{(5)}^{H}  &= \big[ \cB_{12,34}^{H} + \cB_{34,12}^{H}  + {\rm cyc}(2,3,4) \big] + (5\leftrightarrow 1,2,3,4)
\notag
\end{align}
where the  combinations of the type  $\cG(1,2) \cG(1,2)$ yield,\footnote{Throughout, a vertical bar with permutations
of the vertex labels following an integral function will indicate the permutation to be performed on the
entries of the integrals as they were originally defined in (\ref{defJint}), (\ref{defFint}), (\ref{defGint}), (\ref{defHint}).}
\begin{align}
\cB_{12}^{F} =
\int_{\Sigma^5}  { {\cal Q}_{12} \tilde {\cal Q}_{12} {\rm KN}_{(5)} \over i \pi (\det Y)^2}
&= S_{1;2|3|4,5} \tilde S_{1;2|3|4,5} ( F_1 \big|_{3\leftrightarrow 4} )
+ S_{1;2|3|4,5} \tilde S_{1;2|4|3,5} ( F_2 \big|_{3\leftrightarrow 4} ) \notag \\
&+ S_{1;2|4|3,5} \tilde S_{1;2|3|4,5} F_2 + S_{1;2|4|3,5} \tilde S_{1;2|4|3,5} F_1 
 \notag \\
&+S_{2;1|3|4,5} \tilde S_{2;1|3|4,5} ( F_1 \big|^{1\leftrightarrow 2}_{3\leftrightarrow 4} )
+ S_{2;1|3|4,5} \tilde S_{2;1|4|3,5} ( F_2 \big|^{1\leftrightarrow 2}_{3\leftrightarrow 4} )
 \notag \\
&+ S_{2;1|4|3,5} \tilde S_{2;1|3|4,5} ( F_2  \big|^{1\leftrightarrow 2} )
+ S_{2;1|4|3,5} \tilde S_{2;1|4|3,5} (F_1  \big|^{1\leftrightarrow 2})
\label{splitB.4}
\\
&+S_{1;2|3|4,5} \tilde S_{2;1|3|4,5} ( F_3 \big|_{3\leftrightarrow 4} )
+ S_{1;2|3|4,5} \tilde S_{2;1|4|3,5} ( F_4 \big|_{3\leftrightarrow 4} ) \cr
&+ S_{1;2|4|3,5} \tilde S_{2;1|3|4,5} F_4 
+ S_{1;2|4|3,5} \tilde S_{2;1|4|3,5} F_3
\notag \\
&+S_{2;1|3|4,5} \tilde S_{1;2|3|4,5} ( F_3 \big|^{1\leftrightarrow 2}_{3\leftrightarrow 4} )
+ S_{2;1|3|4,5} \tilde S_{1;2|4|3,5} ( F_4 \big|^{1\leftrightarrow 2}_{3\leftrightarrow 4} )
\notag \\
&+ S_{2;1|4|3,5} \tilde S_{1;2|3|4,5} ( F_4  \big|^{1\leftrightarrow 2} )
+ S_{2;1|4|3,5} \tilde S_{1;2|4|3,5} (F_3  \big|^{1\leftrightarrow 2}) 
\notag 
\end{align}
the combinations of the type $\cG(1,2) \cG(1,3)$ yield,
\begin{align}
\cB_{1;23}^{G} =\int_{\Sigma^5}  { {\cal Q}_{12} \tilde {\cal Q}_{13} {\rm KN}_{(5)} \over i \pi (\det Y)^2}&=
S_{1;2|4|3,5} \tilde S_{1;3|4|2,5} ( G_2 \big|_{4\leftrightarrow 5} )
+ S_{1;2|4|3,5} \tilde S_{1;3|5|2,4}( G_1 \big|_{4\leftrightarrow 5} ) \notag \\
&+ S_{1;2|5|3,4} \tilde S_{1;3|4|2,5}  G_1 
+ S_{1;2|5|3,4} \tilde S_{1;3|5|2,4} G_2 
 \notag \\
&+S_{1;2|4|3,5} \tilde S_{3;1|4|2,5} ( G_4 \big|_{4\leftrightarrow 5} )
+ S_{1;2|4|3,5} \tilde S_{3;1|5|2,4}( G_3 \big|_{4\leftrightarrow 5} ) \notag \\
&+ S_{1;2|5|3,4} \tilde S_{3;1|4|2,5}  G_3
+ S_{1;2|5|3,4} \tilde S_{3;1|5|2,4} G_4
\label{splitB.6} \\
&+S_{2;1|4|3,5} \tilde S_{1;3|4|2,5} (  \overline G_4 \big|^{2\leftrightarrow 3}_{4\leftrightarrow 5} )
+ S_{2;1|4|3,5} \tilde S_{1;3|5|2,4} (  \overline G_3 \big|^{2\leftrightarrow 3} ) \notag \\
&+ S_{2;1|5|3,4} \tilde S_{1;3|4|2,5} ( \overline G_3\big|^{2\leftrightarrow 3}_{4\leftrightarrow 5} )
+ S_{2;1|5|3,4} \tilde S_{1;3|5|2,4} ( \overline G_4\big|^{2\leftrightarrow 3} )
 \notag \\
&+S_{2;1|4|3,5} \tilde S_{3;1|4|2,5} ( G_5 \big|_{4\leftrightarrow 5} )
+ S_{2;1|4|3,5} \tilde S_{3;1|5|2,4}( G_6 \big|_{4\leftrightarrow 5} ) \notag \\
&+ S_{2;1|5|3,4} \tilde S_{3;1|4|2,5}  G_6
+ S_{2;1|5|3,4} \tilde S_{3;1|5|2,4} G_5 
\notag
\end{align}
and the combinations of the type $\cG(1,2) \cG(3,4)$ yield,
\begin{align}
\cB_{12,34}^{H} =\int_{\Sigma^5}  { {\cal Q}_{12} \tilde {\cal Q}_{34} {\rm KN}_{(5)} \over i \pi (\det Y)^2}&=
-S_{1;2|3|4,5} \tilde S_{3;4|1|2,5} H_1
- S_{1;2|3|4,5} \tilde S_{3;4|2|1,5} H_4 \notag \\
&- S_{1;2|4|3,5} \tilde S_{3;4|1|2,5}  H_3
- S_{1;2|4|3,5} \tilde S_{3;4|2|1,5} H_2  \notag \\
&- S_{2;1|3|4,5} \tilde S_{3;4|1|2,5} ( H_4 \big|_{1\leftrightarrow 2} )
- S_{2;1|3|4,5} \tilde S_{3;4|2|1,5}( H_1 \big|_{1\leftrightarrow 2} )  \notag \\
&- S_{2;1|4|3,5} \tilde S_{3;4|1|2,5}  ( H_2 \big|_{1\leftrightarrow 2} )
- S_{2;1|4|3,5} \tilde S_{3;4|2|1,5} ( H_3  \big|_{1\leftrightarrow 2} )
\label{splitB.8} \\
&- S_{1;2|3|4,5} \tilde S_{4;3|1|2,5} ( H_3 \big|^{3\leftrightarrow 4} )
- S_{1;2|3|4,5} \tilde S_{4;3|2|1,5}( H_2 \big|^{3\leftrightarrow 4})  \notag \\
&- S_{1;2|4|3,5} \tilde S_{4;3|1|2,5}  ( H_1  \big|^{3\leftrightarrow 4} )
- S_{1;2|4|3,5} \tilde S_{4;3|2|1,5} ( H_4   \big|^{3\leftrightarrow 4} )
\notag \\
&- S_{2;1|3|4,5} \tilde S_{4;3|1|2,5} ( H_2 \big|^{3\leftrightarrow 4}_{1\leftrightarrow 2} )
- S_{2;1|3|4,5} \tilde S_{4;3|2|1,5}( H_3 \big|^{3\leftrightarrow 4}_{1\leftrightarrow 2} )  \notag \\
&- S_{2;1|4|3,5} \tilde S_{4;3|1|2,5}  ( H_4  \big|^{3\leftrightarrow 4}_{1\leftrightarrow 2} )
- S_{2;1|4|3,5} \tilde S_{4;3|2|1,5} ( H_1  \big|^{3\leftrightarrow 4}_{1\leftrightarrow 2} )
\notag 
\end{align}
One can readily recast the expressions above in terms of manifestly BRST-invariant building blocks
valid for all external states via $(S_{a;b|c|d,e} ,T^m_{a,b,c|d,e} )\rightarrow (s_{ab}C_{a;b|c|d,e},C^m_{a,b,c|d,e} )$.
When truncating to the bosonic component sector one may use the
effective building blocks of section~\ref{sec:eff} as
$(S_{a;b|c|d,e} ,T^m_{a,b,c|d,e} )\rightarrow (S^{\rm eff}_{a;b|c|d,e},T^{m,{\rm eff}}_{a,b,c|d,e} )$.

\newpage

\section{The $\ap$ expansion of genus-two  amplitudes}
\setcounter{equation}{0}
\label{secAp}

In this section we shall combine the expansions of the integrals studied in section \ref{sec:3} in order to extract the low energy expansion of  the genus-two five-point amplitude \eqref{LRcontractA}. As a warm-up, we first
consider the low energy expansion of the genus-two four-point amplitude, studied in \cite{D'Hoker:2013eea,DHoker:2014oxd,DHoker:2017pvk,DHoker:2018mys}.
We will follow the normalization conventions based on the first-principles computations in the
non-minimal pure spinor formalism \cite{Gomez:2009qd,Gomez:2010ad,Gomez:2015uha}.

\subsection{The four-point amplitude}
\label{4ptsubsec}

The four-point amplitude at two loops in the pure spinor formalism is given by \cite{Gomez:2010ad,Gomez:2015uha}\footnote{In equation (\ref{4pt2loop}) we have absorbed an overall factor of $2^{20}3^65^2$ coming from $\cB_{(4)}(k_i| \Omega)$ into the normalization of the four-point amplitude given in \cite{Gomez:2015uha}.}
\bea
\label{4pt2loop}
\cA_{(4)}^{\rm genus-2} =  
\delta(k) \, \cN_{(4)} 
\int_{\cM_2} d\mu_2\,  \cB_{(4)}(k_i| \Omega)
\eea
The normalization factor $\cN_{(4)}$ is given by\footnote{Alternatively, $\cN_{(4)}$ may be presented in terms of the 10-dimensional Newton constant $\kappa_{10}^2$ and the vacuum expectation value of the dilaton $\phi$, canonically normalized in Type IIB \cite{DHoker:2005jhf, D'Hoker:2013eea}, as follows $\cN_{(4)} = 2^{-6}  \pi\kappa_{10}^2 \, e^{2 \phi}$.},
\bea
\cN_{(4)} =  {\kappa^4 e^{2\l}\over 2^{25}\pi^6} \, \halfap5
\eea
in terms of  the normalization constant of the massless vertex operators $\kappa$ \cite{DHoker:2005jhf},
and the bare expectation values of the dilaton $\phi$. The S-duality analysis of \cite{Gomez:2015uha} relates,
\be
 e^{2\lambda}=2^6 \pi^4 e^{2\phi}
\ee
The integrand on $\cM_2$ in turn is given by the integral $\cB_{(4)}(k_i| \Omega)$ over the four vertex points defined in (\ref{defB4A}). With the expression \eqref{defK4} for the left chiral correlator $\cK_{(4)}$, the integrand of (\ref{4pt2loop}) can be expressed straightforwardly in terms of the $I$-integrals defined in \eqref{defIint} as 
\bea
\cB_{(4)}(k_i| \Omega)  &=&  I_1  |T_{1,4|2,3}|^2  
+ \big( I_1 \, \big|_{2\leftrightarrow 4} \big) |T_{1,2|3,4}|^2
 \no \\ &&
+  I_2
 (T_{1,2|3,4} \tilde T_{1,4|2,3} + T_{1,4|2,3} \tilde T_{1,2|3,4}) 
\eea
Using the symmetry property $T_{1,3|2,4} = - T_{1,4|2,3} - T_{1,2|3,4}$ of (\ref{symmT}),  momentum conservation, 
as well as  the expansions \eqref{4ptexp} we obtain the following expansion for the integrand, 
\bea
\label{cK4sq}
\cB_{(4)}(k_i| \Omega)  
&= & 32\big(|T_{1,2|3,4}|^2 + |T_{1,3|2,4}|^2 + |T_{1,4|2,3}|^2\big) 
\no \\ &&
+ 64\varphi\big( s_{12}|T_{1,2|3,4}|^2 + s_{13}|T_{1,3|2,4}|^2 + s_{14}|T_{1,4|2,3}|^2 \big)
\no \\ &&
+ 8A_1\big( s_{12}^2|T_{1,2|3,4}|^2 + s_{13}^2|T_{1,3|2,4}|^2 + s_{14}^2|T_{1,4|2,3}|^2 \big)
\no \\ &&
+ 16 A_2 \big|s_{14}T_{1,2|3,4} - s_{12}T_{1,4|2,3}\big|^2 + \cdots
\eea
where the terms in the ellipsis feature ${\cal O}(s_{ij}^{3})$ along with $|T_{a,b|c,d}|^2$, and
we have defined the following combinations, 
\bea
\label{A1A2def}
A_1 & = & \cZ_1 - 2 \cZ_2 + \cZ_3
\no \\  
A_2 & = &  \varphi^2 + \cZ_1  +  \cZ_2 + \half \cZ_4 = 2 \f^2 - \cZ_3
\eea
As shown in \cite{Mafra:2008ar}, the components $\langle T_{1,2|3,4}\rangle_0$  of the genus-two building block are given by, 
\begin{equation}
\label{2loop1loop}
\langle T_{1,2|3,4}\rangle_0 = 16 \invhalfap{} s_{12} \langle V_1T_{2,3,4}\rangle_0
\end{equation}
where $K= \langle V_1T_{2,3,4}\rangle_0$ is the usual  one-loop kinematic factor, which is permutation-symmetric 
and reduces to the usual $t_8F^4=t_8(f_1,f_2,f_3,f_4)$ combination for external gauge fields. Substituting  \eqref{2loop1loop} into \eqref{cK4sq},  the last line cancels  and one arrives at,
\begin{equation}
\label{cb4}
\cB_4(k_i| \Omega) = 2^{13} K\tilde K\invhalfap2\Big[
\sigma_2
+ 2\varphi \, \sigma_3
+ \frac{1}{4} A_1 \sigma_4 + {\cal O}(s_{ij}^5)
\Big]
\end{equation}
where $\sigma_k$ are the usual symmetric polynomials in four-point kinematic variables,
\begin{equation}
\label{defsigma}
\sigma_k = s_{12}^k+s_{13}^k+s_{14}^k
\end{equation}
Substituting \eqref{cb4} into \eqref{4pt2loop} and integrating over $\cM_2$, one finally obtains the 
low energy expansion of the amplitude in ten dimensions, 
\begin{align}
\cA_{(4)}^{\rm 2-loop} &=
\delta(k) \halfap3\!{\kappa^4 e^{2\l}\over 2^{12}\,\pi^6}K\tilde K
\int_{\cM_2}  {d\mu_2}
\Big[ \sigma_2 + 2\varphi \,\sigma_3 + \frac{1}{4} A_1\, \sigma_4+{\cal O}(s_{ij}^5) \Big]\no\\
&=\delta(k) \halfap3\!{\kappa^4 e^{2\l}\over 2^{10}\,3^3\,5\pi^3}K\tilde K
\Big[\sigma_2 + 3\sigma_3 +\frac14 c_1(\Lambda) \, \sigma_4 +{\cal O}(s_{ij}^5)\Big]
\label{4pt2loopfin}
\end{align}
where we used the formula \eqref{defmu2} for the volume of $\cM_2$, and the identities 
\bea
\label{intphiKZ}
\int_{\cM_2} d\mu_2 \, \varphi = \frac32 \, {\rm Vol}_2 
\hskip 1in
\int_{\cM_2(\Lambda)} d\mu_2 \,A_1 = c_1(\Lambda) \, {\rm Vol}_2
\eea
The coefficient $3/2$ in \eqref{intphiKZ} was computed in \cite{DHoker:2014oxd} and shown to be consistent with predictions from S-duality. The coefficient $c_1(\Lambda)$ depends on an infrared 
sliding scale $\Lambda$  which is necessary to disentangle the non-local part of the amplitude, which is governed entirely by exchange of massless particles with momentum less than $\Lambda$, from stringy corrections \cite{Green:2010sp,Pioline:2018pso}.

\subsection{The five-point amplitude}

The genus-two five-point amplitude  is given by \cite{DHoker:2020prr,Gomez:2015uha}
\begin{equation}
\label{5pt2loop}
\cA_{(5)}^{\rm genus-2} =
\delta (k) \cN_{(5)} 
\int_{\cM_2}  d\mu_2\, \cB_{(5)}(k_i| \Omega)
\end{equation}
with $\cB_{(5)}(k_i| \Omega)$ defined by (\ref{defB5}). The overall normalization was obtained using the pure spinor formalism in \cite{Gomez:2015uha}
\bea
\cN_{(5)} = \halfap5 {\kappa^5 e^{2\l}\over 2^{11}\pi^5}
\eea
As shown in section \ref{sec:decB5},  the integrand decomposes as a sum of 4 different types of integrals,
\be
\cB_{(5)} = \cB_{(5)}^J + \cB_{(5)}^F + \cB_{(5)}^G + \cB_{(5)}^H 
 \label{defB5again}
\ee
each one including its own kinematic factor.

\subsubsection{Terms of order $D^2 \cR^5$}

At leading order $D^2 \cR^5$, the $G$ and $H$ integrals do not contribute, and the remaining integrals 
are constants, independent of the period matrix $\Omega$, 
\bea
\begin{array}{l}
J_{1,1} = 128 + {\cal O}(s_{ij}) \ , \\  J_{1,2} = 32+ {\cal O}(s_{ij}) \Big. \ , \\ J_{1,3} = -64 + {\cal O}(s_{ij})  \ ,
\end{array}
\ \ \ \ \ \
\begin{array}{l}
\displaystyle F_{1} = {128 \over s_{12}} + {\cal O}(s_{ij}^0) \Bigg.\ , \\
\displaystyle  F_{2} = {64 \over s_{12}}  + {\cal O}(s_{ij}^0)\Bigg. \ , 
  \end{array}
  \ \ \ \ \ \
\begin{array}{l} 
\displaystyle F_{3} = -{128 \over s_{12}} + {\cal O}(s_{ij}^0) \Bigg.  \\
 \displaystyle F_{4} = -{64 \over s_{12}} + {\cal O}(s_{ij}^0) \Bigg.
\end{array}
\eea
Upon using the kinematic identity $T^m_{1,2,3|4,5}= T^m_{3,4,5|1,2}+{\rm cycl}(1,2,3)$,
the  low energy limit of the contraction $|\cV_I^m|^2$  in (\ref{splitB.2})  can be rewritten as 
\bea
\cB_{(5)}^J \big|_{D^2 \cR^5}
= 64\halfap{} \, T^m_{3,4,5|1,2} \tilde T^m_{3,4,5|1,2} +(1,2|1,2,3,4,5)   
\eea
where the notation $+(1,2|1,2,\ldots,k)$ in the first line instructs to sum over all possibilities
to exchange $(1,2)$ by a different pair $(i,j)$ from $i,j\in \{1,2,\ldots,k\}$ for
a total of ${ k \choose 2}$ terms. Similarly, the integrals in (\ref{splitB.4}) to (\ref{splitB.8}) produce,
\bea
\cB_{(5)}^F =
{ 64 \over s_{12} }\halfap2 \, \big[ | T_{12,3|4,5} |^2  
 + | T_{12,4|3,5} |^2 + | T_{12,5|3,4} |^2  \big]  +(1,2|1,2,3,4,5) 
 \label{thisexp}
 \eea
 and the contributions from the integrals $G$ and $H$ vanish,
 \bea
 \cB_{(5)}^G\big|_{D^2 \cR^5} =
 \cB_{(5)}^H \big|_{D^2 \cR^5} =0
\eea
The expression (\ref{thisexp}) for $\cB_{(5)}^F$ highlights the benefit of using the OPE-like representation for the second equation in (\ref{defcv}) since then all terms proportional to $S_{i;j|k|l,m}$ building blocks trivially cancel as they correspond to non-singular terms on the surface. Hence, at leading order, the integrand \eqref{defB5again}  reduces to,
\bea
\label{FT}
{\cal B}_{(5)} \big|_{D^2 \cR^5}  & = & 
64 \halfap2 \,  {  | T_{12,3|4,5} |^2{+} | T_{12,4|3,5} |^2  {+} | T_{12,5|3,4} |^2  \over s_{12}}  
\no \\ &&
+64 \halfap{} \, T^m_{3,4,5|1,2} \tilde T^m_{3,4,5|1,2}   +(1,2|1,2,\ldots,5) 
\eea
where the permutations $+(1,2|1,2,\ldots,5) $ apply to the entire right side. 

\begin{figure}
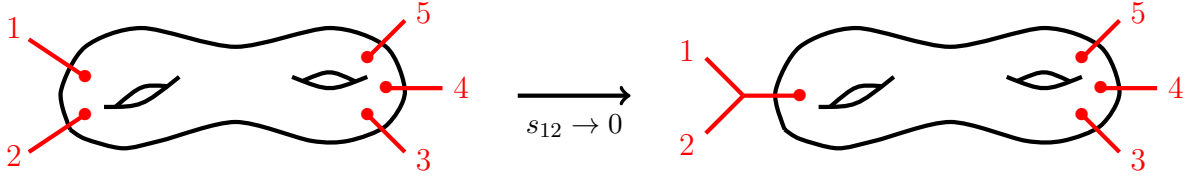

\begin{center}
\tikzpicture[scale=0.5]
\scope[xshift=0cm,yshift=0cm]
\draw[ultra thick] plot [smooth] coordinates {(0.6,0.6) (0.35,1.6) (1,2.8) (2.5, 3.3)  (5,2.8) (7.5, 3.3) (9,2.8) (9.5,1.6) (9,0.7) (8,0.3) (7, 0.3) (5,0.8) (3, 0.3) (2,0.1) (1,0.35) (0.6,0.6) };
\draw[ultra thick] (1.5,1.2) .. controls (2.5, 1.2) .. (3.5,2);
\draw[ultra thick] (1.8,1.2) .. controls (2.5, 1.8) .. (3.2,1.75);
\draw[ultra thick] (6.5,2) .. controls (7.5, 1.6) .. (8.5,2);
\draw[ultra thick] (6.8,1.9) .. controls (7.5, 2.2) .. (8.2,1.87);
\draw[ultra thick, red] (-0.5,0) -- (1, 1);
\draw[red]  (1,1) node{$\bullet$};
\draw[ultra thick, red] (-0.5,3) -- (1, 2);
\draw[red]  (1,2) node{$\bullet$};
\draw[ultra thick, red] (8.5,1) -- (9.5, 0);
\draw[red]  (8.5,1) node{$\bullet$};
\draw[ultra thick, red] (8.5,2.5) -- (9.5, 3.5);
\draw[red]  (8.5,2.5) node{$\bullet$};
\draw[ultra thick, red] (9,1.7) -- (10.5, 1.7);
\draw[red]  (9,1.7) node{$\bullet$};

\draw[red] (-0.9,3.3) node{$1$};
\draw[red] (-0.9,-0.1) node{$2$};
\draw[red] (10,-0.2) node{$3$};
\draw[red] (11,1.7) node{$4$};
\draw[red] (10,3.7) node{$5$};

\draw[ultra thick,->] (12.5,1.5) -- (15.5,1.5);
\draw[ultra thick] (14,0.7) node{{\small $ s_{12} \to 0$}};
\endscope
\scope[xshift=19cm,yshift=0cm]
\draw[ultra thick] plot [smooth] coordinates {(0.6,0.6) (0.35,1.6) (1,2.8) (2.5, 3.3)  (5,2.8) (7.5, 3.3) (9,2.8) (9.5,1.6) (9,0.7) (8,0.3) (7, 0.3) (5,0.8) (3, 0.3) (2,0.1) (1,0.35) (0.6,0.6) };
\draw[ultra thick] (1.5,1.2) .. controls (2.5, 1.2) .. (3.5,2);
\draw[ultra thick] (1.8,1.2) .. controls (2.5, 1.8) .. (3.2,1.75);
\draw[ultra thick] (6.5,2) .. controls (7.5, 1.6) .. (8.5,2);
\draw[ultra thick] (6.8,1.9) .. controls (7.5, 2.2) .. (8.2,1.87);
\draw[ultra thick, red] (-0.5,1.5) -- (1, 1.5);
\draw[red]  (1,1.5) node{$\bullet$};
\draw[ultra thick, red] (-0.5,1.5) -- (-1.5, 2.5);
\draw[ultra thick, red] (-0.5,1.5) -- (-1.5, 0.5);

\draw[ultra thick, red] (8.5,1) -- (9.5, 0);
\draw[red]  (8.5,1) node{$\bullet$};
\draw[ultra thick, red] (8.5,2.5) -- (9.5, 3.5);
\draw[red]  (8.5,2.5) node{$\bullet$};
\draw[ultra thick, red] (9,1.7) -- (10.5, 1.7);
\draw[red]  (9,1.7) node{$\bullet$};

\draw[red] (-2,2.7) node{$1$};
\draw[red] (-2,0.2) node{$2$};
\draw[red] (10,-0.2) node{$3$};
\draw[red] (11,1.7) node{$4$};
\draw[red] (10,3.7) node{$5$};

\endscope
\endtikzpicture
\caption{Factorization of the genus-two five-point amplitude onto a massless intermediate state in the $s_{12}$-channel into the tree-level 3-point function and the genus-two 4-point function with massless external states. \label{fig:3}}
\end{center}
\end{figure}

\sm

This is in agreement with the result obtained in \cite{Gomez:2015uha}, and corresponds to a $D^2\cR^5$ interaction in the low
energy effective action, which is expected to be related to $D^4\cR^4$ by non-linear supersymmetry. The residue of the pole in
$s_{12}$ is precisely given by the two-particle superfields $|T_{12,i|j,k}|^2$ as is expected from factorization of the 5-point
amplitude on a massless external state of two massless states, as shown schematically in figure \ref{fig:3}.

\subsubsection{Terms of order $D^4 \cR^5$}

At next to leading order $D^4 \cR^5$, all integrals are proportional to the Kawazumi-Zhang invariant $\f$ {in (\ref{KZ2})}. The expansions in (\ref{expJs}) receive contributions from the vector block, and give rise to,
\bea
\label{assemBsub}
\cB_{(5)}^J\big|_{D^4 \cR^5}
&=& 128 \f\,  \Big(
s_{35} (T^m_{5,1,2|3,4} \tilde T^m_{1,2,3|4,5} 
+ T^m_{1,2,3|4,5} \tilde T^m_{5,1,2|3,4} ) 
\no\\ &&  \hskip 0.4in 
+ (s_{45}{-}s_{13}) (T^m_{5,1,2|3,4} \tilde T^m_{2,3,4|5,1} + T^m_{2,3,4|5,1} \tilde T^m_{5,1,2|3,4} )   
\no\\ && \hskip 0.4in 
 -(s_{23}{+}s_{45}) T^m_{5,1,2|3,4} \tilde T^m_{5,1,2|3,4} + {\rm cycl}(1,2,3,4,5) \Big) 
\no\\
&=& 128 \f\, s_{12} T^m_{3,4,5|1,2} \tilde T^m_{3,4,5|1,2} + (1,2|1,2,3,4,5)
\eea
The expansions in (\ref{expFs}) produce all the singular terms, and give rise to,
\bea
\cB_{12}^F \big|_{D^4 \cR^5}
&=& 
{128 \f  \over s_{12}} \Big( s_{34} |T_{12,5|3,4} |^2 + s_{35} |T_{12,4|3,5} |^2 + s_{45} |T_{12,3|4,5} |^2 \Big)
\no \\ && 
- 32 \f  \Big( |S_{1;2|3|4,5}|^2 +  |S_{1;2|4|3,5}|^2 +  |S_{1;2|5|3,4}|^2 
\no \\ && \hskip 0.4in 
+ |S_{2;1|3|4,5}|^2+ |S_{2;1|4|3,5}|^2+ |S_{2;1|5|3,4}|^2 \Big)  
 \label{permsum1}
\eea
Finally, the contributions in (\ref{expGs}) and (\ref{expHs}) are analytic in $s_{ij}$ and receive contributions from the scalar block, and give rise to,
\bea
\cB_{1,23}^G  \big|_{D^4 \cR^5}
&=&-32 \f ( S_{1;2|4|3,5}  \tilde S_{1;3|4|2,5} + S_{1;2|5|3,4}  \tilde S_{1;3|5|2,4}  ) 
\no\\
\cB_{12,34}^H \big|_{D^4 \cR^5}
 &=& 0 
 \label{permsum2}
\eea
Whenever possible, the expressions have been simplified by repeatedly applying the relations $T_{12,3|4,5}=S_{1;2|3|4,5}-S_{2;1|3|4,5}$ and $S_{1;2|3|4,5}+{\rm cycl}(3,4,5)\cong0$.

\sm

Adding up these contributions according to (\ref{splitB.5}), we get,
\begin{align}
\cB_{(5)}\big|_{D^4 \cR^5} &= 
128\varphi \Big[  
\halfap{} s_{45} T^m_{1,2,3|4,5} \tilde T^m_{1,2,3|4,5}  + (4,5|1,2,3,4,5) \label{lowenHcomp} \\
&+ \halfap2 \Big({ s_{12}\over s_{45}} |T_{45,3|1,2} |^2 + {s_{13}\over s_{45}} |T_{45,2|1,3} |^2
+ {s_{23}\over s_{45}} |T_{45,1|2,3} |^2  +(4,5|1,2,3,4,5) \Big)\no\\
& - \halfap2 \Big(|S_{1;2|3|4,5}+ S_{1;3|2|4,5}|^2+
|S_{1;2|4|3,5}+ S_{1;4|2|3,5}|^2\no\\
&\qquad{}\ \ \ + |S_{1;3|4|2,5}+ S_{1;4|3|2,5}|^2 +(1\leftrightarrow
2,3,4,5)\Big)  \Big]\no
\end{align}
where the last two relations of (\ref{multi.10}) have been used to simplify the sums over permutations of
(\ref{permsum1}) and (\ref{permsum2}).

\subsubsection{Terms of order $D^6 \cR^5$}

In contrast to the lower-order terms in the previous subsections,
the low energy expansion of the genus-two amplitude at the order $D^6\cR^5$
features linearly independent MGFs
 $\cZ_1,\cZ_2,\cZ_3,\cZ_5$ and $\varphi^2$ defined in \eqref{defZs}.
Their respective coefficients are given by BRST-invariant linear combinations
of the building blocks that compose the correlator. As usual in such situations,
the resulting expressions that arise after expanding the integrals are not
necessarily in the most compact form. However, utilizing various cohomology
manipulations in pure spinor superspace as done in \eqref{lowenHcomp}, it may be possible
to simplify the answer after trial and error.

First, it is beneficial to rewrite the five MGFs
in terms of the linear combinations $A_1$ and $A_2$ identified in
the four-point calculations in section~\ref{4ptsubsec}, along with
$\cZ_1, \cZ_5$ and $\varphi^2$. In doing so the number of terms reduce by approximately $10\%$.
We shall now display the coefficients of $A_1$ and $\cZ_5$.

Curiously, the BRST-invariant coefficient of $A_1$ at $D^6\cR^5$ turns out to be closely related to the
coefficient of $\varphi$ at order $D^4\cR^5$ given in \eqref{lowenHcomp}. In fact, one can show that
\begin{align}
\cB_{(5)}\big|_{A_1} =
16 &\Big[
\halfap{} s_{45}^2 T^m_{1,2,3|4,5} \tilde T^m_{1,2,3|4,5}  + (4,5|1,2,3,4,5) \label{coeffA1} \\
&\;{}+ \halfap2 \Big({ s_{12}^2\over s_{45}} |T_{45,3|1,2} |^2 + {s_{13}^2\over s_{45}} |T_{45,2|1,3} |^2
+ {s_{23}^2\over s_{45}} |T_{45,1|2,3} |^2  +(4,5|1,2,3,4,5) \Big)\no\\
&\;{} - \halfap2 \Big( (s_{23}+s_{45})|S_{1;2|3|4,5}+ S_{1;3|2|4,5}|^2+
(s_{24}+s_{35})|S_{1;2|4|3,5}+ S_{1;4|2|3,5}|^2\no\\
&\qquad{}\ \ \ + (s_{34}+s_{25})|S_{1;3|4|2,5}+ S_{1;4|3|2,5}|^2 +(1\leftrightarrow
2,3,4,5)\Big)  \Big]\no
\end{align}

Given that $\cZ_5$ defined by the fifth line of (\ref{defZs}) only appears in the
$G_2$ integral (\ref{expGs}), its overall coefficient is easily assembled from
(\ref{splitB.6}) and (\ref{splitB.5}):
\begin{align}
\cB_{(5)}\big|_{\cZ_5} &= \sum_{2\leq i <j}^5 (\cB_{1;ij}^{G} + \cB_{1;ji}^{G})  \big|_{\cZ_5}  + (1\leftrightarrow 2,3,4,5)
\notag \\
&= 4 s_{1 2} (S_{1; 3| 5| 2, 4} \tilde S_{1; 4| 5| 2, 3} + 
    S_{1; 4| 5| 2, 3} \tilde S_{1; 3| 5| 2, 4} - 
    S_{1; 3| 2| 4, 5} \tilde S_{1; 4| 2| 3, 5} - 
    S_{1; 4| 2| 3, 5} \tilde S_{1; 3| 2| 4, 5}) \notag\\
    & + 
 4 s_{1 3} (S_{1; 2| 5| 3, 4} \tilde S_{1; 4| 5| 2, 3} + 
    S_{1; 4| 5| 2, 3} \tilde S_{1; 2| 5| 3, 4} - 
    S_{1; 2| 3| 4, 5} \tilde S_{1; 4| 3| 2, 5} - 
    S_{1; 4| 3| 2, 5} \tilde S_{1; 2| 3| 4, 5}) \notag \\
    & + 
 4 s_{1 4} (S_{1; 2| 5| 3, 4} \tilde S_{1; 3| 5| 2, 4} + 
    S_{1; 3| 5| 2, 4} \tilde S_{1; 2| 5| 3, 4} - 
    S_{1; 2| 4| 3, 5} \tilde S_{1; 3| 4| 2, 5} - 
    S_{1; 3| 4| 2, 5} \tilde S_{1; 2| 4| 3, 5})  \notag \\
    &    + (1\leftrightarrow 2,3,4,5) \label{coefU1}
\end{align}
Alternatively, one can recast \eqref{coefU1} in terms of Yang-Mills tree amplitudes as detailed in
the next subsection. To do so one uses
the conversion to genus-two BRST invariants $S_{a;b|c|d,e}\rightarrow s_{ab}C_{a;b|c|d,e}$, and the cohomology
identity \eqref{BRST.12} to rewrite $C_{a;b|c|d,e}$ in terms of genus-one BRST invariants, and finally use \eqref{BRST.13}
to convert to Yang-Mills tree amplitudes. In doing this one
obtains that the coefficient of $\cZ_5$ is proportional to $\cBnottree{7'}$ to be defined in \eqref{nottrees}, 
\begin{equation}
\label{U1asM7prime}
\cB_{(5)}\big|_{\cZ_5} = - 2560\cBnottree{7'}
\end{equation}
This result is interesting because the representation \eqref{coefU1} is
manifestly local while the locality of $\cBnottree{7'}$ is not evident. In addition, the above steps can be used
to derive the expression of the $2\times 2$ matrix $M_7'$ in (\ref{nottrees}) algorithmically, and it 
would be rewarding to look for similar derivations of other $M'_n$ matrices.

\sm

The coefficients of $A_2, \cZ_1$ and $\varphi^2$ can be brought into a form similar to
\eqref{coeffA1} and \eqref{coefU1} and can be downloaded from \cite{PSSsite}. Their components for external Type IIA and IIB states will be further simplified in the next
subsections.

\subsection{Components in Type IIB}
\label{sec.IIB}

Here we shall express the components of the genus-two amplitude of the ten-dimensional Type IIB superstring in terms of color-ordered tree-level amplitudes $A_{\rm YM}(1,2,3,4,5)$ of ten-dimensional SYM. For this
 it is convenient to use the representation \eqref{defcvCm} of the genus-two correlator written in terms of the BRST invariants $C^m_{1,2,3|4,5}$ and $C_{1;2|3|4,5}$. As reviewed in section~\ref{sec:Cs}, these genus-two invariants can be expressed in terms of the genus-one five-point BRST invariants $C^m_{1|2,3,4,5}$ and $C_{1|23,4,5}$. The scalar genus-one invariants $C_{1|23,4,5}$ occurring in the $|\cW|^2$ part of the genus-two correlator were shown in \cite{Mafra:2014oia} to be equivalent to SYM tree amplitudes, see (\ref{BRST.13}). This relation holds for the entire massless multiplets of both Type IIB and IIA.
 
 \sm

To relate the vector invariants $C^m_{1|2,3,4,5}$ occurring in the $|\cV^m_I|^2$ part of the amplitude
to SYM tree-level amplitudes, we use an observation from \cite{Green:2013bza}, which holds only for  Type IIB: even though an individual genus-one invariant $C^m_{1|2,3,4,5}$ cannot be written in terms of   $A_{\rm YM}$, the left-right holomorphic square $\langle |C^m_{1|2,3,4,5}|^2\rangle_0 $ can in fact be written in terms of $A_{\rm YM}\tilde A_{\rm YM}$ provided the external states are five gravitons or four gravitons and one dilaton of Type IIB.
More explicitly \cite{Green:2013bza} (up to an overall normalization), we have,
\begin{align}
&\langle  C^m_{1|2,3,4,5} \tilde C^m_{1|2,3,4,5} +\big[ s_{23} C_{1|23,4,5} \tilde C_{1|23,4,5}
+ (2,3|2,3,4,5) \big] \rangle_0 \, \big|_{{\rm IIB}} \notag \\
&= \tilde A^T_{54} \! \cdot \! S_0   \! \cdot \!  M_3   \! \cdot \! A_{45}
\times \begin{cases} \ \ \, 1 & : \hbox{five gravitons}\cr
-{1\over3} & : \hbox{four gravitons and one dilaton}
\end{cases}\label{BRST.19}
\end{align}
where the two options depend on the total $U(1)_R$ charge of the external states, 0 or $\pm 2$, respectively.\footnote{Due to symmetry under worldsheet parity, which acts by $(-1)^Q$ on a field with $U(1)_R$ charge $Q$, the $U(1)_R$ symmetry can only be violated by an even number. In particular, the four-graviton, one Kalb-Ramond amplitude would violate $U(1)_R$ by $\pm 1$ units and therefore must vanish at all genera, as explained in  \cite{Vafa:1995fj}. We have verified that this 
is indeed the case at genus-two up to order $D^6 \cR^5$. \label{fooparity}} By linearized supersymmetry, the kinematic relation (\ref{BRST.19}) extends to the remaining massless Type IIB state configurations
with the same $U(1)_R$ charges. For instance, the first line also applies to three gravitons and two gravitini of opposite $U(1)_R$ charges, and the second one  to three gravitons, one gravitino and dilatino whose $U(1)_R$ charges have the same sign.

\sm

Since the scalar invariants $\langle |C_{1|23,4,5}|^2\rangle_0$ can be expanded in terms of $A_{\rm YM}\tilde A_{\rm YM}$ it follows that for the genus-two five-graviton and four-graviton-one-dilaton amplitudes in Type IIB,  
$\langle |C^m_{1|2,3,4,5}|^2\rangle_0$, can also be written in terms of SYM tree amplitudes. Such a relation does 
not exist for five-point amplitude of gravitons and dilatons in Type IIA. 

\sm

In \eqref{BRST.19} we have used the following notation for the two-component vectors
of SYM amplitudes that form bases of $\tilde A_{\rm YM}$ and $A_{\rm YM}$
under BCJ relations \cite{Bern:2008qj}
\begin{align}
 \tilde A_{54} = \begin{pmatrix}\tilde A_{\rm YM}(1,2,3,5,4)\cr
 \tilde A_{\rm YM}(1,3,2,5,4)\end{pmatrix} 
 \hskip 0.6in
  A_{45} = \begin{pmatrix} A_{\rm YM}(1,2,3,4,5)\cr A_{\rm YM}(1,3,2,4,5)\end{pmatrix}
\label{SYMvec}
\end{align}
Furthermore, (\ref{BRST.19}) features the field-theory momentum kernel of~\cite{BjerrumBohr:2010hn},
\begin{align}
S_0 =
\begin{pmatrix}
(k_1\cdot k_2)(k_{12} \cdot k_3)  &
(k_1\cdot k_2)(k_1\cdot k_3)\cr
(k_1\cdot k_2)(k_1\cdot k_3) & (k_1\cdot k_3) (k_{13} \cdot k_2)
\end{pmatrix}
\label{Skern}
\end{align}
with $k_{ij}=k_i+k_j$, while the matrix $M_3$ encoding the $\alpha'^3$ corrections
to open- and closed-superstring tree-level amplitudes is given by \cite{Schlotterer:2012ny}, 
\bea
M_3 = \begin{pmatrix} m_{11} &m_{12}  \\ m_{12} \big|_{2\leftrightarrow 3} &m_{11} \big|_{2\leftrightarrow 3} \end{pmatrix} 
\eea
where the permutation inequivalent components are given as follows, 
\bea
\label{BRST.21}
m_{11} & = & s_{34}\big[ s_{45}^2 + s_{34}s_{45} - s_{12}( s_{12}{+}2s_{23}{+}s_{34})\big] + s_{12}s_{15}(s_{12}{+}s_{15}) 
\no \\
m_{12} & = & - s_{13}s_{24}(s_{12}+s_{23}+s_{34}+s_{45}+s_{15})
\eea
Based on (\ref{BRST.13}) and (\ref{BRST.19}), the entire polarization dependence of massless five-point genus-two amplitudes in Type IIB superstrings can be reduced to products 
$A_{\rm YM} \tilde A_{\rm YM}$ as for tree level \cite{Mafra:2011nv, Schlotterer:2012ny}  and genus one \cite{Green:2013bza}.
The relative factor of $-\frac{1}{3}$ in (\ref{BRST.19}) between the $U(1)_R$-conserving 
and $U(1)_R$-violating components plays a crucial role for S-duality \cite{Green:2013bza}, and we will elaborate on its genus-two analogue in section \ref{sec:4}. 

\subsubsection{Five-point tree-level amplitudes of SYM}
\label{sec:sym5pt}

We shall now review a compact way of representing the polarization dependence
of the color-ordered five-point SYM amplitudes in (\ref{SYMvec}). Following the
recursive strategy of Berends and Giele \cite{Berends:1987me}, five-point SYM amplitudes
can be efficiently organized in terms of two-particle polarizations $\ep_{12}^m,
f_{12}^{mn}$ and $\chi_{12}^\alpha$,
\begin{align}
\ep_{12}^m &= i \ep_2^m(k_2 \cdot \ep_1) - i \ep_1^m(k_1 \cdot \ep_2)
 +\frac{i}{2}(k_1^m-k_2^m)(\ep_1 \cdot \ep_2) + (\chi_1 \gamma^m \chi_2)
\notag \\
f_{12}^{mn} &= \ep_{12}^m k_{12}^n - \ep_{12}^n k_{12}^m + i k_1\cdot k_2 (\ep_1^m \ep_2^n - \ep_1^n \ep_2^m)
\label{asym.1}
\\
\chi_{12}^\alpha &= \frac{i}{2} k_{12}^m \gamma_m^{\alpha \beta} \big[ \ep_1^p (\gamma_p \chi_2)_\beta - \ep_2^p (\gamma_p \chi_1)_\beta \big] \notag
\end{align} 
In terms of these data, the five-point SYM amplitude obtained from the superspace expression \cite{Mafra:2010jq} is given by \cite{Mafra:2015vca},
\bea
A_{\rm YM}(1,2,3,4,5) &=&
 \frac{i}{2 s_{12} s_{34} } \big[ \ep_{12}^m f^{mn}_{34} \ep_5^n 
+\ep_{34}^m f^{mn}_{5} \ep_{12}^n +\ep_{5}^m f^{mn}_{12} \ep_{34}^n  \big]  
\notag \\ &&
+  \frac{1}{s_{12} s_{34} } \big[ (\chi_{12} \gamma_m \chi_{34}) \ep_{5}^m 
+ (\chi_{34} \gamma_m \chi_{5}) \ep_{12}^m 
+ (\chi_{5} \gamma_m \chi_{12}) \ep_{34}^m    \big] \label{asym.2} 
\no \\ &&
+ {\rm cycl}(1,2,3,4,5)
\eea
Note that both lines of \eqref{asym.2} contribute to gluino amplitudes
since $\ep_{ij}^m$ and $f_{ij}^{mn}$ contain a term bilinear in $\chi^\alpha_i,\chi^\beta_j$.

\subsection{Components in Type IIA}
\label{sec.IIA}

We now turn to the case of Type IIA superstrings, where the Weyl spinors of the left and right movers have 
opposite chirality. While the previous relation (\ref{BRST.19}) between vector blocks and SYM tree amplitudes no longer works, the difference between the coefficients of these blocks in Type IIA
and Type IIB has a very simple structure, which amounts to flipping the sign
of the ten-dimensional Levi-Civita symbol $\epsilon_{10}$ appearing in $ \tilde T^m_{1,2,3|4,5}$ \cite{PSSsite}:
\bea
\langle \tilde T^m_{1,2,3|4,5} \rangle_0 &=&
 {1\over 360}  s_{45}
\Big[ \tilde \ep_1^m t_8(\tilde f_2,\tilde f_3,\tilde f_4,\tilde f_5) + (1\leftrightarrow 2,3,4,5) \Big]
+ \sum_{j=2}^5 k_j^m \tilde T_j 
\no \\ &&
-{1\over 5760} s_{45} \epsilon_{10}^m(\tilde \ep_1, \tilde f_2, \tilde f_3, \tilde f_4, \tilde f_5) \times \left\{ \begin{array}{rl}
{}+1    & \hbox{(Type IIB)} \\
{}-1   & \hbox{(Type IIA)} \end{array} \right.
\label{typeIIA.5}
\eea
Here, we have used the shorthand $t_8(  f_2,  f_3,  f_4,   f_5)$ as in (\ref{famous}) as well as,
\begin{align}
\epsilon_{10}^m(\ep_1,f_2,f_3,f_4,f_5) &=  (\epsilon_{10})^{m} {}_{np_2q_2 p_3q_3 p_4q_4 p_5q_5} \ep_1^n
f^{p_2q_2}_2 f^{p_3q_3}_3 f^{p_4q_4}_4 f^{p_5q_5}_5
\label{typeIIA.2}
\end{align}
where  $f_i$ denotes the linearized field-strength $f_i^{mn}=\ep_i^m k_i^n -  k_i^m \ep_i^n$ of external state~$i$.
The form of the scalar terms $\tilde T_j$ in the first line of (\ref{typeIIA.5})  will not be relevant in the discussions below due to the vanishing contraction $k_j^m \epsilon_{10}^m (\tilde \ep_1,\tilde f_2,\tilde f_3,\tilde f_4, \tilde f_5)  = 0$ for $j=1,2,3,4,5$.

\sm

Therefore, the difference between Type IIA and Type IIB correlators may be 
inferred from the following simple relation,
\begin{equation}
\label{iiaT}
\langle\tilde T^m_{1,2,3|4,5} \rangle_0 \big|_{\rm IIA} =
\langle\tilde T^m_{1,2,3|4,5} \rangle_0 \big|_{\rm IIB}
+ {s_{45}\over 2880}\epsilon_{10}^m(\tilde \ep_1,\tilde f_2,\tilde f_3,\tilde f_4, \tilde f_5)
\end{equation}
and we obtain, 
\begin{align}
 &\langle \cW \, \tilde \cW - \halfap{}\pi Y^{IJ} \, \cV^m _I \,\tilde \cV^m _J \rangle_0 \, \big|_{\rm IIB}
 -  \langle \cW \, \tilde \cW - \halfap{}\pi Y^{IJ} \, \cV^m _I \,\tilde \cV^m _J \rangle_0 \, \big|_{\rm IIA} \label{typeIIA.3} \\
 &=  {1\over2880} \halfap{}\pi Y^{IJ} \langle {\cal V}^m_I \rangle_0
 \epsilon_{10}^m (\tilde \ep_1,\tilde f_2,\tilde f_3,\tilde f_4, \tilde f_5)  
  \big[ s_{45}   \oom_J(2) \overline{ \Delta (3,4)\Delta (5,1)}  + \hbox{ cycl}(1,2,3,4,5)\big]   
\notag
\end{align}
see (\ref{defcv}) for the cyclic permutations of $T^m_{1,2,3|4,5}$ entering $ {\cal V}^m_I $.
As will be detailed below, the contributions from $\langle {\cal V}^m_I \rangle_0
\epsilon_{10}^m (\tilde \ep_1,\tilde f_2,\tilde f_3,\tilde f_4, \tilde f_5) $ take
different forms depending on the type of external NSNS states. We will show that,
both for five gravitons and for four gravitons and one Kalb-Ramond $B$-field,
the difference between Type IIB and Type IIA amplitudes is proportional to the integral,
\begin{align}
\cJ_{{\rm IIA}}= 
\frac{i}{2} \int _{\Sigma ^5} { \KN_{(5)} \over (\det Y)^2} Y^{IJ}  
  &\big[ s_{45}   \om_I(2)   \Delta (3,4)\Delta (5,1)  + \hbox{ cycl}(1,2,3,4,5)\big]    \notag \\
 \times &\big[ s_{45} \oom_J(2) \overline{ \Delta (3,4)\Delta (5,1)}  + \hbox{ cycl}(1,2,3,4,5)\big]   
\label{typeIIA.4}
\end{align}
This integral is invariant under permutations of all external legs, and can be expressed in terms
of the $J$-integrals defined in \eqref{defJint} 
\bea
\cJ_{{\rm IIA}}  = 
J_{1,1} s_{34}^2+ ( J_{1,2}+ \overline{J_{1,2}}) s_{34}s_{45} + (J_{1,3}+\overline{J_{1,3}}) s_{34} s_{15}   +  {\rm cycl}(1,2,3,4,5)  \label{typeIIA.0} 
\eea
The low energy expansion of this integral follows immediately from (\ref{expJs}), 
\begin{align}
\cJ_{{\rm IIA}}  &= 32 \sum_{1\leq i<j}^5 s_{ij}^2 + 64 \varphi \sum_{1\leq i<j}^5 s_{ij}^3
+ \frac{8}{3} (-\cZ_1 -  8 \cZ_2 - \cZ_3 - \cZ_4 + 10 \varphi^2) \sum_{1\leq i<j}^5 s_{ij}^4 \notag \\
& \ \ \ + \frac{2}{3} ( 4 \cZ_1 + 2 \cZ_2 + 4 \cZ_3 + \cZ_4 - 10 \varphi^2 ) \bigg( \sum_{1\leq i<j}^5 s_{ij}^2 \bigg)^2 +{\cal O}(s_{ij}^5)\label{typeIIA.X}  \\
&= 32 P_2  + 64 \varphi \, P_3 
+ \frac{8}{3} (\cZ_1 - 6 \cZ_2 + \cZ_3 + 8 \varphi^2) P_4 
 + \frac{4}{3} ( \cZ_1 +  \cZ_3 - 4 \varphi^2) P_2^2 +{\cal O}(s_{ij}^5) \no
\end{align}
where $P_n$ denotes the symmetric homogeneous polynomials\footnote{As pointed out in \cite{Boels:2013jua}, the ring of symmetric polynomials in the $s_{ij}$'s subject to the momentum conservation constraint is generated by the polynomials $P_2,P_3,\dots, P_9$ along with an additional degree 6 generator, which we shall not encounter at the order that we work in this paper.}
\begin{equation}
\label{defPn}
P_n = \sum_{1\leq i<j}^5s_{ij}^n
\end{equation}

\subsubsection{Five gravitons in Type IIA}

For five external gravitons (or more generally for external states with $\ep_i=\tilde\ep_i$), the kinematic factors 
$\langle {\cal V}^m_I \rangle_0  \epsilon_{10}^m (\tilde \ep_1,\tilde f_2,\tilde f_3,\tilde f_4, \tilde f_5)$ in (\ref{typeIIA.3}) 
 only receive contributions from,
 \begin{align}
 \langle T^m_{1,2,3|4,5} \rangle_0   \epsilon_{10}^m (\tilde \ep_1,\tilde f_2,\tilde f_3,\tilde f_4, \tilde f_5)
 \rightarrow -{1\over 5760}s_{45} \epsilon_{10}^m ( \ep_1,  f_2,  f_3,  f_4,   f_5)\epsilon_{10}^m ( \ep_1,  f_2,  f_3,  f_4,   f_5) 
 \label{typeIIA.31}
  \end{align}
since terms of the form $\epsilon_{10}(\ep_j,\tilde \ep_1,\tilde f_2,\tilde f_3,\tilde f_4, \tilde f_5)$
from the first line of (\ref{typeIIA.5}) vanish due to the 
 symmetry of the graviton polarization tensors under
  $\tilde \ep_j \leftrightarrow \ep_j$. 
 Hence, the difference between Type IIB and Type IIA 
integrands  reduces to, 
 \begin{align}
\cB_{(5)}  \big|^{h^5}_{{\rm IIB}} - 
\cB_{(5)}  \big|^{h^5}_{{\rm IIA}} = \halfap{}{2\over 5760^2}
\epsilon_{10}^m ( \ep_1,  f_2,  f_3,  f_4,   f_5)\epsilon_{10}^m ( \ep_1,  f_2,  f_3,  f_4,   f_5)
\cJ_{{\rm IIA}} 
\label{typeIIA.32}
\end{align}
Since $\cJ_{{\rm IIA}}$ behaves as  $32 \sum_{1\leq i<j}^5 s_{ij}^2$ in the low energy  limit, the expression
\eqref{typeIIA.32} reproduces, up to an overall constant, the result (5.47) of \cite{Gomez:2015uha}.
The complete Type IIA five-graviton amplitude can be assembled from (\ref{typeIIA.32}) and from the Type IIB components that are expressible in terms of SYM tree amplitudes by the discussion in section \ref{sec.IIB}.
The same conclusion holds for any five-point amplitude involving gravitons and dilatons, since it only depends on the symmetry property under  $\ep_j\leftrightarrow \tilde\ep_j$.

\subsubsection{Four gravitons and one $B$-field in Type IIA}

For Type IIA amplitudes with four external gravitons and one $B$-field in the first leg
with polarization $B_1^{mn}= {1\over 2}(\ep_1^m\tilde \ep_1^n- \ep_1^n\tilde \ep_1^m)$, 
the kinematic factors $\langle {\cal V}^m_I \rangle_0
 \epsilon_{10}^m (\tilde \ep_1,\tilde f_2,\tilde f_3,\tilde f_4, \tilde f_5)$ in (\ref{typeIIA.3}) 
 only receive contributions from
 \begin{align}
 \langle T^m_{1,2,3|4,5} \rangle_0   \epsilon_{10}^m (\tilde \ep_1,\tilde f_2,\tilde f_3,\tilde f_4, \tilde f_5)
 \rightarrow 
 {1\over 360}  
 s_{45} \epsilon_{10} ( B_1,  f_2,  f_3,  f_4,   f_5)t_8(  f_2,  f_3,  f_4,   f_5)
 \label{typeIIA.33}
  \end{align}
with shorthand
\bea
\epsilon_{10} ( B_1,  f_2,  f_3,  f_4,   f_5) = (\epsilon_{10})_{mn p_2q_2 p_3q_3 p_4q_4 p_5q_5} B_1^{mn}
f^{p_2q_2}_2 f^{p_3q_3}_3 f^{p_4q_4}_4 f^{p_5q_5}_5
\eea 
and the $t_8$-tensor in (\ref{famous}).

\sm

Since the Type IIB amplitude involving four gravitons and one $B$-field  vanishes to all orders in $\ap$ (see footnote \ref{fooparity}), we conclude from \eqref{typeIIA.3} that the integrand in Type IIA is,
 \begin{align} 
\cB_{(5)}  \big|^{Bh^4}_{{\rm IIA}} = -\halfap{} 
 {1\over 360\cdot 2880} 
 \epsilon_{10} ( B_1,  f_2,  f_3,  f_4,   f_5)t_8(  f_2,  f_3,  f_4,   f_5) 
 \cJ_{{\rm IIA}}
 \label{typeIIA.36}
  \end{align}

\subsection{Type IIB 5-point amplitudes up to genus-two}
\label{sec:amps}

In this section we verify that the $\alpha'$ expansion of the Type IIB five-point  amplitude at genus two leads to the same kinematic factors appearing in the expansion of the same amplitude at tree level and genus one.

\sm

The properly normalized five-point amplitudes at tree level, genus one and genus two are given by \cite{Gomez:2015uha}\footnote{In the conventions  of \cite{Gomez:2015uha} the $n$-point amplitudes have no length dimension independently of loop order; $[\cA_{(n)}]=0$. Note that $[\ap]=2$, $[\kappa]=-2$, $[\delta^{10}(k)]=10$, $[k^m]=-1$ and $[\varepsilon_i^m]=0$. In addition we absorbed a common factor of $2^9 5^2(2/\ap)^2$ in the expressions for the various interactions in $\cB^{\rm  genus-1}$ into the overall coefficient of the genus-one amplitude from \cite{Gomez:2015uha}, namely $(\ap/2)^3{\kappa^5 \over 2^{14}\,5^2\pi}$. Similarly a factor of $2^{34} 3^6 5^2(2/\ap)^4$ from $\cB^{\rm genus-2}$ was absorbed into the overall coefficient $(\ap/2)^5 {\kappa^5 e^{2\l}\over
2^{45}\,3^6\,5^2\pi^5}$ of the genus-two amplitude. See appendix \ref{app:convert} for more details on the normalization of the genus-two amplitude.}
\begin{align}
\cA^{\rm tree}_{(5)} &= \delta(k) \halfap{}\; \kappa^5 e^{-2\l}(2\pi)^2
\cB_{(5)}^{\rm tree}(k_i)\,, \\
\cA_{(5)}^{\rm genus-1} &= \delta(k)
\halfap{}{\kappa^5 \over 2^{5}\pi}
\int_{\cM_1} d\mu_1 \,
\cB^{\rm genus-1}_{(5)}(k_i| \tau)\,,\label{oneloopfive}\\
\cA_{(5)}^{\rm genus-2} &= \delta(k)
\halfap{} {\kappa^5 e^{2\l}\over 2^{11}\pi^5}
\int_{\cM_2} d\mu_2\, 
\cB^{\rm genus-2}_{(5)}(k_i| \Omega)
\label{twoloopfive}
\end{align}
where  $[\cB_{(5)}^{\rm tree}] = [\cB_{(5)}^{\rm genus-1}] =[\cB_{(5)}^{\rm genus-2}]=-2$,
\begin{align}
\label{cbtree}
\cB_{(5)}^{\rm tree}(k_i) &= \cBtree{0}
+ \cBtree{3}\zeta_3
+ \cBtree{5}\zeta_5
+ \cBtree{3,3}\zeta_3^2
+ \cBtree{7}\zeta_7
+ \cdots\\
\cB^{\rm genus-1}_{(5)}(k_i| \tau) &=
\cBoneloop{3}
+\cBoneloop{5}
+ \cBoneloop{3,3}
+ \cBoneloop{7}
+ \cBoneloop{7'} + \cdots\\
\cB_{(5)}^{\rm genus-2}(k_i|\Omega) &=
\cBtwoloop{5}
+ \cBtwoloop{3,3}
+ \cBtwoloop{7}
+ \cBtwoloop{7'} + \cdots
\label{cBtwo}
\end{align}
and the terms in the ellipsis take the schematic form $s_{ij}^m A_{\rm YM} \tilde A_{\rm YM}$ with $m\geq 8$.
As will become clear below, the notation for the subscripts of
$\cB_{\{\ldots \}}^{\rm tree}, \cB_{\{\ldots \}}^{\rm genus-1}$ and $\cB_{\{\ldots \}}^{\rm genus-2}$
indicates the polynomial dependence on $s_{ij}$ occuring at different genera.

\subsubsection{Tree level}

In writing the expansion \eqref{cbtree} we defined the shorthands
\begin{align}
\label{M7s}
\cBtree{n} = 2 \tilde A^T_{54}\cdot S_0 \cdot M_n \cdot A_{45} \, , \ \ \ \ \ \ 
\cBtree{3,3} = 2 \tilde A^T_{54}\cdot S_0 \cdot M_3^2 \cdot A_{45}
\end{align}
in terms of the two-component vectors  $\tilde A^T_{54}$ and $A_{45}$ of SYM tree-amplitudes
and the momentum kernel $S_0$ of (\ref{SYMvec}) and (\ref{Skern}), respectively. Here $M_n$ are $2\times 2$ matrices with entries  composed of degree $n$ polynomials in kinematic invariants \cite{Schlotterer:2012ny} generalizing \eqref{BRST.21},
explicit results are available for download on \cite{mzvwebsite}.
For uniformity we set $M_0 ={1\over2} {\rm Id}$ and note that $\cBtree{0} = 
 \tilde A^T_{54}\cdot S_0  \cdot A_{45}$ corresponds to the Kawai--Lewellen--Tye
representation of the supergravity tree-level amplitude \cite{Kawai:1985xq}.
In addition,
\begin{equation}
\label{nottrees}
\cBnottree{n'} = 2 \tilde A^T_{54}\cdot S_0 \cdot M'_n \cdot A_{45}
\end{equation}
where $M'_n$ are similar $2\times 2$ matrices with entries  composed of degree $n$ polynomials in $s_{ij}$, but which only start to contribute at genus one \cite{Green:2013bza}.
The explicit form of $M_7'$ can be found in the ancillary files of \cite{Green:2013bza}.
The notation $\cBnottree{n'}$ is meant to convey both the analogy with the
matrices $M_n$ appearing at tree level, and to emphasize its absence at that level.

\subsubsection{Genus one}

Collecting the results from \cite{Green:2013bza} and \cite{Gomez:2015uha} we get, for 5 Type IIB gravitons
\begin{align}
\cBoneloop{3}\big|_{h^5} &=  -\cBtree{3}\big|_{h^5} &&\hbox{$(\cR^5/k^2)$}  \notag\\
\cBoneloop{5}\big|_{h^5} &= 2E_2\, \cBtree{5}\big|_{h^5} &&\hbox{$(D^2\cR^5)$} \notag\\
\cBoneloop{3,3}\big|_{h^5} &= (5 E_3 + \zeta_3) \cBtree{3,3}\big|_{h^5}&&\hbox{$(D^4\cR^5)$} \label{cons1}\\
\cBoneloop{7}\big|_{h^5} &= (4 C_{2,1,1} + 2 E_2^2 - 2 E_4)\cBtree{7}\big|_{h^5}&&\hbox{$(D^6\cR^5)$} \notag\\
\cBoneloop{7'}\big|_{h^5} &=
\Big(\frac{15}{4} C_{2,1,1} - \frac{ 25}{8} E_2^2 + \frac{57}{8} E_4 \Big)\cBnottree{7'}\big|_{h^5}&&\hbox{$(D^6\cR^5)'$} \notag
\end{align}
and for 4 gravitons and 1 dilaton
\begin{align}
\cBoneloop{3}\big|_{\phi h^4} &= {1\over3}\,\cBtree{3}\big|_{\phi h^4} &&\hbox{$(\phi \cR^4)$} \notag\\
\cBoneloop{5}\big|_{\phi h4} &= {2\over5}E_2\,\cBtree{5}\big|_{\phi h^4} &&\hbox{$(\phi D^4\cR^4)$}\notag\\
\cBoneloop{3,3}\big|_{\phi h^4} &= {1\over3}(5 E_3 + \zeta_3)\cBtree{3,3}\big|_{\phi h^4}&&\hbox{$(\phi D^6\cR^4)$}\label{vio1}\\
\cBoneloop{7}\big|_{\phi h^4} &=
\frac{3}{7}(4 C_{2,1,1} + 2 E_2^2 - 2 E_4)\cBtree{7}\big|_{\phi h^4} &&\hbox{$(\phi D^8\cR^4)$}\notag\\
\cBoneloop{7'}\big|_{\phi h^4} &=
\Big(\frac{43}{4} C_{2,1,1} - \frac{61}{8} E_2^2 + \frac{93}{8} E_4\Big) \cBnottree{7'}\big|_{\phi h^4}&&\hbox{$(\phi D^8\cR^4)'$}\notag
\end{align}
Here, $E_k$ is the standard non-holomorphic Eisenstein series, defined  for $k \geq 2$ on a torus with modulus $\tau$ and momentum lattice $\Lambda = \ZZ + \tau \ZZ$ by, 
\begin{align}
E_k(\tau) =  \sum_{p \in \Lambda '} \frac{\tau_2^k }{ \pi^k |p|^{2k}} 
\label{defEis}
\end{align}
where $\Lambda ' = \Lambda \setminus \{ 0 \}$, while $C_{2,1,1}$ is the two-loop MGF \cite{DHoker:2015wxz} defined by,
\begin{align}
C_{2,1,1}(\tau) =  \sum_{p_1,p_2,p_3 \in \Lambda '} 
\frac{\tau_2^4 \, \delta (p_1+p_2+p_3) }{ \pi^4 |p_1|^4 \,  |p_2|^{2} \, |p_3|^2}
\label{defC211}
\end{align}
In view of (\ref{cons1}) and (\ref{vio1}), the five-point genus-one amplitude \eqref{oneloopfive} in Type IIB
becomes
\begin{align}
\cA_{(5)}^{\rm genus-1} &= \delta(k) \halfap{}{\kappa^5\over 2^5\pi}\int_{\cM_1} d\mu_1
\times \begin{cases} \cB_{(5)}^{\rm genus-1}\big|_{h^5} & : \hbox{five gravitons}\cr
\cB_{(5)}^{\rm genus-1}\big|_{\phi h^4} & : \hbox{four gravitons, one dilaton}
\end{cases}
\label{also1loop}
\end{align}
where \cite{Green:2013bza}
\begin{align}
\cB_{(5)}^{\rm genus-1}|_{h^5} &=
-\cBtree{3}\big|_{h^5}
+ 2E_2\, \cBtree{5}\big|_{h^5}
+(5 E_3 + \zeta_3) \cBtree{3,3}\big|_{h^5} \label{allcon1}\\
& + (4 C_{2,1,1} + 2 E_2^2 - 2 E_4)\cBtree{7}\big|_{h^5}
+ \Big(\frac{15}{4} C_{2,1,1} - \frac{ 25}{8} E_2^2 + \frac{57}{8} E_4 \Big)\cBnottree{7'}\big|_{h^5} + \cdots\no\\
\cB_{(5)}^{\rm genus-1}|_{\phi h^4} &=
{1\over3}\,\cBtree{3}\big|_{\phi h^4}
+ {2\over5}E_2\,\cBtree{5}\big|_{\phi h^4}
+{1\over3}(5 E_3 + \zeta_3)\cBtree{3,3}\big|_{\phi h^4} \label{allvio1}\\
&+ \frac{3}{7} (4 C_{2,1,1} + 2 E_2^2 - 2 E_4) \cBtree{7}\big|_{\phi h^4}
+ \Big(\frac{43}{4} C_{2,1,1} - \frac{61}{8} E_2^2 + \frac{93}{8} E_4\Big) \cBnottree{7'}\big|_{\phi h^4}+\cdots\no
\end{align}
with terms of order $s_{ij}^{m\geq 8} A_{\rm YM} \tilde A_{\rm YM}$ in the ellipsis.
Apart from the last term $\cBnottree{7'}$, these $\ap$-corrections involve the same polynomial
dependence on the $s_{ij}$ as the coefficients of $\zeta_3,\zeta_5,\zeta_3^2$ and $\zeta_7$ in
the tree-level amplitude.

\subsubsection{Genus two}

Explicit pure spinor superspace component evaluations of the genus-two kinematic factors
for 5 external Type IIB graviton states yield
\begin{align}
\cBtwoloop{5}\big|_{h^5} &= {-}\cBtree{5}\big|_{h^5} &&\hbox{$(D^2\cR^5)$} \notag\\
\cBtwoloop{3,3}\big|_{h^5} &= 3 \varphi \cBtree{3,3}\big|_{h^5} &&\hbox{$(D^4\cR^5)$}\label{ratioD2R5}\\
\cBtwoloop{7}\big|_{h^5} &= -{1\over4}A_1 \cBtree{7}\big|_{h^5} &&\hbox{$(D^6\cR^5)$}\notag\\
\cBtwoloop{7'}\big|_{h^5} &=  A_3 \cBnottree{7'}\big|_{h^5} &&\hbox{$(D^6\cR^5)'$}\notag
\end{align}
while for 4 gravitons and one dilaton,
\begin{align}
\cBtwoloop{5}\big|_{\phi h^4} &= {3\over5}\cBtree{5}  &&\hbox{$(\phi D^4\cR^4)$}\notag\\
\cBtwoloop{3,3}\big|_{\phi h^4} &= - \varphi \cBtree{3,3} &&\hbox{$(\phi D^6\cR^4)$}\label{ratioD4R5}\\
\cBtwoloop{7}\big|_{\phi h^4} &= {1\over 28}A_1 \cBtree{7} &&\hbox{$(\phi D^8\cR^4)$}\notag\\
\cBtwoloop{7'}\big|_{\phi h^4} &= A_4 \cBnottree{7'}&&\hbox{$(\phi D^8\cR^4)'$}\notag
\end{align}
where
\bea
A_3  &=& {5\over 64}\big(A_1 - 40 A_2 - 2\cZ_5\big)\label{A3def}\\
A_4  &=&
{45\over 64}A_1 - {17\over8}A_2 - {5\over 32}\cZ_5 + \cZ_3 + 2\cZ_2 - 6\varphi^2 
\label{A4def} 
\eea
Therefore, the five-point genus-two amplitude for Type IIB external states is given by,
\begin{align}
\cA_{(5)}^{\rm genus-2} &= \delta(k) \halfap{}{\kappa^5e^{2\l}\over 2^{11}\pi^5}\int_{\cM_2} d\mu_2
\times \begin{cases} \cB_{(5)}^{\rm genus-2}\big|_{h^5} & : \hbox{five gravitons}\cr
\cB_{(5)}^{\rm genus-2}\big|_{\phi h^4} & : \hbox{four gravitons, one dilaton}
\end{cases}
\label{also2loop}
\end{align}
where
\begin{align}
\label{twoh5}
\cB_{(5)}^{\rm genus-2}|_{h^5} &=
{-}\cBtree{5}\big|_{h^5}
+ 3\varphi \cBtree{3,3}\big|_{h^5}
- {1\over4} A_1\cBtree{7}\big|_{h^5}
+  A_3 \cBnottree{7'}\big|_{h^5} + \cdots\\
\label{twophih4}
\cB_{(5)}^{\rm genus-2}|_{\phi h^4} &=
{3\over5}\cBtree{5}\big|_{\phi h^4}
-\varphi \cBtree{3,3}\big|_{\phi h^4}
+ {1\over28} A_1 \cBtree{7}\big|_{\phi h^4}
+  A_4\cBnottree{7'}\big|_{\phi h^4}
+ \cdots
\end{align}
with terms of order $s_{ij}^{m\geq 8} A_{\rm YM} \tilde A_{\rm YM}$ in the ellipsis.
The relative factor $-3/5$ between the ratios of the genus-two to tree-level amplitudes for the $h^5$ and $\phi h^4$
components at the order of $D^2\cR^5$ agrees with the S-duality analysis of \cite{Gomez:2015uha}, while the factors $-1/3$ and
$-1/7$ for the $D^4\cR^5$ and $D^6\cR^5$ interactions are new. In the next section we will explain these relative coefficients
from the point of view of modular forms.

\sm

Note that the results of this section can be adapted to the entire massless Type IIB multiplet
upon replacing $h^5$ or $\phi h^4$ by state 
configurations with the same $U(1)_R$ charges.

\newpage

\section{Consistency with supergravity and S-duality}
\label{sec:4}

In this section, we shall check that our results for the low energy expansion of the genus-two five-point amplitudes are consistent with the structure of UV divergences in supergravity, in particular the absence of $U(1)_R$-violating divergences in the supergravity limit, and with predictions from S-duality.

\subsection{R-symmetry violation and UV divergences in supergravity}

When the five external states are in a configuration which violate the supergravity
$U(1)_R$ symmetry the corresponding interaction is local \cite{Boels:2012zr}. For the specific
case of $\phi h^4$ in Type IIB theory, the local interactions at different $\ap$ orders can be written
as $K_{\phi h^4}\sum_m a_m  \cO_m$ where\footnote{In $D=10$ we have 
 \[
\cBtree{3}|_{\phi h^4} = \frac{5 \phi_1}{32768} \bigg\{ t_8(f_2,f_3,f_4,f_5) t_8(f_2,f_3,f_4,f_5) 
        - \frac{ 1}{512} \eps_{10}^{mn}(f_2,f_3,f_4,f_5) \eps_{10}^{mn}(f_2,f_3,f_4,f_5)
        \bigg\}
        \] 
However, this representation depends on the dimension of spacetime as there is a 
contraction between left- and right movers. That is why
we chose the dimension-agnostic representation \eqref{t8t8sympol}.
\label{foophihfour}}
\begin{equation}
\label{t8t8sympol}
K_{\phi h^4} = \cBtree{3}|_{\phi h^4}
\end{equation}
$a_m$ are rational coefficients and $\cO_m$ are symmetric polynomial in the kinematic invariants $s_{ij}$,
which can be expressed in terms of the polynomials $P_n$ defined in \eqref{defPn}.
More specifically, the kinematic factors in (\ref{vio1}) and (\ref{ratioD4R5}) are
related to (\ref{t8t8sympol}) via
\begin{align}
\cBtree{5}|_{\phi h^4}
&= {5\over 12} \cO_2 K_{\phi h^4}
\
%
&\cBtree{3,3}|_{\phi h^4}
&= - {1\over 3}  \cO_3 K_{\phi h^4}
\notag\\
%
\cBtree{7}|_{\phi h^4}
&
 = { 7  \over 16} {\cal O}_{5,1} K_{\phi h^4}
 &\cBnottree{7'}|_{\phi h^4}
&
 = - { 1 \over 9} {\cal O}_{5,2} K_{\phi h^4} 
\label{U1ms} 
\end{align}
where \cite{Green:2019rhz}  
\begin{equation}
\label{congkaoGreen}
\cO_2 = P_2\,,\qquad
\cO_3 = P_3\,,\qquad
\cO_{5,1} = P_4+{1\over12}P_2^2\,,\qquad
\cO_{5,2} = P_4-{1\over4}P_2^2
\end{equation}
We will in fact evaluate $R$-symmetry violating IIB amplitudes with a $D$-dimensional dilaton state
with polarization 
\begin{equation}
\label{dilatonid}
\varepsilon_1 \cdot \tilde\varepsilon_1 = (D-2)\phi_1
\end{equation}
rather than the standard ten-dimensional dilaton. In those cases, the coefficients of the quantities
in (\ref{U1ms}) in $\cBoneloop{n}\big|_{\phi h^4}$ and
$\cBtwoloop{n}\big|_{\phi h^4} $ become $D$-dependent
and reduce to the expressions (\ref{vio1}) and (\ref{ratioD4R5}) if $D\rightarrow 10$.

\subsubsection{$D$-dimensional dilatons at genus two}

By evaluating the Type IIB components of the genus-two kinematic factors
with four gravitons and one $D$-dimensional dilaton state (\ref{dilatonid}), one arrives at
\begin{equation}
\label{UVsugra}
\cB_{(5)}^{\rm genus-2}|_{\phi h^4} =  {1\over10} K_{\phi h^4}\Big[{5\over6}(D-7) P_2 + {5\over3}(D-8)\varphi P_3
+ {5\over32}(D-9)A_1 \cO_{5,1} + W_D\cO_{5,2} +{\cal O}(s_{ij}^5) \Big]
\end{equation}
where $K_{\phi h^4}=\cBtree{3}$ and,
\begin{equation}
\label{weird}
W_D=(27-D){5\over36}A_2
+(1-D){25\over288}A_1
+(D-2)\Big({-}{5\over18}\cZ_2
-{5\over36}\cZ_3
+{5\over6}\varphi^2\Big)
+{25\over144}\cZ_5
\end{equation}

These results allow for sharp tests of our expressions for the low energy expansion.
Indeed, as discussed in \cite{Green:2010sp,Pioline:2018pso}, the $\ap$ expansion of the genus-two superstring amplitude must reproduce the logarithmic divergences of two-loop supergravity in various dimensions. For four-graviton scattering, UV divergences proportional 
to $D^4\cR^4$, $D^6\cR^4$ and $D^8\cR^4$ arise in $D=7,8,9$, respectively \cite{Bern:1998ug}, and the coefficient is precisely reproduced by the tropical limit of the string integrand \cite{Pioline:2018pso}. At the five-point level, the UV divergences for 5 gravitons have not yet been computed
in supergravity, but the UV divergences for 4 gravitons and one dilaton must certainly be absent, since supergravity amplitudes preserve R-symmetry. Indeed, from
\eqref{UVsugra} it is apparent that the divergences proportional to $P_2, P_3$ and $\cO_{5,1}$ in $D=7,8,9$ cancel as they should. This is not obvious, however, for the term proportional to 
$\cO_{5,2}$, which is potentially divergent in $D=9$. In this dimension, the coefficient evaluates to 
\be
W_9 = \frac{25} {144} \left[ -4 A_1 + {72\over5}A_2 - {56\over5}\cZ_2 - {28\over5}\cZ_3 + {168\over5}\varphi^2 + \cZ_5 \right]
\ee
Using the results in appendix \ref{sec:trop}, one finds that in the tropical limit $V\to 0$,
\be
\label{Atrop}
W_9 \sim -\frac{5}{36}  \frac{32\pi^2}{V^2} \left[ 0 -\frac{1}{63} A_{0,2} + \frac{25}{99} A_{1,1} -\frac{91}{120} A_{2,0} \right] + \cO(V)
\ee
As explained in \cite[(B.16)]{Pioline:2018pso}, the regularized integrals of the local modular forms
$A_{0,2}, A_{1,1}, A_{2,0}$ (see section 5.3 of \cite{DHoker:2018mys})
over the complex modulus $S$ parametrizing the Schwinger parameters $L_1,L_2,L_3$ at fixed 
discriminant $\det  Y = L_1 L_2 +L_2 L_3 + L_3 L_1$ vanish, so that the only UV divergence comes
from the integral over $A_{0,0}=1$, whose coefficient vanishes in the combination~\eqref{Atrop}. 

\subsubsection{$D$-dimensional dilatons at genus one}

The genus-one  analogue of (\ref{UVsugra}) can be obtained by promoting the results
of \cite[\S 5.3]{Green:2013bza} to a $D$-dimensional dilaton state,
\begin{align}
\label{1loopUV}
\cB_{(5)}^{\rm genus-1}|_{\phi h^4} &=   K_{\phi h^4}
\Big[{1\over 6} (D-8) + { E_2 \over 12} (D-12) P_2 - { (5 E_3 + \zeta_3) \over 36} (D-14)  P_3
 \notag \\
&
+ \frac{1}{16} (2 C_{2,1,1} + E_2^2 - E_4)
 (D-16){\cal O}_{5,1} + W'_D {\cal O}_{5,2} + {\cal O}(s_{ij}^5) \Big]
\end{align}
where
\bea
 W'_D =
 \Big( {2\over 9} + {7 D \over 72} \Big) C_{2,1,1}
+ \Big( {2\over 3} + {D \over 16} \Big) E_4  -  \Big( {2 \over 9} + {D \over 16}  \Big) E_2^2
\label{weirder}
\eea
The contributions of $P_2, P_3$ and ${\cal O}_{5,1}$ to (\ref{1loopUV}) correspond to one-loop
UV divergences proportional to $D^4 \cR^4, D^6\cR^4$ and $D^8\cR^4$ which occur in $D=12,14$
and $D=16$, respectively. Again, the R-symmetry violation by these UV divergences is prevented by
the prefactors $(D-12), (D-14)$ and $(D-16)$ in (\ref{1loopUV}), and the coefficient (\ref{weirder}) 
of ${\cal O}_{5,2}$ requires closer inspection in the critical dimension $D=16$:
\begin{align}
W'_{16} &= \frac{1}{9} (16 C_{2,1,1} +   15 E_4 - 11 E_2^2)  \label{wp16}\\
&={-} \frac{2 \pi \tau_2 \zeta_3}{135}
+ \frac{20 \zeta_5}{27 \pi \tau_2}
 - \frac{5 \zeta_3^2}{3 \pi^2 \tau_2^2} 
 + \frac{ 49 \zeta_7}{24 \pi^3 \tau_2^3} + {\cal O}(e^{-2\pi \tau_2} ) \notag
\end{align}
In passing to the second line, we have inserted the asymptotics of the modular graph functions
around the cusp \cite{D'Hoker:2015foa} which is captured by Laurent polynomials in $\tau_2$. The order of $\tau_2^4$
which is present in the individual $ C_{2,1,1} , E_2^2 $ and $E_4$ drops out from the particular 
combination in $W'_{16}$ and signals the absence of a 16-dimensional UV divergence in supergravity
as expected. 

\sm

Note that the classes of multiple zeta values in the Laurent expansion of modular 
graph functions as in (\ref{wp16}) are under active investigation in both the physics and mathematics literature \cite{Zerbini:2015rss, DHoker:2015wxz, DHoker:2019xef, Zagier:2019eus, Gerken:2020yii, Vanhove:2020qtt}. By
comparing with the multiple zeta values in the tree-level effective action of the Type IIB and IIA theories \cite{Schlotterer:2012ny}, one can associate the leftover terms in (\ref{wp16}) with UV divergences due to loop diagrams
with insertions of $D^{2k}\cR^n$ operators with $n\geq 4$ \cite{Green:2010sp,Pioline:2018pso}.

\subsection{S-duality analysis}

According to the standard S-duality conjecture in Type IIB string theory, the low energy effective action  must be invariant under the action of  $SL(2,\ZZ)$. In Einstein frame, $SL(2,\mathbb Z)$ acts by fractional linear transformations on the axion-dilaton field $\tau=a + i/g_s^2$, and by $U(1)_R$ rotations on the other fields, leaving the metric invariant.  
Thus, the coefficients of effective interactions violating $U(1)_R$ symmetry by $2q$ units
 must  transform with modular weight $(q,-q)$ under S-duality.  Typically, these interactions are related
to $U(1)_R$-preserving interactions by non-linear supersymmetry, so that their coefficients are obtained by acting repeatedly with a covariant derivative operator $\cD=\tau_2 \partial_\tau -\frac{iw}{2}$, which maps modular forms of weight $(w,\bar w)$ to modular forms of weight $(w+1,\bar w-1)$. An example of this is the dilatino vertex $\Lambda^{16}$, which violates $U(1)_R$ by 24 units 
and is  related to the $\cR^4$ coupling by acting with $\cD^{12}$ \cite{Green:1997me}.

\sm

At the four-point level, the amplitudes must conserve the $U(1)_R$ charge, and are all related to four-graviton scattering by supersymmetry.  The  expansion of the analytic part of the four-graviton all-genus amplitude in Einstein frame
takes the form \cite{Green:1999pv},
\bea
{\cal A}_{(4)} \, \big|_{\rm analytic} = |t_8(f_1{,}f_2{,}f_3{,}f_4)|^2 
\Big[ \frac{g_s^{-2}}{\sigma_3} + {\cal E}_{(0,0)} + \sigma_2 {\cal E}_{(1,0)}  + \sigma_3 {\cal E}_{(0,1)}
+ \tfrac{\sigma_2^2}{2} {\cal E}_{(2,0)} + {\cal O}(s_{ij}^5)  \Big]
\label{fourexp}
\eea
At each order,
the coefficient  ${\cal E}_{(p,q)}$ of the term $\tfrac{\sigma_2^p \sigma_3^q}{p! q!}$ (where 
$\sigma_k$ are the symmetric polynomials in (\ref{defsigma})) must be a modular function of $\tau$
under the action of S-duality. The coefficients ${\cal E}_{(0,0)}$ and ${\cal E}_{(1,0)}$
of the first interactions $\cR^4$ and $D^4 \cR^4$ beyond supergravity are well-known
to be captured by the non-holomorphic Eisenstein series $E_{3/2}$ and 
$E_{5/2}$ defined in  \eqref{defEis}  \cite{Green:1997tv,Green:1999pu}, 
 whereas the next term $D^6 \cR^4$ involves a more complicated
type of automorphic function $\cE_{(0,1)}$ constructed in \cite{Green:2005ba,Green:2014yxa}. 

\sm

For five-particle scattering, the $U(1)_R$ symmetry is violated by at most 2 units, e.g in 
the scattering of one dilaton and 4 gravitons. We expect that the 5-graviton interaction $D^2\cR^5$
is related by non-linear supersymmetry \cite{Green:2013bza} to the 4-graviton $D^4\cR^4$ interaction, governed by the automorphic form $\cE_{(1,0)}$ 
with weak coupling expansion, 
\begin{equation}
\label{E52}
\cE_{(1,0)} = 2\zeta_5 e^{-5\phi/2} + 0 +  {8\over 3}\zeta_4e^{3\phi/2} + \cdots
\end{equation}
corresponding to the tree-level, vanishing genus-one and non-vanishing genus-two contributions, plus instanton corrections indicated by the dots. By linear supersymmetry, it follows that the $\phi D^4 \cR^4$ interaction between one dilaton and 4 gravitons at the same order in the derivative expansion should be controlled by,  
\begin{equation}
\cD \cE_{(1,0)} \propto 
-5 \zeta_5 e^{-5\phi/2} +0 + 4 \zeta_4e^{3\phi/2} +\cdots
\end{equation}
where we use the fact that $\cD$ maps  $e^{q\phi} \rightarrow q e^{q\phi}$. This predicts that the ratio
of the genus-two and tree-level contributions to $\phi D^4 \cR^4$ is modified by a factor $-3/5$ compared to the ratio of the genus-two and tree-level contributions to the $D^2 \cR^5$ coupling, in perfect agreement with  \eqref{ratioD2R5}, as noted already in \cite{Gomez:2015uha}. 

\sm

By the same logic, the $D^4\cR^5$ coupling is expected to be related by non-linear supersymmetry
to the $D^6\cR^4$ coupling, governed by the automorphic function 
$\cE_{(0,1)}$ \cite{Green:2005ba, Green:2014yxa} with weak 
coupling expansion 
\begin{equation}
\label{E3232}
\cE_{(0,1)}=
4\zeta_3^2 e^{-3\phi}
+ 8\zeta_2\zeta_3e^{-\phi}
+ {48\over5}\zeta_2^2e^\phi
+ {8\over9}\zeta_6 e^{3\phi} + \ldots
\end{equation}
corresponding to tree-level up to genus-three contributions, plus instanton and anti-instanton corrections
indicated by the dots.   By linear supersymmetry, it follows that the $\phi D^6 \cR^4$ interaction
between one dilaton and 4 gravitons  should be controlled by,  
\begin{equation}
\label{DE3232}
\cD\cE_{(0,1)} \propto 
-12\zeta_3^2 e^{-3\phi}
-8\zeta_2\zeta_3e^{-\phi}
+ {48\over5}\zeta_2^2e^\phi
+ {8\over3 }\zeta_6 e^{3\phi} + \dots 
\end{equation}
This predicts that the ratio of the
genus-$(1,2,3)$ to the tree-level contributions to $\phi D^6 \cR^4$ is modified by
factors $(1/3,-1/3,-1)$ compared to the ratio of the genus-(1,2,3) to the tree-level contributions to the $D^4 \cR^5$ coupling. At genus one, the factor $1/3$ was checked in \cite[\S 5.3]{Green:2013bza}, and at genus two, the factor $-1/3$ is again in perfect agreement with \eqref{ratioD4R5}. 

\sm

Assuming that the coefficient of the $D^6\cR^5$ interaction, related to $D^8 \cR^4$ by non-linear supersymmetry, has the weak coupling expansion (where the dots now stand for additional perturbative and non-perturbative corrections), 
\bea
\cE_{(2,0)} = a_0 e^{-\frac72\phi} + a_1 e^{-\frac32\phi} + a_2 e^{\frac12\phi} + \cdots 
\label{schem20}
\eea
the $\phi D^8\cR^4$ interaction following from linear supersymmetry is then accompanied by, 
\bea
\cD \cE_{(2,0)} \propto  
-{7\over2}a_0 e^{-\frac72\phi} -{3\over2} a_1 e^{-\frac32\phi} + {1\over2}a_2 e^{\frac12\phi} + \cdots 
\eea
predicting a factor $-1/7$ between the ratios of the genus-two and tree-level contributions to $D^6\cR^5$ and $\phi D^8\cR^4$, respectively. This is indeed in agreement with \eqref{twoh5} and \eqref{twophih4}.

\sm

Note that the genus-two interactions $(D^6\cR^5)'$ and $(\phi D^8\cR^4)'$ proportional to $A_3$ and $A_4$ in \eqref{A3def} and
\eqref{A4def} do not have any corresponding interactions at tree level since the tree-level coefficient of $\cBnottree{7'}$
vanishes. Instead, we should consider the ratio of the genus-two and genus-one contributions. From
\cite{Green:2013bza}\footnote{In \cite[Eq. (5.4)]{Green:2013bza}, the statements $\int_{\cM_1} d\mu_1 (D_4,D_2^2,D_{211}) = 0$ turn out to be incorrect; instead, one can use $\int_{\cM_1} d\mu_1 D_{1111}=0$, $\int_{\cM_1} d\mu_1 D_{211} =-\frac12 \int_{\cM_1} d\mu_1 D_2^2$ and the identity $D_4 = 24 D_{211}+3 D_2^2-18 D_{1111}$ from \cite{DHoker:2015sve} to express all integrals in terms of a
single one, leading to $\Xi_{7’} {=} - \frac{5 \pi}{2} \int_{\cM_1} d\mu_1 D_2^2$ and $\hat\Xi_{7’} {=} - \frac{9 \pi}{2} \int_{\cM_1}
d\mu_1 D_2^2$.}, we find that the ratio of the one-loop contributions to $\phi h^4$ compared to $h^5$ is $9/5$.
Defining, in analogy with \eqref{intphiKZ}, the regularization-dependent coefficients $c_3,c_4$ by,
\begin{equation}
\label{defc34}
\int_{\cM_2(\Lambda)} d\mu_2 A_3 = c_3(\Lambda)\,  \, {\rm Vol}_2 \,,\qquad
 \int_{\cM_2(\Lambda)} d\mu_2 A_4 = c_4(\Lambda)\,  \, {\rm Vol}_2
\end{equation}
we predict that $c_3/c_4=-3/(9/5)=-5/3$. Indeed, using the results  in appendix \ref{sec:degen}, we find that the $\cO(t^2)$ coefficients in the  minimal non-separating degeneration  for $A_3$ and $A_4$, responsible for logarithmic divergences in $D=10$, are in the ratio $-5/3$. In particular, the combination  $c_3+\frac53c_4$ is infrared finite in $D=10$. 


\appendix

\newpage

\section{Functions on Riemann surfaces}
\label{sec:func}

In this appendix, we shall collect definitions, notations, and conventions for holomorphic forms, the period matrix, bi-holomorphic forms, the Arakelov Green function, and present some of the basic formulas needed in this paper  for integrals involving these quantities.

\subsection{Convention for forms}
\label{sec:conv}

Throughout, we shall follow the conventions of \cite{DHoker:2020prr} and display only the coefficient functions of differentials on a Riemann surface $\Sigma$ in a system of local complex coordinates $(z,\bar z)$ on  $\Sigma$. In this convention, a $(1,0)$ form $\om dz$ will be referred to as $\om$ and its integral along a curve $\cC$ will be abbreviated  $\int _\cC dz \, \om \rightarrow \int _\cC \om$, while a $(1,1)$-form $v \, dz\wedge d\bar z $ will be referred to as $v$  and its integral on $\Sigma$ will be abbreviated $\int _\Sigma dz \wedge d \bar z \, v \rightarrow \int _\Sigma v$. In the particular case of interest here the $(1,1)$ form may be the result of a wedge product between a $(1,0)$ form $\om \, dz$ and a $(0,1)$ form $\bar \psi \, d\bar z$, in which case the convention is $\int _\Sigma \om \, dz \wedge \bar \psi \, d\bar z \to \int _\Sigma \om \bar \psi = \int _\Sigma \bar \psi \om  $ because the component functions $\om$ and $\bar \psi$ commute with one another. We shall also use the abbreviation $\int _u = \int _{\Sigma_u}$ to indicate the integration over $\Sigma$ in the variable $u$.

\subsection{Holomorphic 1-forms and the period matrix}

We choose a canonical basis of $\mA_I$ and $\mB_I$ cycles  in $H_1(\Sigma , \ZZ)$ for which the intersection pairing $\mJ$ takes the form of the standard symplectic matrix, $\mJ(\mA_I, \mA_J)  = \mJ(\mB_I, \mB_J)  =  0$ and  $\mJ(\mA_I, \mB_J)  =  \delta _{IJ}$ for $I,J =1,2$. A canonical basis of holomorphic Abelian differentials $\om_I$ for $H^{(1,0)} (\Sigma)$ may be normalized  on $\mA$-cycles, and we have, 
\bea
\label{omnorm}
\oint _{\mA_I} \om_J = \delta _{IJ} 
\hskip 1in 
\oint _{\mB_I} \om_J = \Omega _{IJ} 
\eea
By the Riemann relations, the period matrix $\Omega$ is symmetric, and has positive definite imaginary part $Y$ 
as a result of the following pairing relation, 
\bea
\label{ompair}
\int_\Sigma \omega_I  \overline{\omega_J} = -2i\, Y_{IJ} \hskip 1in Y = \Im \Omega >0
\eea
The Siegel upper half space $\cH_2$ may be defined as the space of all $2\times 2$ complex-valued symmetric matrices whose imaginary part is positive definite. Alternatively, a more geometrical definition is  $ \cH_2 = Sp(4,\RR)/ \left ( SU(2) \times U(1) \right )$. The presence of the $U(1)$ factor  implies that $\cH_2$ is a K\"ahler manifold and its $Sp(4,\RR)$-invariant K\"ahler metric is given as follows,\footnote{Throughout, summation over pairs of repeated upper and lower indices will be implied.}
\bea
ds^2 = Y^{IK} Y^{JL} d\Omega _{IJ} d\bar \Omega _{KL}
\eea
where $(Y^{-1})^{IJ} = Y^{IJ} $ are the components of the inverse of the matrix $Y$. The moduli space $\cM_2$ may be identified with $\cM_2 = Sp(4,\ZZ)\backslash \cH_2$ provided we remove from $\cH_2$  all elements which correspond to disconnected surfaces. 

\sm

The Jacobian variety $J(\Sigma)= \CC^2/(\ZZ^2 + \Omega \ZZ^2)$ supports the canonical K\"ahler form~$K $,
\bea
K = {i \over 2} Y^{IJ} d\zeta_I \, d\bar \zeta_J
\eea
where $\zeta_I$ are local complex coordinates on the flat torus $J(\Sigma)$.
The form $ \kap$ is  the pull-back of $K$  from $J(\Sigma)$ to $\Sigma$ under the Abel-Jacobi map and, for a compact Riemann surface $\Sigma$,  may be  normalized to unit volume,
\bea
\label{muh}
\kap   = { i \over 4}  Y^{IJ} \omega_I \, \overline{ \omega_J}
\hskip 1in 
\int _\Sigma \kappa =1
\eea
The form $\kappa$ is conformal invariant as it is constructed solely out of the conformal invariant Abelian differentials.

\subsection{The bi-holomorphic forms $\Delta$ and $\nu$}

We define the bi-holomorphic form $\Delta$ by the anti-symmetric combination of $(1,0)$ forms, 
\bea
\label{Deltaxy}
\Delta (x,y)  = \ep ^{IJ} \om_I(x)  \om_J(y) = - \Delta(y,x)
\eea
where $\ep ^{IJ} = - \ep ^{JI}$ and $\ep ^{12}=1$.  Moreover, the ubiquitous anti-hermitian combination $\nu(x,y)$ of $(1,0)$ and $(0,1)$ forms is  defined in (\ref{nu0}). We shall list useful relations between the forms $\kappa$, $\Delta$ and $\nu$ in the remainder of this subsection, and give useful integral relations between these forms in the next subsection.

\sm

The identity $\ep ^{IJ}\ep ^{KL}+\ep ^{IK}\ep ^{LJ}+\ep ^{IL}\ep ^{JK}=0$ implies,
\bea
\label{omdel}
\om _I(x)  \Delta (y,z) + \om_I(y)  \Delta (z,x) + \omega_I(z)  \Delta (x,y) & = & 0
\no \\
\Delta (w,x)  \Delta(y,z) + \Delta (w,y)  \Delta (z,x) + \Delta (w,z)  \Delta (x,y) & = & 0
\eea
The form $\nu$ obeys simple relations with $\Delta$ and $\kappa$, 
\bea
\label{Delnu}
\Delta(x,y) \overline{ \Delta (w,z)} & = &  4 (\det Y) \Big (  \nu(x,z) \nu(y,w)  -  \nu(x,w)  \nu(y,z)  \Big )
\no \\
\Delta(x,y) \overline{ \Delta (y,z)} & = & 4  (\det Y) \Big ( 2  \nu(x,z) \kappa(y) -  \nu(x,y) \nu(y,z) \Big )
\no \\
\Delta(x,y) \overline{\Delta(y,x)} & = & 4 (\det Y) \Big ( 4 \kappa (x) \kappa (y) - \nu(x,y) \nu(y,x) \Big )
\eea
where the second equation follows from the first by setting $w=y$ and the third follows from the second by setting $z=x$. The following formula is being used to establish these results,
\bea
\ep^{IJ} \ep ^{KL} = (\det Y) (Y^{IK} Y^{JL} - Y^{IL} Y^{JK})
\eea
The cyclic identities (\ref{omdel}) imply further relations between $\kappa$, $\nu$ and $\Delta$,
\bea
\label{cycnudel}
\nu(x,w) \Delta(y,z) + \nu(y,w) \Delta (z,x) + \nu(z,w) \Delta (x,y) & = & 0
\no \\
2 \kappa (x) \Delta (y,z) + \nu(y,x) \Delta (z,x) + \nu(z,x) \Delta (x,y) & = & 0
\eea
Note that $\kappa$ and $\nu$ may be defined for arbitrary genus, but $\Delta$ exists only for genus two.

\subsection{Some useful integrals}

Useful integrals involving $\Delta$ are as follows,
\bea
\label{int1}
\int_u \om_I(u) \overline{\Delta(u,y)} & = & - 2 i Y_{IJ} \ep ^{JK}  \oom_K(y)
\no \\
\int _u \Delta (x,u) \overline{ \Delta (u,y)} & = & 4 (\det Y) \, \nu (x,y)
\eea
Useful integrals involving  $\nu$ and $\Delta$ are as follows,
\bea
\label{nu1}
\int_u \nu (x,u) \, \om_I(u) & = & \om_I(x)
\no \\
\int _u \nu (x,u) \, \nu (u,y) & = &  \nu (x,y)
\no \\
\int _u \nu (x,u) \, \Delta (u,y) & = & \Delta (x,y)
\eea
The following double integrals will also come in handy, 
\bea
\int _u \int _v \nu(u,v)\nu(v,u) & = & 2 
\no \\
\int _u \int_v  \Delta (u,v) \overline{\Delta (v,u) } & = & 8 \, \det Y
\no \\
\int _u \int_v  \Delta (x,u) \, \overline{\Delta (u,v)} \, \Delta (v,y) & = &
4 \, (\det Y) \, \Delta (x,y)
\eea
They may all be derived by making use of (\ref{ompair}) to carry out the integrals, and then using algebraic relations between $Y$ and $\ep$ to express the result in simplified form.

\subsection{The Arakelov Green function}
\label{sec:arakapp}

The Arakelov Green function $\GA(x,y)$ is a real-valued symmetric function  on $\Sigma \times \Sigma$ which provides an inverse to the scalar Laplace operator on $\Sigma$ with the canonical metric associated with $\kappa$, on the space of functions orthogonal to constants. In terms of local complex coordinates $(z,\bar z)$ and the convention stated in subsection \ref{sec:conv}, we have,
\bea
\label{AraG}
\p_z \,  \pbz  \, \GA(z,y)  & = &   - \pi \, \delta^{(2)} (z,y)  - 2 \pi i \kappa (z)
\hskip 0.9in
 \int _\Sigma \kap (z) \, \GA(z,y) = 0
 \no \\
 \p_z \,  \pby  \,  \GA(z,y)  & = &   \pi \, \delta^{(2)} (z,y) + 2 \pi i \nu(z,y)  
\eea
where  $\delta ^{(2)}(z,y)$ is the coordinate Dirac $\delta$-function normalized by,
\bea
{ i \over 2} \int _\Sigma  dz \wedge d \bar z \, \delta ^{(2)} (z,y)= 1
\eea 
Note that the right side of the first equation in (\ref{AraG})  integrates to zero in $z$ against constants, while the right side of the second equation integrates to zero in $y$ against the holomorphic forms $\om_I(y)$ and in $z$ against the anti-holomorphic forms $\oom_I(z)$. An explicit expression for $\GA$ may be obtained by relating it to the Green function $G$ which is often used in string theory, as reviewed for example in \cite{DHoker:2017pvk}.

\subsection{Reducing integrals of Arakelov Green functions}
\label{sec:A7}

Beyond the basic integrals $\cZ_1,\dots, \cZ_5$ defined in \eqref{defZs}, in expanding
the five-point amplitude up to order $D^6\cR^5$ we encounter various other  integrals
which can be easily reduced to the ones above, along with  the square of the Kawazumi-Zhang invariant,
 \footnote{The second, third, and fourth integrals were denoted by $\cB_5^{(2,0)},\cB_6^{(2,0)},\cB_7^{(2,0)}$ in \cite{Basu:2018bde}.}
\bea
\label{nuint}
&& \int _{\Sigma ^3} \kappa (1) \nu(2,4) \nu(4,2)
 \cG(1,2) \cG(1,4)  = - { 1 \over 4} \cZ_2
\no \\ &&
\int _{\Sigma ^3} \nu(1,2) \nu(2,4) \nu(4,1) \cG(1,2) \cG(1,4) = - {1 \over 4} \cZ_2
\no \\ &&
\int _{\Sigma ^4} \nu(1,2) \nu(2,3) \nu (3,4) \nu(4,1) \cG(1,2) \cG(3,4) = \half \f^2
\no \\ &&
\int _{\Sigma ^4} \nu(1,2) \nu(2,3) \nu (3,4) \nu(4,1) \cG(1,3) \cG(2,4) =  \half \cZ_3-  \half \f^2 
\no \\ &&
\int_{\Sigma^4} {  \Delta(1,2) \overline{ \Delta(2,3) } \Delta(3,4) \overline{\Delta(4,1)} \over (\det Y)^2}   \cG(1,3) \cG(2,4)  = 8\cZ_3 - 8\f^2
\no \\ &&
\int_{\Sigma^5} { \cG(1,4) \over (\det Y)^2}\,\partial_1 \cG(1,2)\, \bar \partial_1 \cG(1,3)\,
\, \Delta(2,4)\, \Delta(3,5)\, \overline{  \Delta(2,5) } \,\overline{  \Delta(3,4) } = i \pi \cZ_5
\no \\ &&
\int_{\Sigma^5} { \cG(1,3) \over (\det Y)^2}\,\partial_1 \cG(1,2)\, \bar \partial_1 \cG(1,4)\,
\, \Delta(2,4)\, \Delta(3,5)\, \overline{  \Delta(2,5) } \,\overline{  \Delta(3,4) }
= - i \pi \cZ_5
\eea
The first two lines of (\ref{nuint}) follow from using the last line of (\ref{Delnu}), and the last line of (\ref{cycnudel}) on the combination $\kappa( 1) \Delta (2,4)$, respectively, in the definition (\ref{defZs}) of $\cZ_2$.
To derive the third line of (\ref{nuint}), we use the second line of (\ref{omdel}) on the product $\Delta(1,3) \Delta(2,4)$ and its complex conjugate in the third line of (\ref{defZs}), cancel the $\cZ_3$ contribution, and express the remainder in terms of $\f$. To derive the last line of (\ref{nuint}), we use the second line of (\ref{omdel}) on the product $\Delta(1,3) \Delta(2,4)$ but not on its complex conjugate in (\ref{defZs}), and recast one of the integrals in terms of $\f^2$, giving the desired integral.

\newpage

\section{Expanding the integrals}
\label{app:exp}

In this appendix, we spell out intermediate steps in obtaining the $\alpha'$ expansion of the five-point integrals $J_{1,i},F_j,G_j, H_j$ defined in (\ref{defJint}) to (\ref{defHint}). The integrals $J_{1,i},G_j, H_j$ admit convergent Taylor series expansions at $s_{ij}=0$, while each $F_j$ has a simple poles in $s_{12}$. Since the integrals $H_j$ and $G_j$ will be needed only to order $s_{ij}$ they are the simplest and will be carried out first. The integrals $J_{1,i}$ will be needed to order $s_{ij}^2$ and are carried out next, finishing with  $F_j$ which may be expressed in terms of the integrals $J_{1,i}$ and $G_j$.

\subsection{The $H$-integrals}
\label{sec:expH}

The $H$-integrals defined in \eqref{defHint} have a  convergent Taylor series expansion at $s_{ij}=0$.
The contributions of order $\cO(s_{ij}^0)$ clearly vanish. For the contributions of order $\cO(s_{ij})$, only the term proportional to $s_{13}$ is non-vanishing, so that we get,
\bea
H_j ={ s_{13}  \over i \pi} \int_{\Sigma^5}  \p_1 \cG(1,2) \, \bar \p_3 \cG(3,4) \, \cG(1,3) \, \frac{ \Xi_j }{(\det Y)^2} + {\cal O}(s_{ij}^2)
\eea
where $\Xi_j$ is a shorthand for the combination of $\Delta$ and $\overline{\Delta}$ in (\ref{defHint}).
Upon integrations by parts, the $\p_1$ and $\bar \p_3$ differentials can be made to both act on $\cG(1,3)$
which gives 
\bea
H_j =  s_{13} \int_{\Sigma^4} \cG(1,2) \cG(3,4) \big[{-}i\delta^{(2)}(z_1,z_3) +2\nu(1,3) \big] \int_{z_5} \frac{ \Xi_j }{(\det Y)^2} + {\cal O}(s_{ij}^2)
\eea
based on (\ref{AraG}).
With the $\Delta$ and $\overline{\Delta}$ in (\ref{defHint}), integration over point 5 yields,
\bea
\int_{z_5} \frac{ \Xi_1 }{(\det Y)^2}& = & - 4 \nu(3,1) \frac{ \Delta(2,4) \overline{\Delta(2,4)} }{\det Y}    \rightarrow - 16 \nu(2,4) \nu(4,2) \nu(3,1)
\no \\
\int_{z_5} \frac{\Xi_2 }{(\det Y)^2}& = & - 4 \nu(4,2) \frac{ \Delta(2,3) \overline{\Delta(1,4)} }{\det Y} \rightarrow 16 
\nu(4,2) \big[  \nu(2,1)  \nu(3,4)  -\nu(2,4)  \nu(3,1)  \big]
\no \\
\int_{z_5} \frac{ \Xi_3 }{(\det Y)^2}& = & - 4 \nu(4,1) \frac{ \Delta(2,3) \overline{\Delta(2,4)} }{\det Y} \rightarrow - 16 \nu(3,2)  \nu(2,4) \nu(4,1)
\no \\
\int_{z_5} \frac{ \Xi_4 }{(\det Y)^2}& = & - 4 \nu(3,2) \frac{ \Delta(2,4) \overline{\Delta(1,4)} }{\det Y} \rightarrow - 16 \nu(3,2) \nu(2,4) \nu(4,1)
\eea
where terms involving $\kappa(j)$ have been dropped in the step marked by the
arrow since they integrate to zero in presence of $\cG(1,2) \cG(3,4) $.
Decomposing the remaining integrals via (\ref{nuint}), we find the results in (\ref{expHs}).

\subsection{The $G$-integrals}
\label{sec:expG}

The $G$-integrals defined in \eqref{defGint} also admit a convergent Taylor series expansion at $s_{ij}=0$. Integrating by parts the factor  $\bar \p_1 \cG(1,3)$ one readily sees that they vanish at leading order in $s_{ij}$,
except for $G_2$ which turns out to be proportional to the Kawazumi-Zhang invariant:
\bea
G_2&=&
-{1\over i\pi} \int_{\Sigma^5} { \partial_1 \bar \partial_1 \cG(1,2) \over (\det Y)^2} \,  \cG(1,3)\,
\, \Delta(2,4)\, \Delta(3,5)\, \overline{  \Delta(2,5) } \,\overline{  \Delta(3,4) } + {\cal O}(s_{ij})
\no\\
&=& - 32 \varphi + {\cal O}(s_{ij})\, ,
\eea
Similar manipulations may be used to obtain the results (\ref{expGs}) to order $s_{ij}^2$ included in terms of the functions $\f$, $\cZ_1, \cZ_2, \cZ_3, \cZ_4$, plus the additional integral  $\cZ_5$ defined in (\ref{defZs}).

\subsection{The $J$-integrals}
\label{sec:expJ}

The $J$-integrals were defined in \eqref{defJint}.
It is straightforward to evaluate their leading $s_{ij}$ contributions $J^{(0)}_{r,s}$, 
\bea
J^{(0)} _{1,1} =128
\hskip 0.7in 
J^{(0)} _{1,2} =32
\hskip 0.7in 
J^{(0)} _{1,3} =-64
\eea 

\subsubsection{First order in $s$}

Evaluating the first order corrections, given by the sum over $s_{ij} \cG(i,j)$ for $i<j$, for $J_{1,1}$ we see that the contributions where $i=1$ vanish by (\ref{AraG}), so that we may integrate over $z_1$,
\bea
J_{1,1} ^{(1)}& = &  { 2 \over (\det Y)^2} \, \int _{\Sigma ^4}  \Delta (2,3) \Delta (4,5)
\overline{ \Delta (2,3) \Delta (4,5) } \sum_{2\leq i < j \leq 5} s_{ij} \cG(i,j) 
\eea
The contributions proportional to $s_{23}$ and $s_{45}$ are equal to one another. The contributions from $s_{24},s_{25}, s_{34}, s_{35}$ are also equal to one another, so that we get, 
\bea
J_{1,1} ^{(1)}& = &  2{  s_{23} + s_{45}  \over (\det Y)^2} \, \int _{\Sigma ^4}  \Delta (2,3) \Delta (4,5)
\overline{ \Delta (2,3) \Delta (4,5) } \cG(2,3)
\no \\ &&
+ 2 {  s_{24} + s_{25} +s_{34} + s_{35}  \over (\det Y)^2} \, \int _{\Sigma ^4}  \Delta (2,3) \Delta (4,5)
\overline{ \Delta (2,3) \Delta (4,5) } \cG(2,4)
\eea
The integral over the point $3$ in the second line is proportional to $\kappa(2)$, whose integral against $\cG(2,4)$ vanishes, so that the second line vanishes. Integrating over the points $4,5$ in the first line and using the formula for $\f$, we find,
\bea
J_{1,1} ^{(1)}=  -64 (  s_{23} + s_{45} )\, \f 
\eea
The first order contributions $J_{1,2}^{(1)}$ are given by,
\bea
J_{1,2} ^{(1)} = { 1 \over (\det Y)^2} \int _{\Sigma ^5}  \nu(1,2) \Delta (2,3) \overline{ \Delta (3,4)}  \Delta (4,5)
\overline{ \Delta (5,1) } \sum _{i<j} s_{ij} \cG(i,j)
\eea
For each $i<j$, we integrate over the three points that are different from $i,j$ using only the formulas of (\ref{nu1}). The remaining  integrals over $i,j$ are then evaluated using one of the representations of $\varphi$ in (\ref{KZ2}). Noting that the contributions of $s_{1j}$ and $s_{2j}$ for $j=3,4,5$ are equal to one another; that the contributions of $s_{24}$ and $s_{35}$ are equal to one another; and that the contributions of $s_{23}, s_{25},s_{34},s_{45}$ are equal to one another, we find, 
\bea
J_{1,2}^{(1)} = 64  \, s_{35} \, \f
\eea
where we have used momentum conservation to obtain the final result. Finally, 
\bea
J_{1,3} ^{(1)} =  {1 \over (\det Y)^2}  \int _{\Sigma ^5}  \nu(1,3)  \Delta (2,3) \Delta (4,5)
\overline{ \Delta (4,5) \Delta (1,2) } \sum _{i<j} s_{ij} \cG(i,j)
\eea
When $j=5$ and $i \not=4$, as well as when $j=4$ and $i <4$, the integrals vanish because by integrating out one of the variables different from $i$ and $j$ they result in an integration of $\cG$ against $\kappa$ which vanishes. Thus the only remaining contributions involve $s_{12}, s_{23}, s_{13}, s_{45}$, and they are readily evaluated,
\bea
J^{(1)} _{1,3} = 32 (s_{12} -s_{13} +s_{23} +s_{45})\, \f  = 64 (s_{45} - s_{13}) \, \f 
\eea

\subsubsection{Second order in $s$}

To second order, we have, 
\bea
J_{1,1} ^{(2)} & = & \int_{\Sigma ^5}  \kappa(1) { |\Delta(2,3) \Delta(4,5)|^2  \over (\det Y)^2}  
\sum_{i<j} \sum _{k<\ell} s_{ij} s_{k\ell} \, \cG(i,j) \cG(k,\ell)
\eea
Each sum has 10 terms, so the total has 100 terms. However, all terms involving $s_{1j} s_{k\ell}$ with $j,k,\ell \not=1$ cancel, reducing the number of terms to $16+36=52$. To organize these, we proceed by evaluating first the perfect squares, 
\bea
\int _{\Sigma ^5}  \kappa(1) { |\Delta(2,3) \Delta(4,5)|^2  \over (\det Y)^2}  
\sum_{i<j} s_{ij} ^2 \, \cG(i,j)^2 
\eea
for which $s_{12}^2, s_{13}^2, s_{14}^2,s_{15}^2$ all have the same coefficient $8 \cZ_1$. Moreover, $s_{24}^2, s_{25}^2, s_{34}^2, s_{35}^2$ also all have the same coefficient $8 \cZ_1$, and $s_{23}^2, s_{45}^2$ also have the same coefficients $ 16 \cZ_1 + 8 \cZ_4$. Next, we evaluate the terms involving $s_{1j}s_{1\ell}$ with $\ell \not=j$ (to which we refer as ``angles"). For $(j,\ell)=(2,3) ,(4,5)$, the coefficient is $16 \cZ_1$, while for $(j,\ell) = (2,4), (2,5),(3,4), (3,5)$ the coefficient vanishes because the integral over the remaining point produces a $\kappa$ which integrates to zero against $\cG$. Finally, for $2 \leq i<j \leq 5$ and $2 \leq k< \ell \leq 5$, the coefficients of the terms $s_{23}$ times $s_{24}, s_{25}, s_{34}, s_{35}$ vanish as do their mirror images $s_{45} $ times $s_{24}, s_{25}, s_{34}, s_{35}$. The remaining terms are readily evaluated, and we find, 
\bea
J_{1,1} ^{(2)} & = & 8 \cZ_1 \sum _{i<j} s_{ij}^2 + 8 (\cZ_1+\cZ_4) (s_{23}^2 + s_{45}^2) 
+ 16 \cZ_3 (s_{24} s_{35} + s_{25 } s_{34} ) + 32 \f^2 s_{23} s_{45}
\no \\ &&
+ 16 \cZ_2 \Big ( s_{12} s_{13} + s_{14} s_{15} + s_{24} s_{25} + s_{24} s_{34} + s_{34} s_{35} + s_{25}s_{35} \Big )
\eea
Recasting the expression in terms of the cyclic variables $s_{i, i+1}$ we obtain  (\ref{expJs}).

\sm

Next we evaluate,
\bea
J_{1,2} ^{(2)} =  \half \int _{\Sigma ^5}  \nu(1,2) { \Delta (2,3) \overline{ \Delta (3,4)} \Delta (4,5)
 \overline{ \Delta (5,1) }  \over (\det Y)^2} \, \sum_{i<j} \sum _{k<\ell} s_{ij} s_{k\ell} \, \cG(i,j) \cG(k,\ell)
\eea
The sum over $i,j,k,\ell$ again involves 100 terms. To take advantage of symmetries, we decompose the continuous products $\Delta \bar \Delta$ into $\nu$ and $\kappa$ using the second equation of (\ref{Delnu}). This will multiply the number of terms by 4, but we can handle them using symmetry arguments, and all integrals become mechanical. We organize the calculation as follows, 
\bea
J_{1,2} ^{(2)} = X_1 + X_2 +X_3 +X_4
\eea
with
\bea
X_1 & = & 32 \int _{\Sigma ^5} \nu(1,2) \nu(2,4) \nu(4,1) \kappa (3) \kappa(5) \sum_{i<j} \sum _{k<\ell} s_{ij} s_{k\ell} \, \cG(i,j) \cG(k,\ell)
\no \\
X_2 & = &- 16 \int _{\Sigma ^5} \nu(1,2) \nu(2,3) \nu(3,4) \nu(4,1)  \kappa(5) \sum_{i<j} \sum _{k<\ell} s_{ij} s_{k\ell} \, \cG(i,j) \cG(k,\ell)
\no \\
X_3 & = &- 16 \int _{\Sigma ^5} \nu(1,2) \nu(2,4) \nu(4,5) \nu(5,1)  \kappa(3) \sum_{i<j} \sum _{k<\ell} s_{ij} s_{k\ell} \, \cG(i,j) \cG(k,\ell)
\no \\
X_4 & = & 8 \int _{\Sigma ^5} \nu(1,2) \nu(2,3) \nu(3,4) \nu(4,5) \nu(5,1) \sum_{i<j} \sum _{k<\ell} s_{ij} s_{k\ell} \, \cG(i,j) \cG(k,\ell)
\eea 
In $X_1$, the points $3,5$ enter either into two or zero Green functions. This significantly reduces the number of contributions, and we find,
\bea
X_1 & = & 8 \cZ_1 ( s_{31}^2 + s_{32}^2 + s_{34}^2 + s_{35}^2 + s_{51}^2+s_{52}^2+s_{54}^2) 
- 8 \cZ_4 (s_{12}^2+ s_{24}^2+s_{41}^2) 
\\ &&
-16 \cZ_2  ( s_{31} s_{32} + s_{31} s_{34} +s_{32}s_{34} + s_{51} s_{52} + s_{51}s_{54} 
\no \\ && \qquad ~
+ s_{52} s_{54} + s_{12} s_{14} +s_{21} s_{24} + s_{42}s_{41}  )
\no 
\eea 
In $X_2$, either two or zero Green functions $\cG$  involve the point $5$, and the integration measure is cyclic symmetric in the remaining points.  For squares, all terms where point $5$ occurs twice have the same coefficient, while for the terms independent on point $5$, we distinguish between contributions where $\cG$ connects contiguous points or not. For angles, we distinguish whether the angle is anchored at the point $5$ or not and whether $\cG$ connect contiguous points or not. The disconnected contributions do not involve the point $5$. The result is as follows, 
\bea
X_2 & = & 4 \cZ_4 \sum _{i<j} s_{ij}^2 - 4(\cZ_1 +\cZ_4) (s_{51}^2+s_{52}^2+s_{53}^2+s_{54}^2)
\no \\ &&
+ 8 \cZ_2 ( s_{51}s_{52} +s_{51}s_{53} +s_{51}s_{54} +s_{52}s_{53} +s_{52}s_{54} +s_{53}s_{54} 
\no \\ && \qquad ~
+ s_{12} s_{13}  + s_{12} s_{14}  + s_{13}s_{14} +s_{21} s_{23} +s_{23} s_{24} +s_{21} s_{24}
\no \\ && \qquad ~
 +s_{31} s_{34} +s_{32}s_{34}+ s_{32}s_{31} +s_{41} s_{42} + s_{41} s_{43}  +s_{43}s_{42} )
\no \\ && 
- 16 \f^2 (s_{12} s_{34} + s_{23} s_{14} - s_{13}s_{24}) - 16 \cZ_3 s_{13} s_{24}
\eea
The calculation of $X_3$ is analogous, but the special point is now $3$, and we find,
\bea
X_3 & = & 4 \cZ_4 \sum _{i<j} s_{ij}^2 - 4(\cZ_1 +\cZ_4) (s_{31}^2+s_{32}^2+s_{34}^2+s_{35}^2)
\no \\ &&
+ 8 \cZ_2 ( s_{31}s_{32} +s_{31}s_{34} +s_{31}s_{35} +s_{32}s_{34} +s_{32}s_{35} +s_{34}s_{35} 
\no \\ && \qquad ~
+ s_{12} s_{15}  +s_{14} s_{15} +s_{12} s_{14}  +s_{21} s_{24} +s_{21} s_{25} +s_{24}s_{25} 
\no \\ && \qquad ~
+s_{41}s_{42}+s_{42} s_{45} +s_{41} s_{45} + s_{51} s_{54} +s_{52}s_{54} + s_{51}s_{52} )
\no \\ && 
- 16 \f^2 (s_{12} s_{45} - s_{14} s_{25} + s_{15}s_{24}) - 16 \cZ_3 s_{14} s_{25}
\eea
For $X_4$, we  exploit the cyclic symmetry of the integrand. The squares of nearest neighbors have the same coefficient, and so do the squares of next-to-nearest neighbors; there are 4 classes  of angles depending on the relative position of the vertex of the angle and the two other points; and there are three classes of disconnected contributions. In total we get, 
\bea
X_4 & = & - 2 \cZ_4 \sum _{i<j} s_{ij}^2 - 4 \cZ_2  \sum _i \sum _{j<k} s_{ij} s_{ik} 
\no \\ &&
+ 8 \cZ_3 (s_{13} s_{25} + s_{24} s_{31} + s_{35} s_{42}  + s_{41} s_{53} + s_{52} s_{14}) 
\no \\ &&
+ 8 \f^2( s_{12} s_{34} + s_{12} s_{35} + s_{12} s_{45} - s_{13}s_{24} - s_{13} s_{25} +s_{13} s_{45} + s_{14}s_{23} 
- s_{14} s_{25}
\no \\ && \qquad ~
- s_{14} s_{35} +s_{15} s_{23} + s_{15} s_{24} + s_{15} s_{34}+ s_{23} s_{45} - s_{24} s_{35} + s_{25} s_{34} )
\eea
Putting all together, we have,
\bea
J_{1,2}^{(2)} & = & 
 (4\cZ_1 +2\cZ_4) (s_{51}^2+s_{52}^2+s_{54}^2+s_{31}^2+s_{32}^2+s_{34}^2)
 - 2 \cZ_4 (s_{12}^2+s_{14}^2+s_{24}^2+s_{35}^2)
\no \\ &&
+ 4 \cZ_2 ( s_{12}s_{13} -s_{12}s_{14} +s_{12}s_{15} +s_{13}s_{14} - s_{13}s_{15} + s_{14}s_{15} 
\no \\ && \qquad ~
+ s_{23} s_{24}  - s_{23} s_{25} +s_{23} s_{21} +s_{24} s_{25} - s_{24}s_{21}+s_{25} s_{21} 
\no \\ && \qquad ~
+ s_{34}s_{35} -s_{34}s_{31} -s_{34}s_{32} + s_{35}s_{31} +s_{35}s_{32} - s_{31}s_{32} 
\no \\ && \qquad ~
 +s_{45} s_{41} + s_{45} s_{42} - s_{45} s_{43} -s_{41}s_{42}+s_{41} s_{43} + s_{42} s_{43}  
 \no \\ && \qquad ~
 -s_{51} s_{52} + s_{51} s_{53} - s_{51} s_{54} +s_{52}s_{53} - s_{52} s_{54} + s_{53} s_{54}  
 \no \\ && 
- 8 \f^2 (s_{12} s_{34} + s_{23} s_{14} - s_{13}s_{24}+ s_{12} s_{45} - s_{14} s_{25} + s_{15}s_{24} 
- s_{12} s_{35}  + s_{13} s_{25} 
\no \\ && \qquad ~
- s_{13} s_{45}  + s_{14} s_{35} - s_{15} s_{23} - s_{15} s_{34} - s_{23} s_{45} + s_{24} s_{35} - s_{25} s_{34} )
\no \\ &&
- 8 \cZ_3 ( s_{24} s_{31} + s_{52} s_{14} - s_{35} s_{42}  - s_{41} s_{53}  - s_{13} s_{25}) 
\eea
Recasting the expression in terms of the cyclic variables $s_{i, i+1}$ we obtain  (\ref{expJs}).
\sm

Next we evaluate,
\bea
J_{1,3} ^{(2)} =  \half \int _{\Sigma ^5}  \nu(1,3) { \Delta (3,2) \overline{ \Delta (2,1)} |\Delta (4,5)|^2
 \over (\det Y)^2} \, \sum_{i<j} \sum _{k<\ell} s_{ij} s_{k\ell} \, \cG(i,j) \cG(k,\ell)
\eea
We proceed in analogy with $J^{(2)}_{1,3}$ and decompose as follows, 
\bea
J_{1,3} ^{(2)} = Y_1 + Y_2 +Y_3 +Y_4
\eea
with
\bea
Y_1 & = & -64 \int _{\Sigma ^5} \nu(1,3) \nu(3,1) \kappa(2) \kappa (4) \kappa(5)  \sum_{i<j} \sum _{k<\ell} s_{ij} s_{k\ell} \, \cG(i,j) \cG(k,\ell)
\no \\
Y_2 & = & 32 \int _{\Sigma ^5} \nu(1,3) \nu(3,2) \nu(2,1) \kappa(4) \kappa(5) \sum_{i<j} \sum _{k<\ell} s_{ij} s_{k\ell} \, \cG(i,j) \cG(k,\ell)
\no \\
Y_3 & = & 16 \int _{\Sigma ^5} \nu(1,3)\nu(3,1) \nu(4,5)\nu(5,4)  \kappa(2) \sum_{i<j} \sum _{k<\ell} s_{ij} s_{k\ell} \, \cG(i,j) \cG(k,\ell)
\no \\
Y_4 & = & -8 \int _{\Sigma ^5} \nu(1,3) \nu(3,2) \nu(2,1) \nu(4,5) \nu(5,4) \sum_{i<j} \sum _{k<\ell} s_{ij} s_{k\ell} \, \cG(i,j) \cG(k,\ell)
\eea 
The combinatorics is similar to $J_{1,3}^{(2)}$ and the integrals are readily recognized,
\bea
Y_1 & = & 16 (\cZ_1 + \cZ_4 )s_{13}^2 - 16 \cZ_1\sum_{i<j} s_{ij}^2
+ 32 \cZ_2( s_{21} s_{23} + s_{41} s_{43} + s_{51}s_{53})
\no \\
Y_2 & = & -8(\cZ_1+\cZ_4) ( s_{12}^2 + s_{23}^2 + s_{31}^2) + 8 \cZ_1 \sum _{i<j} s_{ij}^2 
\no \\ &&
-16 \cZ_2 (s_{12} s_{13} + s_{21} s_{23} + s_{31} s_{32} +s_{41} s_{42} + s_{41} s_{43} + s_{42} s_{43} 
+s_{51} s_{52} + s_{51} s_{53} + s_{52} s_{53} )
\no \\
Y_3 & = & - 8(\cZ_1+\cZ_4) (s_{13}^2 + s_{45}^2) + 8 \cZ_1 \sum _{i<j} s_{ij}^2
+ 32 \f^2 s_{13} s_{45} + 16 \cZ_3 (s_{14} s_{35} + s_{15} s_{34})
\no \\ &&
- 16 \cZ_2 ( s_{12} s_{23} + s_{14} s_{15} + s_{24} s_{25}  + s_{34} s_{35} +s_{14} s_{34} + s_{15} s_{35})
\no \\
Y_4 & = & 4 \cZ_4( s_{12}^2 + s_{13}^2+s_{23}^2+s_{45}^2) 
- 4 \cZ_1 (s_{14}^2+s_{15}^2+s_{24}^2+s_{25}^2+s_{34}^2+s_{35}^2) 
\no \\ &&
-16 \f^2 (s_{12} +s_{23}+s_{31})s_{45} 
- 8 \cZ_3 (s_{14}s_{25}+s_{14}s_{35}+s_{24} s_{15} +s_{24} s_{35} + s_{34} s_{15} +s_{34} s_{25} )
\no \\ &&
+ 8 \cZ_2 ( s_{12} s_{13} + s_{13} s_{23} + s_{12} s_{23} + s_{41} s_{42} + s_{42} s_{43} + s_{43} s_{41} 
\no \\ && \qquad
+ s_{51} s_{52} + s_{52} s_{53} + s_{53} s_{51} + s_{14} s_{15} + s_{24} s_{25} + s_{34} s_{35})
\eea
Assembling all contributions, we find the following equivalent of the $s_{ij}^2$ order in (\ref{expJs}), 
\bea
J^{(2)}_{1,3} & = & 
- 4 \cZ_1 (2s_{12}^2 + 2 s_{23}^2 + 2 s_{45}^2 +s_{14}^2 +s_{15}^2 +s_{24}^2 +s_{25}^2 +s_{34}^2 +s_{35}^2 ) 
\no \\ && 
 - 4 \cZ_4 (s_{12}^2 + s_{23}^2 -s_{31}^2+s_{45}^2 )
-16 \f^2 (s_{12} +s_{23} - s_{13}) s_{45} 
\no \\ &&
- 8 \cZ_3 (s_{14}s_{25}-s_{14}s_{35}+s_{24} s_{15} +s_{24} s_{35} - s_{34} s_{15} +s_{34} s_{25} )
\no \\ && 
+  8 \cZ_2 ( - s_{12} s_{13} - s_{13} s_{23} + s_{12} s_{23} - s_{41} s_{42} - s_{42} s_{43} + s_{43} s_{41} 
\no \\ && \qquad ~
- s_{51} s_{52} - s_{52} s_{53} + s_{53} s_{51} - s_{14} s_{15} - s_{24} s_{25} - s_{34} s_{35})
\eea
One readily verifies that the result is symmetric in $1,3$ as well as in $4,5$. Recasting the expression in terms of the cyclic variables $s_{i, i+1}$ we obtain  (\ref{expJs}).

\subsection{The $F$-integrals}
\label{sec:expF}

Finally, we turn to the $F$-integrals defined in \eqref{defFint}. As stressed in section \ref{sec:intexp}, these integrals
have a simple pole at $s_{12}=0$, which can be exposed by means of the identity \eqref{KNbyparts}.
Applying the same method as for \eqref{F1dec}, we get,
\bea
\label{F2dec}
F_2 & = & 
- { 1 \over i \pi } \sum_{k=3}^5 {s_{1k} \over s_{12}} \int _{\Sigma ^5} { \KN_{(5)} \over (\det Y)^2}
\p_1 \cG(1,2) \bar \p_1 \cG(1,k) \Delta (2,3) \Delta (4,5)  \overline{\Delta (2,4)}  \overline{\Delta(3,5)} 
\no \\ &&
+ { 2 \over s_{12} } \int _{\Sigma ^5} { \KN_{(5)} \over (\det Y)^2}
\kappa(1) \Delta (2,3) \Delta (4,5)  \overline{\Delta (2,4)}  \overline{\Delta(3,5)} 
\\
\label{F3dec}
F_3 & = & 
- { 1 \over i \pi }  \sum_{k=3}^5 {s_{1k}  \over s_{12}} \int _{\Sigma ^5} { \KN_{(5)} \over (\det Y)^2}
\p_1 \cG(1,2) \bar \p_2 \cG(2,k) \Delta (2,3) \overline{\Delta(1,3)} |\Delta (4,5)|^2 
\no \\ &&
- { 2 \over s_{12} } \int _{\Sigma ^5} { \KN_{(5)} \over (\det Y)^2}
\nu(1,2) \Delta (2,3) \overline{\Delta(1,3)} |\Delta (4,5)|^2 
\\
\label{F4dec}
F_4 & = & 
- { 1 \over i \pi }  \sum_{k=3}^5 {s_{1k}  \over s_{12}} \int _{\Sigma ^5} { \KN_{(5)} \over (\det Y)^2}
\p_1 \cG(1,2) \bar \p_2 \cG(2,k) \Delta (2,3) \Delta (4,5)  \overline{\Delta (1,4)} \overline{\Delta(3,5)} 
\no \\ &&
- { 2 \over s_{12} } \int _{\Sigma ^5} { \KN_{(5)} \over (\det Y)^2}
\nu(1,2) \Delta (2,3) \Delta (4,5)  \overline{\Delta (1,4)}  \overline{\Delta(3,5)} 
\eea
%
For each $F_a$, the first term on the right side may be expressed as a linear combination of the functions $G_1$ and $G_2$, whose $\alpha'$ expansion was computed in section \ref{sec:expG} and given in (\ref{expGs}). The second term on the right may be expressed as a linear combination of the functions $J_{r,s}$, whose $\alpha'$ expansion was computed in section \ref{sec:expJ} and given in (\ref{expJs}). As a result of the integration-by-parts relations, we obtain,
\begin{align}
\label{FFFF}
s_{12} F_1  &=  J_{1,1} -  s_{13} \big( G_1 + G_1\big|_{4\leftrightarrow 5} - G_2 - G_2\big|_{4\leftrightarrow 5}  \big)  
- s_{14} \big( G_1 \big|_{3\leftrightarrow 4}  \big) - s_{15} \Big( G_1 \Big| \begin{smallmatrix} 3,4,5 \\ \downarrow \\ 5,3,4\end{smallmatrix} \Big) 
\notag \\
s_{12} F_2  &=  J_{1,1} - J_{1,2} - J_{1,5} -  s_{13} \big( G_1 - G_2 \big|_{4\leftrightarrow 5}  \big) 
 \notag \\ &\ \ \ \ 
- s_{14} \big( G_1 \big|_{3\leftrightarrow 4}  - G_2 \big|_{3\leftrightarrow 4} \big) 
- s_{15}   \Big( G_2 \Big| \begin{smallmatrix} 3,4,5 \\ \downarrow \\ 5,3,4\end{smallmatrix} \Big) 
\no \\
s_{12}F_3 &= -2 J_{1,1} -2 J_{1,3}  
- s_{23}\Big( \overline{G}_3 \Big| \begin{smallmatrix} 1,2,3 \\ \downarrow \\ 2,3,1\end{smallmatrix} +\overline{G}_3 \Big| \begin{smallmatrix} 1,2,3,4,5 \\ \downarrow \\ 2,3,1,5,4\end{smallmatrix} -\overline{G}_4 \Big| \begin{smallmatrix} 1,2,3 \\ \downarrow \\ 2,3,1\end{smallmatrix} -\overline{G}_4 \Big| \begin{smallmatrix} 1,2,3,4,5 \\ \downarrow \\ 2,3,1,5,4\end{smallmatrix} \Big) 
\notag \\
&\ \ \ \ - s_{24} \Big( \overline{G}_3 \Big| \begin{smallmatrix} 1,2,3,4,5 \\ \downarrow \\ 2,4,1,5,3\end{smallmatrix} \Big)
- s_{25}\Big( \overline{G}_3 \Big| \begin{smallmatrix} 1,2,3,4,5 \\ \downarrow \\ 2,5,1,4,3\end{smallmatrix} \Big)
\notag \\
s_{12} F_4 &= - 2 J_{1,1} -2 J_{1,3} +2  J_{1,2} 
- s_{23} \Big( \overline{G}_3 \Big| \begin{smallmatrix} 1,2,3,4,5 \\ \downarrow \\ 2,3,1,5,4\end{smallmatrix}
- \overline{G}_4 \Big| \begin{smallmatrix} 1,2,3,4,5 \\ \downarrow \\ 2,3,1,5,4\end{smallmatrix} \Big) 
\notag \\
& \ \ \ \ - s_{24} \Big( \overline{G}_3 \Big| \begin{smallmatrix} 1,2,3,4,5 \\ \downarrow \\ 2,4,1,5,3\end{smallmatrix}
- \overline{G}_4 \Big| \begin{smallmatrix} 1,2,3,4 \\ \downarrow \\ 2,4,1,3\end{smallmatrix} \Big)
- s_{25} \overline{G}_4 \Big| \begin{smallmatrix} 1,2,3,4,5 \\ \downarrow \\ 2,5,1,3,4\end{smallmatrix}
\end{align}
which yields the expansions of $F_j$ in (\ref{expFs}). The transpositions and permutations annotated on the
right of the $|_{\ldots}$ act on the external momenta in the $\alpha'$ expansion of the respective integrals.
We note that $J_{1,5}$ may be evaluated in terms of $J_{1,2}$ by using the relation,
\bea
J_{1,5} = \overline{J}_{1,2} \Big| \begin{smallmatrix} 1,2,3,4,5 \\ \downarrow \\ 5,1,2,3,4 \end{smallmatrix}
\eea
The complex conjugations on $G$ in the expressions for $F_3, F_4$ and $J_{1,2}$ do not complex conjugate the kinematic variables $s_{ij}$.

\newpage

\section{Degenerations of genus-two modular graph functions}
\label{sec:degen}

In this appendix, we shall obtain the non-separating degeneration, the separating degeneration, and the tropical limit of the modular graph functions which are needed to order $D^6 \cR^5$ in the analysis of the 5-point amplitude. Standard mathematical references on degenerations of Riemann surfaces are  \cite{fay73, Schiffer}. Here we shall briefly review the methods developed in \cite{DHoker:2017pvk,DHoker:2018mys} to obtain the non-separating and tropical degenerations of higher genus modular graph functions, restricted here to the application to genus two.

\subsection{The non-separating degeneration}

To describe the non-separating degeneration of a genus-two surface $\Sigma$ to a genus-one surface, it is useful to parametrize the period matrix $\Omega$ of the Riemann surface $\Sigma$ as follows, 
\be
\Omega = \begin{pmatrix} \tau & v \\ v & \sigma_1 + i (t + \tau_2 u_2^2)  \end{pmatrix}
\hskip 1in 
Y = \Im \Omega = \begin{pmatrix} \tau_2 & \tau_2 u_2 \\ \tau_2 u_2 &t + \tau_2 u_2^2  \end{pmatrix}
\ee
where $\tau = \tau_1 + i \tau_2$ and $v = u_1 + \tau u_2$ 
with $\tau_1, \tau_2, u_1, u_2, \sigma _1, t \in \RR$ and $\tau_2, t >0$.  The non-separating degeneration corresponds to the limit $t \to \infty$ keeping $\tau,v$ and $\sigma_1$ fixed.

\begin{figure}[h]
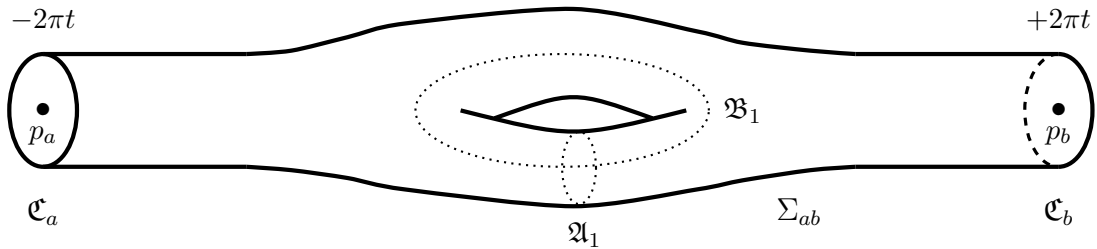

\begin{center}
\tikzpicture[scale=1.5]
\scope[xshift=-5cm,yshift=0cm]
\draw [ultra thick] (1.8,-1) arc (-90:90:0.3 and 0.5);
\draw [ultra thick] (1.8,0) arc (90:270:0.3 and 0.5);
\draw [ultra thick] (1.8,0) -- (3.6,0);
\draw [ultra thick] (1.8,-1) -- (3.6,-1);
\draw[ultra thick] (5.5,-0.5) .. controls (6.5, -0.75) .. (7.5,-0.5);
\draw[ultra thick] (5.8,-0.57) .. controls (6.5, -0.32) .. (7.2,-0.57);
\draw [ultra thick] (10.8,-1) arc (-90:90:0.3 and 0.5);
\draw [very thick, dashed] (10.8,0) arc (90:270:0.3 and 0.5);
\draw [ultra thick] (9,0) -- (10.8,0);
\draw [ultra thick] (9,-1) -- (10.8,-1);
\draw[ultra thick] plot [smooth] coordinates {(3.6,0) (4,0.03) (4.5, 0.15) (5, 0.27) (6.5,0.4)  (7.6, 0.2) (8, 0.10) (8.6,0.03) (9,0)};
\draw[ultra thick] plot [smooth] coordinates {(3.6,-1) (4,-1.03) (4.5, -1.10)  (5, -1.2) (6.5,-1.35) (7.6, -1.2) (8, -1.12) (8.6,-1.03) (9,-1)};
\draw [thick, dotted] (7.7,-0.5) arc (0:360:1.3 and 0.5);
\draw [thick, dotted] (6.7,-1.01) arc (0:360:0.15 and 0.32);
\draw (1.8,-1.4) node{$\mC_a$};
\draw (1.8, -0.5) node{$\bullet$};
\draw (1.8, -0.72) node{\small $p_a$};
\draw (1.8,0.3) node{\small $-2 \pi t$};
\draw (10.8,-1.4) node{$\mC_b$};
\draw (10.8,0.3) node{\small $+2 \pi t$};
\draw (10.8, -0.5) node{$\bullet$};
\draw (10.8, -0.72) node{\small $p_b$};
\draw (8,-0.5) node{{\small $\mB_1$}};
\draw (6.6,-1.6) node{{\small $\mA_1$}};
\draw (8.5,-1.4) node{{$\Sep$}};
\endscope
\endtikzpicture
\caption{The surface $\Sep$ is obtained from $\Sigma$ in the vicinity of the non-separating degeneration limit  by cutting $\Sigma$ along a cycle homologous to $\mA_2$ and adjusting the position of the cycle so that $\mC_a$ and $\mC_b$  are level sets for $f= \pm 2 \pi t$. \label{fig:2}}
\end{center}
\end{figure}

Actually, to obtain the desired expansion of modular graph functions, we shall be interested not just in the non-separating degeneration limit (which is a genus-one surface with two punctures $p_a, p_b$), but in a small but finite neighborhood of this limit. To parametrize this neighborhood, we reconstruct the genus-two surface $\Sigma$ from a genus-one surface $\Sigma_{ab}$ with two disconnected boundary discs $\mC_a, \mC_b$ as shown in figure \ref{fig:2}. The surface $\Sigma _{ab}$ may be obtained from an underlying compact genus-one surface $\Sigma_1$ with modulus $\tau$ and two marked points $p_a,p_b$ obeying $v=p_b-p_a$ from which the discs $\mC_a,\mC_b$ centered at $p_a,p_b$  of radius $R$ have  been removed. Note that the points $p_a,p_b$ belong to $\Sigma _1$ but not to $\Sigma _{ab}$. The genus-two surface $\Sigma$ is obtained by gluing annular neighborhoods of size $t$ of the boundary curves $\mC_a, \mC_b$  together. The details of the construction may be found in \cite{DHoker:2017pvk,DHoker:2018mys}.

\sm

To describe this construction concretely, it is useful to introduce  the linear combination $\om_t=\om_2 - u_2 \om_1$ of holomorphic $(1,0)$ forms on $\Sigma$, such that the canonical forms defined in \eqref{nu0}, \eqref{Deltaxy} decompose as follows, 
\bea
\kappa (x) & = & { i \over 4 \tau_2} \om_1 (x)  \oom_1(x) + { i \over 4 t} \om_t (x)  \oom_t(x) 
\no \\
\nu(x,y) & = & { i \over 2 \tau_2} \om_1 (x)  \oom_1(y) + { i \over 2 t} \om_t (x)  \oom_t(y) 
\no\\
\label{doubnu}
\Delta(x,y) & = &
\om_1(x)   \om_t(y) -  \om_t(x) \om_1(y)  
\eea
In the limit $t\to \infty$ for a fixed point $x \in \Sigma_{ab}$, the holomorphic $(1,0)$ forms   behave as follows,
\bea
\om _1 = 1 + \cO( e^{ - 2 \pi t})  \hskip 1in
\om _t = { i \over 2\pi} \p_x f (x) + \cO(e^{- 2 \pi t} )
\eea
Here, the real-valued function  $f(x)$ plays the role of a Morse function on $\Sigma_{ab}$ and may be given explicitly  in terms of the genus-one Arakelov Green function $g(x,y)= g(x-y|\tau )$ on $\Sigma_1$ by the following exact formula,
\bea
f(x) = g(x,p_b) - g(x,p_a) \hskip 1in v = p_b -p_a
\eea
The discs $\mC_a$ and $\mC_b$ may  be specified concretely by the conditions $f(\mC_a) = - 2 \pi t$ and $f(\mC_b)= + 2 \pi t$, as shown in figure \ref{fig:2}. For sufficiently large $t$, the discs will be disjoint. 

\sm

The Arakelov Green function $\cG(x,y)$ has an exact asymptotic expansion as $t\to \infty$, for fixed $x,y\in \Sigma_{ab}$,  given  by \cite{DHoker:2017pvk},
\bea
\label{AGtotildeG}
\cG(x,y) = \tilde G(x,y) + \tilde \gamma (x) + \tilde \gamma (y) + \tilde \gamma_0 + \cO(e^{- 2 \pi t}) 
\eea
where the terms in the sum are given by,
\bea
\tilde G(x,y) & = & g(x,y) - { f(x) f(y) \over 4 \pi t}
\no \\
\tilde \gamma (x) & = & - { 1 \over 4} g(x,p_a) - { 1 \over 4} g(x,p_b) + {f(x)^2 \over 16 \pi t}
\no \\
\tilde \gamma _0 & = & { \pi t \over 12} + { 1 \over 4} g(v) - { F_2(v)\over 8 \pi t}
\eea
and $F_k$ is the genus-one elliptic modular graph function defined in (\ref{defFk}). 
For later use, it will be useful to further introduce combinations familiar from  \cite{DHoker:2018mys},
\bea
D_k(\tau) &=&  \int _{\Sigma _1} \kappa_1(z) \, g(z)^k 
\no\\
D_k^{(\ell)}(v\vert \tau) & = & \int _{\Sigma _1} \kappa_1(z) \, g(z,v)^\ell g(z)^{k-\ell} 
\no \\
g_{k+1}(v\vert \tau) & = & \int _{\Sigma _1} \kappa_1(z) \, g(z,v) g_k(z)
\eea
such that $g_1(v|\tau)=g(v|\tau)$, $D_k^{(\ell)}(0|\tau) = D_k (\tau)$, and $g_k(0|\tau)= E_k(\tau)$, where $E_k$ is the non-holomorphic Eisenstein series (\ref{defEis}). Henceforth we shall  suppress the dependence on $v$ and~$\tau$.

\subsubsection{Useful integrals}

To expand the genus-two integrals near the non-separating degeneration, the following simple integrals over $\Sigma_{ab}$ will be needed, 
\bea
\label{oddf}
&&
\int  _{\Sep }\, \om _t \, \oom _1 \, f^n  = \int  _{\Sep } \om _t \, \overline{\om _t } \, f^{2n+1}  = 0
\\ &&
\label{regs}
\int _{\Sep } \om _t \, \overline{\om _t } \, \, f^{2n}  =  - { 2 i \over 2n+1} (2 \pi)^{2n} t^{2 n+1}
\\ &&
\label{gD}
\int _{\Sep }\om_t(y) \, \oom_1(y) \, g_n(y-x)  =   {\tau_2 \over  \pi } \, \p_x  f_{n+1} 
\eea
where we define $f_n (x) = g_n(x-p_b) - g_n(x-p_a)$, such that $f_1=f$. 
For any function  $\psi(x)$ which is smooth on $\Sep $ and whose Laplacian $\p_x\pbx \psi(x)$ is smooth on $\Sep$,  but which does not need to extend to a smooth function at the punctures $x=p_a,p_b$, we have,\footnote{The middle term was omitted in equation (A.21) of  \cite{DHoker:2018mys}, but its effect was correctly included in the subsequent equations in section A.5 of that reference.}
\bea
\label{fintnew}
\int _\Sep \om _t  \overline{\om_t } \, f^n \psi
& = & 
- \,  { i \, (2 \pi t )^{n+1}  \over 4 \pi^2 (n+1)} \int _0 ^{ 2 \pi} d \theta 
\Big ( \psi \left (p_b^\theta   \right ) 
+ (-)^n \psi \left (p_a^\theta  \right ) \Big ) 
\no \\ &&
- { i (2 \pi t)^{n+2} \over 8 \pi^2 (n+1) (n+2)} \int _0 ^{2 \pi} d \theta \, R { \p \over \p R}  \Big ( \psi (p_b ^\theta) + (-)^n \psi (p_a^\theta) \Big )
\no \\ &&
- {i \tau_2 \over 2 \pi^2(n+1)(n+2) }  \int _{\Sigma _{ab}} \kappa_1 (z) \,  f(z)^{n+2} \, \p_z \pbz \psi (z)
\eea
where $p_{a,b}^\theta= p_{a,b} + R e^{i\theta}$ and $\theta$ is the coordinate on the boundary circles $\mC_a, \mC_b$.  
The relation between $R$ and $t$ is given by the behavior of the scalar Green function for nearly coinciding points, 
\bea
g(z-p|\tau) = 2 \pi t - \ln { |z-p|^2 \over R^2} + g(v|\tau) + \cO(z-p)
\eea
for $p=p_a,p_b$.
Using these fundamental formulas, we derive integrals required to evaluate the non-separating asymptotics of $\f, \cZ_1, \cZ_2,\cZ_3, \cZ_4,\cZ_5$ and, abbreviating $g = g(v|\tau)$,  we get, 
\bea
\label{Sepint}
2 \pi i \int _\Sep \om_t (z) \oom_t (z) g(z,p_{a,b}) 
&=& 2 \pi^2 t^2 + 4 \pi t g +F_2 
\no \\
- { i \over 4t} \int _\Sep \om_t (z) \oom_t (z) g(z,p_{a,b})^2  
& = &
- {\pi^2 t^2 \over 3}  - { \pi t \over 2} g - \half g^2 -{ D_3 - D_3 ^{(1)} \over 8 \pi t} - {\Delta _v F_4 \over 16 \pi^2 t} 
\no  \\
- {i \over 4t} \int _\Sep \om_t (z) \oom_t (z) g(z,p_a) g(z,p_b) 
& = &
 - { \pi t \over 2} g - \half g^2 -{ D_3 - D_3 ^{(1)} \over 8 \pi t} - {\Delta _v F_4 \over 16 \pi^2 t} 
\eea
The Laplacian on $v$  is defined by  $\Delta_v = 4\tau_2 \p_v \p_{\bar v}$.

\subsubsection{Non-separating degeneration of $\f$ and $\cZ_{1,2,3}$}

Using the formulae above, and some further identities derived from them, the  asymptotic expansion of the
Kawazumi-Zhang invariant may be derived and gives,
\bea
\varphi = \frac{1}{6} \pi t +\frac{1}{2} g  + \frac{5 F_2}{4 \pi t} + \cO(e^{-2 \pi t})
\eea
while the asymptotics of the integrals $\cZ_{1,2,3}$ in \eqref{defZs} was obtained in equation (3.21) of \cite{DHoker:2018mys},
\bea
\cZ_1
& = &  
\frac{13\pi^2 t^2}{90}  
+ \frac{\pi t}{3} g 
+  \frac{ E_2+ g^2- F_2}{2}
+ \frac{1}{ \pi t} \bigg ( - D_3- D_3^{(1)}   - \half g  F_2 + 2 g_3
 \\ && \hskip 1in
   + 4 \zeta_3 + \frac{\Delta _v F_4}{4\pi}  \bigg )
 + { 1 \over 8 \pi^2 t^2} \Big ( 3F_2 ^2 + 12 F_4 +\cK^c  \Big ) 
 + \cO(e^{-2\pi t})
\no \\
\cZ_2 
&= & 
-\frac{7\pi ^2 t^2}{90}
-\frac{\pi t}{3}  g 
-\frac{2E_2+g^2-F_2}{2}
+ \frac{1}{ \pi t} \bigg ( - 2 D_3   + \half  g F_2 
 + 2 g_3 + 2  \zeta_3 
 \no \\ && \hskip 1in 
 - \frac{\Delta_v \left ( F_2^2 + 2 F_4 \right ) }{16\pi}  \bigg )
 -\frac{ (\Delta_\tau + 5)F_4}{4 \pi ^2 t^2}   + \cO(e^{-2\pi t})
  \no  \\
\cZ_3&=& 
\frac{ (\pi t)^2} {18}
+\frac{\pi t}{3} g 
+\frac{1}{6}   ( F_2+3   g^2)
  +  \frac{1}{ \pi t}  \left( - \half g F_2 + \frac{\Delta_vF_2^2}{8\pi}  \right)
+ \frac{( \Delta_\tau +5)F_2^2 }{ 8 \pi^2 t^2} + \cO(e^{-2\pi t})
\no
\label{Z123exp}
\eea
where the Laplacian on $\tau$ is defined by $\Delta_\tau= 4 \tau_2^2 \p_\tau \p_{\bar\tau}$ while $\cK^c$ in $\cZ_1$ is the (complicated) regularized integral defined in equation (3.40) of \cite{DHoker:2018mys}, which depends on $\tau$ but not on $v$.

\subsubsection{Non-separating degeneration of $\cZ_4$}
\label{secZ4min}

Using (\ref{nu1}), the integral $\cZ_4 $ defined in (\ref{defZs})  can be recast as follows, 
\be
\cZ_4 +2 \cZ_1 =  - \int _{\Sigma^2} \frac{|\Delta(x,y)|^2}{\det Y}  \cG(x,y)^2 
\ee
The computation of the asymptotics of the integral is similar to the one for $\cZ_1$ in \cite{DHoker:2018mys}. Using the last identity in  (\ref{doubnu}) we decompose it into  $\cZ_4+2\cZ_1 = \cZ_- - \cZ_+$, where
\bea
\cZ_+  & = & { 2 \over \tau _2 t} \int _{\Sigma_{ab}} \cG(x,y)^2 \, \om_1 (x) \oom_1 (x) \om_t(y) \oom_t(y)
\no \\
\cZ_-  & = & { 2 \over \tau _2 t} \int _{\Sigma_{ab}} \cG(x,y)^2 \, \om_1 (x) \oom_t (x) \om_t(y) \oom_1(y)
\eea
Substituting \eqref{AGtotildeG} into these equations, we get 
$\cZ_\pm = \cZ_\pm ^{(a)} + \cZ_\pm ^{(b)} + \cZ_\pm  ^{(c)}  +{\cal O}(e^{-2\pi t})$, 
where
\bea
\cZ_+  ^{(a)}  & = & { 2 \over \tau _2 t} \int _{\Sigma_{ab}} \tilde G(x,y)^2 \, \om_1 (x) \oom_1 (x) \om_t(y) \oom_t(y)
\no \\
\cZ_+  ^{(b)}  & = & { 4 \over \tau _2 t} \int _{\Sigma_{ab}}  \tilde  G(x,y) \Big (  \tilde \gamma (x) +  \tilde \gamma (y) +  \tilde \gamma _0 \Big )  \, \om_1 (x) \oom_1 (x) \om_t(y) \oom_t(y)
\no \\
\cZ_+  ^{(c)}  & = & { 2 \over \tau _2 t} \int _{\Sigma_{ab}} \Big (  \tilde \gamma (x) +  \tilde \gamma (y) +  \tilde \gamma _0 \Big ) ^2  \, \om_1 (x) \oom_1 (x) \om_t(y) \oom_t(y)
\eea
and
\bea
\cZ_-  ^{(a)}  & = & { 2 \over \tau _2 t} \int _{\Sigma_{ab}}  \tilde G(x,y)^2 \, \om_1 (x) \oom_t (x) \om_t(y) \oom_1(y)
\no \\
\cZ_-  ^{(b)}  & = & { 4 \over \tau _2 t} \int _{\Sigma_{ab}}  \tilde G(x,y) \Big ( \tilde  \gamma (x) +  \tilde \gamma (y) +  \tilde \gamma _0 \Big )  \, \om_1 (x) \oom_t (x) \om_t(y) \oom_1(y)
\no \\
\cZ_-  ^{(c)}  & = & { 2 \over \tau _2 t} \int _{\Sigma_{ab}} \Big ( \tilde  \gamma (x) + \tilde  \gamma (y) +  \tilde \gamma _0 \Big ) ^2  \, \om_1 (x) \oom_t (x) \om_t(y) \oom_1(y)
\eea
These integrals can be computed using the same techniques as in \cite{DHoker:2018mys}:
\begin{itemize}
\item For $\cZ_+  ^{(a)}$, using \eqref{regs} we get 
\bea
\cZ_+  ^{(a)}  &=& - { 4i \over  t} \int _{\Sigma_{ab}} \kappa_1(x) \Big ( g(x,y)^2 - g(x,y) { f(x) f(y) \over 2 \pi t} + { f(x)^2 f(y)^2 \over 16 \pi^2 t^2} \Big )  \om_t(y) \oom_t(y)
\notag\\
&=& - 8 E_2 + { 8 \over 3} F_2- { 4 F_4 \over \pi^2 t^2} 
\eea
\item For $\cZ_+  ^{(b)}$, using \eqref{gD} and \eqref{fintnew} we get
\bea
\cZ_+  ^{(b)}  &=&  { 2i \over  t} \int _{\Sigma_{ab}} \kappa_1(x) g (x,y) \left (
 g(x,p_a) + g(x,p_b) - {f(x)^2 \over 4 \pi t} \right )    \om_t(y) \oom_t(y)
 \notag\\
 &=&  4 E_2 + 4 g_2 -{ 2 \over \pi t} \big ( D_3 - D_3^{(1)} \big ) +{ 6 F_4 - F_2^2 \over 2 \pi^2 t^2}
\eea
\item For $\cZ_+  ^{(c)}$, expanding in powers of $\tilde\gamma(y)$ we get
\bea
\label{Zpc}
\cZ_+^{(c)} & = & 
- 8 \int _{\Sigma_1} \kappa _1 (x) \big ( \tilde\gamma (x) + \tilde\gamma _0 \big )^2 
- { 8 i \over t} \Big ( \tilde\gamma _0 + { F_2 \over 8 \pi t}  \Big )
\int _{\Sigma _{ab}} \tilde\gamma (y) \, \om_t(y) \oom_t(y)
\no \\ &&
- { 4i \over t} \int _{\Sigma_{ab}} \tilde\gamma (y)^2   \om_t(y) \oom_t(y)
\eea
Using \eqref{Sepint} the three terms evaluate to the three lines below,
\bea
\cZ_+^{(c)}  &=&
 - 8 \tilde\gamma _0^2 - 2 \tilde\gamma _0 { F_2 \over \pi t}- E_2 -g_2
+\frac{1}{2\pi t}(D_3 - D_3^{(1)})  -{  3F_4  \over 4 \pi^2 t^2} 
\no \\ &&
+ \Big( \tilde\gamma _0 + { F_2 \over 8 \pi t}  \Big ) \Big (  {2F_2 \over  \pi t} +8g + { 8 \pi t \over 3} \Big )
\no \\ &&
 - { 4 \pi^2 t^2 \over 15} - { 4 \pi t \over 3} g - 2 g^2  - { \Delta _v F_4 \over 4  \pi^2 t} -\frac{1}{2\pi t}(D_3 - D_3^{(1)})  + { F_4  \over 4 \pi^2 t^2} 
\eea

\item For $\cZ_-  ^{(a)}$, integrating by parts using $\tau_2 \pbx \p_y g(x,y) = \pi \delta (x,y) - \pi$ we get 
\bea
\cZ_-^{(a)} &=&  { 2 \over \tau _2 t} \int _{\Sigma_{ab}} \Big ( g(x,y)^2 - g(x,y) { f(x) f(y) \over 2 \pi t}  \Big ) \, \om_1 (x) \oom_t (x) \om_t(y) \oom_1(y) \notag\\
&=& - { 2 \tau_2 \over \pi^2 t} \int _{\Sigma _1} \kappa_1(x) \kappa_1(y) g(x,y)^2 \pbx f(x) \p_y f(y)
+  { 6 F_4 - F_2^2 \over \pi^2 t^2}  \notag\\
&=& - { 4  \over \pi t} \big ( D_3 - D_3^{(1)}  \big ) +  { 6 F_4 - F_2^2 \over \pi^2 t^2}  
\eea

\item For $\cZ_-  ^{(b)}$, integrating by parts using $\p_x \pbx f_2= -\pi f/\tau_2$, we get
\bea
\cZ_-  ^{(b)}  &=& { 2 \over \tau _2 t} \int _{\Sigma_{ab}} g(x,y) \Big ( 2\tilde\gamma (x)  + \tilde\gamma _0 \Big )  \, \om_1 (x) \oom_t (x) \om_t(y) \oom_1(y) + \hbox{c.c.} \notag\\
&=& 
 - { 2 \tau_2 \over \pi^2 t} \int _{\Sigma_{ab}} \kappa_1(x)
\Big ( \p_x f_2(x)    \pbx f  + \pbx f_2(x) \p_x f(x) \Big ) \Big ( 2\tilde\gamma (x)  + \tilde\gamma _0 \Big )  
\notag\\
&=&  {  \mZ_-^{(b)} \over \pi t}  -  { 4 F_4\over \pi^2 t^2} - 8\tilde\gamma _0 {  F_2 \over \pi t}  
\eea
where the term  $\mZ_-^{(b)}$ originates from the contributions $g(x,p_a)+g(x,p_b)$ inside $\tilde\gamma(x)$. It  can be evaluated again by substituting $f=g(x,p_b) - g(x,p_a)$ and integrating by parts,
\bea
\mZ_-^{(b)} & = &
 { \tau_2 \over \pi} \int \kappa _1(x) \Big ( \p_x f_2 (x) \pbx f(x) + \pbx f_2(x) \p_x f(x) \Big ) 
 \Big ( g(x,p_a) + g(x,p_b) \Big )
 \notag\\
 &=& 2 D_3 - 2 D_3 ^{(1)} + 4 g\, F_2 + 4 E_3 - 4 g_3
\eea

\item For $\cZ_-  ^{(c)}$, using \eqref{gD} and $\tau_2|\p  f_2 (p_a)|^2 = \tfrac{1}{8\pi}\Delta_v (F_2^2) -  g F_2$  we get
\bea
\cZ_-  ^{(c)}  &=& { 4 \over \tau _2 t} \int _{\Sigma_{ab}}  \tilde\gamma (x)  \tilde\gamma (y)  \om_1 (x) \oom_t (x) \om_t(y) \oom_1(y)
\notag\\
&=&  - { \tau_2 \over 4 \pi^2 t} \Big |  \p_{p_a} f_2(p_a) + \p_{p_b} f_2(p_b) \Big |^2
=  - {1 \over 8 \pi^2  t} \left( \Delta_v F_2^2 - 8\pi g F_2\right) 
\eea
\end{itemize}
Collecting all terms, we find 
\bea
\cZ_+ &=& 
-\frac{\pi ^2 t^2}{10}-\frac{\pi t}{3}  g -2  E_2-\frac{F_2}{6}-\frac{1}{2} g^2
-\frac{4  D_3- 4 D_3^{(1)} -  g\, F_2 }{2 \pi  t} 
\notag\\
&& 
-\frac{\Delta_v F_4 }{4 \pi ^2 t} -\frac{  3F_2^2+12 F_4 }{8 \pi ^2   t^2} + \cO(e^{-2\pi t})
\\
 \cZ_- &=& -\frac{2}{3} F_2 
  -{  2D_3-2D_3^{(1)}- 3 g F_2 + 4 g_3 - 4 E_3 \over  \pi t}
 -{ \Delta_v F_2^2  \over 8 \pi^2 t} +\frac{2 F_4}{\pi ^2 t^2} + \cO(e^{-2\pi t})
 \no
\eea
and therefore, using the expansion of $\cZ_1$ in \cite[(3.21)]{DHoker:2018mys},
\bea
   \label{Z4exp} 
\cZ_4 &=& 
-\frac{17 \pi ^2 t^2}{90}-\frac{\pi t}{3}   g -6  E_2+ {F_{2} \over 2} - {g^2 \over 2}
+ \frac{4 D_3+4 D_3^{(1)}+8 E_3+7 g \, F_{2} -16 (g_3+\zeta_3)}  {2  \pi t}
 \no \\ &&  
 - \frac{\Delta_v (F_2^2+2F_4) }  {8  \pi ^2 t}
   -\frac{3 F_{2}^2-4 F_{4}+2 \cK^c}{8 \pi ^2  t^2}  + \cO(e^{-2\pi t})
\eea  
where $\cK^c$ is the regularized integral defined in formula  (3.40) of \cite{DHoker:2018mys}.

\subsubsection{Non-separating degeneration of $\cZ_5$}
\label{secZ5min}

To evaluate the non-separating asymptotics of $\cZ_5$, we start from its defining formula in (\ref{defZs}) as well as a closely related integral we shall denote here by $\cZ_5'$, 
\bea
\cZ_5 ={ 16 \over i \pi} \int _{ \Sigma ^4}  \p_1 \cG(1,2) \bar \p_1 \cG(1,3) \cG(1,4) \nu(2,4) \nu(4,3) \nu(3,2)
\label{defZ5}
\no\\
\cZ'_5 ={ 16 \over i \pi} \int _{ \Sigma ^4}  \p_1 \cG(1,2) \bar \p_1 \cG(1,4) \cG(1,3) \nu(2,4) \nu(4,3) \nu(3,2)
\eea 
Since this is the first example, here and in \cite{DHoker:2018mys}, of the non-separating degeneration of a genus-two modular graph functions which  involves derivatives of the Arakelov Green function, we shall present the computations in detail. We begin by using the identity, 
\bea
\label{triplenu1}
&&
\nu(x,y) \nu(y,z) \nu(z,x) + \nu(x,z) \nu(z,y) \nu(y,x) 
\no \\ && \hskip 0.5in
= - 8 \kappa(x) \kappa(y) \kappa(z) - 2\kappa (x) \nu(y,z) \nu(z,y)
\no \\ && \hskip 0.65in 
- 2 \kappa (y) \nu(x,z) \nu(z,x) -  2\kappa (z) \nu(x,y) \nu(y,x) 
\eea
and the fact that the Arakelov Green function $\cG(x,y)$ integrates to zero against the canonical K\"ahler form $\kappa(x)$, to conclude right away that we have,
\be
\cZ_5+\cZ'_5=0
\ee
To evaluate the difference, we instead use the identity, 
\bea
\label{triplenu2}
&&
\nu(x,y) \nu(y,z) \nu(z,x) - \nu(x,z) \nu(z,y) \nu(y,x) 
\\ && \hskip 0.2in 
=  { \kappa(x) - \kappa_1(x) \over 2 t \tau_2} 
\Big ( \om_t(y) \oom_1(y) \om_1(z) \oom_t(z) 
- \om_1(y) \oom_t(y) \om_t(z) \oom_1(z) \Big ) 
+ \hbox{cycl}(x,y,z)
\no \eea
This leads to \bea
\cZ_5 & = & { 4 i \over  \pi \tau_2 t } \int _{ \Sigma ^4}  \p_1 \cG(1,2) \bar \p_1 \cG(1,3) \cG(1,4) 
\Big ( \kappa_1(2) \om_t(3) \oom_1(3) \om_1(4) \oom_t(4)  
\no \\ && \hskip 1.3in
- \kappa_1(2) \om_1(3) \oom_t(3) \om_t(4) \oom_1(4)  + \hbox{cycl}(2,3,4) \Big )
\eea
For fixed $z_1$, the integrals over  $z_2, z_3, z_4$ reduce to one of the following  integrals,
\bea
K(x) & = & \int _{\Sigma_{ab}} \kappa_1(y) \cG(x,y) = { \pi t \over 12} -{ 1 \over 4} \Big ( g(x,p_a) + g(x,p_b) - g(p_a,p_b) \Big ) 
+{ f(x)^2 \over 16 \pi t }  +{\cal O}(e^{-2\pi t})
\no \\
L(x) & = & \int _{\Sigma_{ab}} \om_t(y) \oom_1(y) \cG(x,y) = { \tau_2 \over \pi} 
\left ( \p_x f_2(x) {-} { 1 \over 4} \p_{p_a} f_2(p_a) {-} { 1 \over 4} \p_{p_b} f_2(p_b) \right ) +{\cal O}(e^{-2\pi t})
\label{KLeqs1}
\eea
to their complex conjugate, or to one of their derivatives,
\bea
\p_x K(x) & = & \int _{\Sigma_{ab}} \kappa_1(y) \p_x \cG(x,y) 
= -{ 1 \over 4}  \p_x g(x,p_a)  -{ 1 \over 4}  \p_x g(x,p_b) +{ f(x) \p_x f(x)  \over 8 \pi t } 
+{\cal O}(e^{-2\pi t})
\no\\
\p_x L(x) & = & \int _{\Sigma_{ab}} \om_t(y) \oom_1(y) \p_x \cG(x,y) 
= { \tau_2 \over \pi}  \p_x^2 f_2(x)
+{\cal O}(e^{-2\pi t})
\no \\
\pbx L(x) & = & \int _{\Sigma_{ab}} \om_t(y) \oom_1(y) \p_{\bar x} \cG(x,y) 
= { \tau_2 \over \pi}  \p_{\bar x} \p_x f_2(x)+{\cal O}(e^{-2\pi t})
= - f(x)+{\cal O}(e^{-2\pi t})
\label{KLeqs2}
\eea
The function $f_2$ and its derivatives $\p f_2, \bar \p f_2$ are regular at $p_a,p_b$, while the singularities of $f, K, \p^2 f_2, \bar \p ^2 f_2$ are powers of logarithms at the worst. As  a result, the above integrals are absolutely convergent term by term and can be extended to  the compact torus $ \Sigma_1$,  
\bea
\cZ_5 & = & { 8 \over  \pi t} \int _{\Sigma_1} \kappa_1(x) \bigg [
 \p_x K(x) \Big (  L(x) \, \pbx L^*(x) -  L^*(x) \pbx L(x)   \Big )
 \no \\ && \hskip 0.9in 
 + \pbx K(x) \Big ( L^*(x) \p_x L(x) - L(x) \p_x  L^*(x)  \Big )
 \no \\ && \hskip 0.9in 
 +  K(x) \Big (  \bar \p_x L^*(x) \pbx L(x) - \p_x L(x) \pbx  L^*(x)  \Big ) \bigg ]
\eea
Integrating the first two lines by part so as to expose $K$ without derivatives, we get, 
\bea
\cZ_5  = \frac{ 24 \tau_2^2 }{ \pi^3 t} \int _{\Sigma_1} \kappa_1(x) K(x) 
\Big ( (\p_x \pbx f_2(x))^2 - \p_x^2 f_2(x)  \, \pbx^2 f_2(x)  \Big )  +{\cal O}(e^{-2\pi t})
\eea
Since we have,
\bea
(\p_x \pbx f_2(x))^2 - \p_x^2 f_2(x)  \, \pbx^2 f_2(x)  =
\half \p_x \left ( \p_x \pbx f_2 \, \pbx f_2 - \pbx^2 f_2 \, \p_x f_2 \right ) + \hbox{c.c.}
\eea
the $x$-independent terms in $K(x)$ give a vanishing contribution. 
The remaining terms may be organized as follows,
\bea
\label{ImUexp}
\cZ_5 = { C_1 \over 2\pi t} + { C_2 \over 2  \pi^2 t^2} + \cO(e^{-2\pi t})
\eea
where $C_1$ and $C_2$ are $t$-independent genus-one elliptic modular functions, defined by,
\bea
\label{defC12} 
C_1 & = & -12 \int _{\Sigma _1} \kappa _1(x) \Big ( g(x-p_a) + g(x-p_b) \Big ) 
\left ( f(x)^2 - { \tau_2^2 \over \pi^2} \, \p_x^2 f_2 \, \pbx^2 f_2 \right )
\no \\
C_2 & = & 3 \int _{\Sigma _1} \kappa _1(x) f(x)^2 
\left ( f(x)^2 - { \tau_2^2 \over \pi^2} \, \p_x^2 f_2 \, \pbx^2 f_2 \right )
\eea
Using formulas (B.33) and (B.35) of \cite{DHoker:2018mys}, in particular,
\bea
\p_x^2 f_2(x) = 2 \pi i \p_\tau f(x) \hskip 1in \p_\tau \p_{\bar \tau} g(x,y)=0
\eea
it is immediate to compute $C_2$,
\bea
\label{C2exp}
C_2 = 72 F_4(v) - 6 \Delta _\tau F_4(v)
\eea
The integral of the term proportional  to $f(x)^2$ in $C_1$ is easily evaluated,  leading to, 
\bea
C_1 & = &  - 24 \big ( D_3 - D_3 ^{(1)}(v) \big )  + 24  \mathcal{C}_1
\no \\ 
\cC _1 & = & { \tau_2^2 \over \pi^2} \int _{\Sigma _1} \kappa _1(x)  g(x,p_a)  \p_x^2 f_2 \, \pbx^2 f_2
\eea
The integral $\mathcal{C}_1$ is computed by integrating by parts and using  $\p_x \pbx f_2(x)=-\frac{\pi}{\tau_2} f(x)$,  
\bea
\cC _1 & = & - { \tau_2^2 \over \pi^2} \int _{\Sigma _1} \kappa _1(x)  \left ( \pbx g(x,p_a)  \p_x^2 f_2 \pbx f_2 - { \pi \over \tau_2}   g(x,p_a)  \p_x f \pbx f_2 \right )
\no\\
 & = & \tilde \cC_1 - { \tau_2 \over \pi} |\p  f_2 (p_a)|^2 + 2 E_3 -2 g_3 - D_3 + D_3^{(1)} 
\eea
where
\bea
\tilde \cC_1 & = &  { \tau_2 \over \pi} \int_{\Sigma _1} \kappa _1(x) g(x,p_a) \Big ( \pbx f_2 \, \p_x f + \p_x f_2 \, \pbx f \Big )
\eea
To evaluate the last integral, we substitute $f=g(x,p_b)-g(x,p_a)$ and integrate by parts, 
\bea
\tilde \cC_1 =  
D_3 - D_3^{(1)}(v)  + { \tau_2 \over \pi} \int_{\Sigma _1} \kappa _1(x) g(x,p_a) \Big ( \pbx f_2 \, \p_x g(x,p_b)  + \p_x f_2 \, \pbx g(x,p_b)  \Big )
\eea
To compute this last integral, we use the following identity,
\bea
\p_x \pbx \Big ( f_2 (x) g(x,p_b) \Big )
& = & 
\pbx f_2 \, \p_x g(x,p_b)  + \p_x f_2 \, \pbx g(x,p_b) 
\no \\ &&
- { \pi \over \tau_2} \Big ( f(x) g(x,p_b) + f_2(x) (\delta (x,p_b) -1) \Big )
\eea
The integral is now readily evaluated and we obtain, 
\bea
\tilde \cC_1 & = &   D_3 - D_3^{(1)}  + 2 g\, F_2 + 2 E_3 - 2 g_3
\no \\
\cC_1 & = &  2 g\,  F_2 + 4 E_3 - 4 g_3  - { \tau_2 \over \pi} |\p  f_2 (p_a)|^2  
\eea
Using $\tau_2|\p  f_2 (p_a)|^2 = \tfrac{1}{8\pi}\Delta_v (F_2^2) -  g F_2$ we find,
\bea
\label{C1exp}
C_1 =  - 24 (D_3 - D_3 ^{(1)})  + 72   g  F_2 + 96 (E_3 - g_3)
- \frac{3}{\pi} \Delta_v F_2^2 
\eea
We conclude that in the minimal non-separating degeneration,
\bea
\label{Uexp}
\cZ_5 = { C_1 \over 2 \pi t} + { C_2 \over 2  \pi^2 t^2} + \cO(e^{-2\pi t})
\eea
where $C_1$ and $C_2$ are given by \eqref{C1exp} and \eqref{C2exp}.

\subsubsection{A novel identity for genus-one elliptic modular graph functions}
\label{sec:F2F4}

Using the large $t$ expansions \eqref{Z123exp} and \eqref{Z4exp}, and choosing coefficients
judiciously, we observe that all terms up to order $\cO(1/t^2)$ cancel in the linear combination, 
\bea
\label{MGIexp}
\cZ_1 + \cZ_2 + \cZ_3+\half \cZ_4 - \varphi^2 = 
\frac{  \Delta_\tau (F_2^2-2F_4) - 6 F_2^2 + 4 F_4 }{8\pi^2 t^2} + \cO(e^{-2\pi t})
\eea
Moreover, using the Laurent expansions computed in \cite[\S C.3.3]{DHoker:2018mys}, we find that the coefficient of the  $\cO(1/t^2)$ in \eqref{MGIexp} is exponentially suppressed near the cusp. This strongly suggests that the left-hand side of \eqref{MGIexp} actually vanishes, motivating the conjectures \eqref{MGI} and \eqref{MGIell}. In appendix \ref{sec:MGIproof}, we shall prove that the genus-two identity \eqref{MGI} indeed holds, and obtain \eqref{MGIell} as a consequence of this fact.\footnote{After the first version of this work, a direct proof of \eqref{MGIell} 
based on genus-one methods has been given in \cite{Basu:2020pey}.}
Without doubt, \eqref{MGIell} is only the first in an infinite family of relations between genus-one elliptic MGFs, 
and systematic methods for deriving such identities are being developed in \cite{DKS}.

\subsection{Separating degeneration}

We shall now consider the separating degeneration, where the genus-two Riemann surface $\Sigma$ degenerates into two genus-one curves $\Sigma_1$ and $\Sigma'_1$, with two marked points $p\in  \Sigma_1, p'\in \Sigma'_1$ joined by a thin tube. We refer to \cite[\S 4]{DHoker:2018mys} for a detailed discussion of this degeneration,
and only recall a few basic facts.

\sm

This limit is obtained by sending to zero the off-diagonal entry of the period matrix $\Omega$,
keeping fixed the diagonal entries $\tau,\sigma$ corresponding to the complex moduli of 
$\Sigma_1$ and $\Sigma'_1$.  In the limit $v\to 0$, the Siegel modular group $Sp(4,\ZZ)$ is broken to  the product $SL(2,\ZZ)_\tau \times SL(2,\ZZ)_\sigma \ltimes \ZZ_2$, where the two $SL(2,\ZZ)$ factors  act by fractional linear transformations of $\tau$ and $\sigma$ and $\ZZ_2$ exchanges these two variables. 
The modulus $|\hat v|$ of the degeneration parameter
\be
\hat v= 2\pi v\, \eta(\tau)^2 \, \eta(\sigma)^2
\ee
stays invariant under the unbroken part of $Sp(4,\ZZ)$.
The Abelian differentials degenerate, up to terms of order $\cO(|\hat v|)$, to
\bea
\label{SDAb}
\omega_1  =  \begin{cases}
\omega(x)& ~~ x\in \Sigma_1 \\
0 &~~ x\in \Sigma'_1 \end{cases}\ ,
\qquad
\omega_2  =  \begin{cases}
0 &~~ x\in \Sigma_1 \\
 \omega'(x) ~~ & ~~ \, x\in \Sigma'_1
 \end{cases}
\eea
where $\omega$ and $\omega'$ are the standard Abelian differentials on $\Sigma_1$ and 
$\Sigma'_1$. The Arakelov Green function becomes, up to terms of order $\cO(|\hat v|)$,  
\bea
\label{SDArak}
\cG(x,y) \sim \begin{cases}
- \half \ln |\hat v|  + g(x-y|\tau) - \half g(x-p|\tau) - \half g(y-p|\tau) & x, y \in \Sigma _1  \\
- \half \ln |\hat v|  + g(x-y|\sigma) - \half g(x-p'|\sigma ) - \half g(y-p'|\sigma ) & x, y \in \Sigma _1'  \\
\half \ln|\hat v| +\half g(x-p|\tau)+\half g(y-p'|\sigma)\ , & x\in\Sigma_1  , y \in\Sigma'_1 
\end{cases}
\eea
The expansion of the modular graph functions $\varphi$ and $\cZ_{1,2,3}$
was computed in \cite[\S 4]{DHoker:2018mys} using these formulae, 
\bea
\varphi &=& -\ln |\hat v| + \cO(|\hat v|) \no\\
\cZ_1  &=& 2 (\ln |\hat v| )^2 + 4 E_2(\tau) + 4 E_2(\sigma) + \cO(|\hat v|) \no\\
\cZ_2  &=& -2 (\ln |\hat v| )^2 - E_2(\tau) - E_2(\sigma) + \cO(|\hat v|) \no\\
\cZ_3  &=& 2 (\ln |\hat v| )^2 + \cO(|\hat v|) 
\label{sepdegZi}
\eea
where $E_k(\tau)$ is the usual non-holomorphic Eisenstein series (\ref{defEis}) of $SL(2,\ZZ)$.

\sm

For the integral $\cZ_4$ defined in \eqref{defZs}, we see that the measure
 $\nu(x,y)\nu(y,x)$ vanishes in the limit $v\to 0$ unless
$x,y$ lie on the same elliptic curve, say $\Sigma_1$,  in which case it  reduces to $-\kappa_1(x)
\kappa_1(y)$. Hence the integral reduces to 
\bea
-4\int_{\Sigma_1^4 } \left[ - \half \ln |\hat v|  + g(x-y|\tau) - \half g(x-p|\tau) - \half g(y-p|\tau)  \right]^2 \kappa_1(x) \kappa_1(y) + (\Sigma_1\leftrightarrow \Sigma_1')  \no\\
\eea
Observing that the crossproducts integrate to zero, this evaluates to 
\bea
\cZ_4 &\sim & -\int_{\Sigma_1} \left[ ( \ln |\hat v|)^2  + 4 g(x-y|\tau)^2 +  g(x-p|\tau)^2 + g(y-p|\tau)^2  \right] \kappa_1(x) \kappa_1(y) + (\Sigma_1\leftrightarrow \Sigma_1') \no\\
  &=& -2 (\ln |\hat v| )^2 - 6E_2(\tau) - 6E_2(\sigma) + \cO(|\hat v|) 
\eea
Using \eqref{sepdegZi} this behavior is indeed 
 consistent with the identity \eqref{MGI}.

\sm

Turning to the integral $\cZ_5$ defined in \eqref{defZs}, we see that the measure  $\nu(2,4) \nu(4,3) \nu(3,2)$ vanishes unless the points 2,3,4 are on the same elliptic curve, say $\Sigma_1$, in which case it reduces to  $\kappa_1(2) \kappa_1(3) \kappa_1(4)$.  When the point 1 is also on $\Sigma_1$, we get 
\bea
\frac{16}{i\pi}  &&\int_{\Sigma_1^4} \left[ \partial_1 g(x_1-x_2)-\frac12 \partial_1 g(x_1-p) \right]
 \left[ \bar\partial_1 g(x_1-x_3)-\frac12 \bar\partial_1 g(x_1-p) \right] \no\\ &&\times 
 \left[ - \half \ln |\hat v|  + g(x_1-x_4) - \half g(x_1-p) - \half g(x_4-p) \right]
 \kappa_1(2) \kappa_1(3) \kappa_1(4)
\eea
In the first line, using translational invariance one can replace $\partial_1 g(x_1-x_2)$ and $\bar\partial_1 g(x_1-x_3)$  by 
$-\partial_2 g(x_1-x_2)$ and  $-\bar\partial_3 g(x_1-x_3)$, which  integrate by parts to zero. The terms proportional to $ g(x_1-x_4)$ or $ g(x_4-p)$ on the second line also integrate to zero. The integrals over $x_2,x_3,x_4$ are then trivial, leading to 
\be
\label{Usep1}
-\frac{2}{i\pi} \int_{\Sigma_1} \partial_1 g(x_1-p) \, \bar\partial_1 g(x_1-p) \, 
 \left[ \ln |\hat v|   + g(x_1-p) \right]
\ee
When the point 1 is  on $\Sigma'_1$, we get instead
\bea
\label{Usep2}
\frac{4}{i\pi} &&\int_{\Sigma'_1\times \Sigma_1^3} \partial_1 g(x_1-p') \bar\partial_1 g(x_1-p')  \left[  \half \ln |\hat v|  + \half g(x_1-p') + \half g(x_4-p) \right]
 \kappa_1(2) \kappa_1(3) \kappa_1(4) \no \\&&
 = \frac{2}{i\pi} \int_{\Sigma_1'} \partial_1 g(x_1-p') \, \bar\partial_1 g(x_1-p') \, 
 \left[ \ln |\hat v|   + g(x_1-p') \right] \qquad
\eea
The  contributions \eqref{Usep1} and \eqref{Usep2} cancel against those  where $\Sigma_1$ and $\Sigma'_1$ are exchanged, so we find that $\cZ_5$ vanishes in the separating degeneration, up to terms of order $|\hat v|$. It is quite remarkable that $\cZ_5$ vanishes both in the separating and
non-separating degenerations.

\subsection{Tropical limit}
\label{sec:trop}
Having obtained the expansion of $\cZ_4$ and $\cZ_5$ in the non-separating degeneration $t\to \infty$ keeping fixed $\tau$ and $v$, we can obtain the tropical limit by further sending $\tau\to i\infty$ keeping $u_2=\Im v/\tau_2$ fixed. The result can be re-expressed in terms of the variables $V,S=S_1+i S_2$ parametrizing the imaginary part of the period matrix via, 
\be
Y = \frac{1}{V S_2} \begin{pmatrix} 1 & S_1 \\ S_1 & |S|^2 \end{pmatrix}
\ee
such that the tropical limit corresponds to $V=(t\tau_2)^{-1/2} \to 0$ keeping $S_1=u_2$ and $S_2=(t/\tau_2)^{1/2}$ fixed.  For the modular graph functions $\varphi$  and $\cZ_{1,2,3}$ in 
\eqref{defZs}, this leads to  \cite[\S 5]{DHoker:2018mys}, 
\bea
   \label{Z123trop}
\varphi &\sim &
    \frac{5\pi}{6V} A_{1,0} +  \frac{5\zeta_3 }{4 \pi^2} \, A_{0,0} V^2 
\no\\
 \varphi^2 &\sim &
\frac{32\pi^2}{V} \left[ - \frac{1}{1512} A_{0,0} +  \frac{1}{1512} A_{0,2} - \frac{5}{1584} A_{1,1} + \frac{25}{1152} A_{2,0} \right]
+ \frac{25\zeta_3}{12\pi} A_{1,0} V + \frac{25\zeta_3^2}{16\pi^4} V^4    
 \no\\  
\cZ_1 &\sim & { 32\pi^2 \over V^2} \left[
-\frac{1}{315} A_{0,0} + \frac{1}{252} A_{0,2}-\frac{1}{792} A_{1,1} + \frac{23}{960} A_{2,0} \right] 
\no\\ &&+  \frac{\zeta_3}{\pi}  \left[ \frac{18}{5}\, A_{0,1}-\frac{1}{2} A_{1,0} \right] V
-\frac{\zeta_5}{2\pi^3} \,  A_{0,1} V^3 + \zeta_3^2  \frac{
(2\beta-3)}{16\pi^4} A_{0,0} V^4
\\
\cZ_2 &\sim& {32\pi^2 \over V^2} \left[
\frac{1}{504} A_{0,0} - \frac{1}{1008} A_{0,2} -\frac{5}{792} A_{1,1} - \frac{17}{960} A_{2,0} \right] 
-   \frac{5\zeta_3}{2\pi}\,  A_{1,0} V -\frac{7\zeta_5 }{4\pi^3} A_{0,1}  V^3 
\no \\
\cZ_3 &\sim & {32\pi^2 \over V^2} \left[
-\frac{11}{7560} A_{0,0} + \frac{1}{1512} A_{0,2}+\frac{1}{792} A_{11} + \frac{17}{576} A_{2,0} \right] 
+ \frac{5\zeta_3}{6\pi}  A_{1,0}  V +  \frac{11 \zeta_3^2}{8\pi^4}  A_{0,0} V^4 
\no
\eea
where $\sim$ indicates equality in the limit $V\to 0$ up to corrections of order $e^{-1/V}$.


Similarly, starting with \eqref{Z4exp} and  using the formulae in appendix C  of \cite{DHoker:2018mys}, 
we arrive at 
   \bea
   \label{Z4trop}
\cZ_4&\sim&
 \frac{32\pi^2}{V^2} \left[ \frac{1}{252} A_{0,0} - \frac{1}{168} A_{0,2} + \frac{5}{792} A_{1,1} -\frac{9}{320} A_{2,0}
\right]
\notag\\
&&+  \zeta_3 \frac{85 A_{1,0}-72 A_{0,1}}{10\pi} V + \frac{9}{2\pi^3} \zeta_5 A_{0,1} V^3 
+ \zeta_3^2\, \frac{
(3-\beta)}{4\pi^4}  A_{0,0} V^4
      \eea
where $A_{i,j}(S)$ are the local modular forms introduced in section 5.3 of 
{\it loc. cit.}, and $\beta$ is the unknown (presumably rational) 
coefficient appearing at order $V^4$ in the
tropical limit of $\cZ_1$. As a strong consistency check on the expansion  \eqref{Z4exp}, we
have reproduced the leading term in \eqref{Z4trop} from a worldline integral using the tropical
Arakelov Green function $\cG^{(sg)}$ along the lines of \cite[section 5.3]{DHoker:2018mys}.
       
\sm

Using \eqref{Z4trop} we obtain the tropical limit of the combinations $A_1$ and $A_2$ defined in \eqref{A1A2def},
\bea
A_1 &\sim& \frac{32\pi^2}{V^2} \left[ -\frac{13}{1512}  A_{0,0} +  \frac{5}{756} A_{0,2}  +\frac{5}{396} A_{1,1} +  \frac{4}{45} A_{2,0} \right]  \notag\\ 
&& + \zeta_3 \left[ \frac{18}{5\pi}  A_{0,1}  + \frac{16}{3\pi} A_{1,0} \right] V+ \frac{3\zeta_5}{\pi^3} A_{0,1} V^3 + \frac{(2 \beta +13) \zeta_3^2}{16 \pi ^4} V^4 
\\
A_2 &\sim& \frac{32\pi^2}{V^2} \left[ \frac{1}{7560}  A_{0,0} +  \frac{1}{1512} A_{0,2}  -\frac{1}{132} A_{1,1} +  \frac{1}{72} A_{2,0} \right] + \frac{10 \zeta_3}{3 \pi}  A_{1,0} V + \frac{7 V^4 \zeta_3^2}{4 \pi ^4}
\notag
\eea
In contrast, we find that the tropical limit of the integral \eqref{defZ5} starts at order $V$,
\be
\cZ_5 \sim \frac{72}{\pi} (A_{1,0} - \frac15 A_{0,1} ) V \zeta_3 + \frac{30}{\pi^3} A_{0,1} V^3 \zeta_5
\ee
The vanishing of  the leading $\cO(1/V^2)$ term is quite remarkable, and follows from the cancellation
of the leading term in the combination $f(x)^2 -\frac{\tau_2^2}{\pi^2} \p_x^2 f_2 \pbx^2 f_2$ appearing on both lines of \eqref{defC12}. We have also confirmed the vanishing at leading order by a wordline computation using the tropical
Arakelov Green function. 

\newpage

\section{Proof of the modular graph function identities}
\label{sec:MGIproof}

In this subsection, we shall prove the identity \eqref{MGI} 
between the genus-two modular graph functions defined in 
(\ref{KZ2}) and (\ref{defZs}).

\sm

Translation invariance and the resulting momentum conservation on the torus provides a fundamental tool in the proof of identities between genus-one modular functions (along with holomorphic subgraph reduction \cite{DHoker:2016mwo, Gerken:2018zcy} or Fay identities \cite{Gerken:2020aju}). The absence of translation invariance  prevents us from using the same techniques for higher genus surfaces. However, the lemma below provides the appropriate alternative tool, valid  for arbitrary genus~$h$.\footnote{Since the first version of this work,
the lemma (\ref{keylemma}) has been further generalized and applied to derive higher-weight identities at arbitrary 
genus \cite{DHoker:2020uid}.}

{\lem
\label{lemma}
On a compact Riemann surface $\Sigma$, with Arakelov Green function $\cG(x,y)$, the following identity holds
for arbitrary $y,z \in \Sigma$,  
\bea
\om_I(y) \, \p_z \GA (y,z) + \om _I (z) \, \p_y \GA (y,z) 
-  \p_z \Phi _I{}^J (z) \,  \om_J(y) -  \p_y \Phi _I{}^J (y) \, \om _J (z) =0
\label{keylemma}
\eea
where the tensor $\Phi_I{}^J (z)= \Phi_{IK} (z) Y^{KJ} $ is given by the following integral in $x \in \Sigma$, 
 \bea
 \label{PhiIJ}
 \Phi_{IJ} (z) = {i \over 2} \int _\Sigma \GA(z,x) \, \om_I(x) \overline{\om _J(x)}
 \eea
The tensor $\Phi$ is Hermitian $\overline{\Phi_{IJ} (z) } = \Phi _{JI} (z)$.}

\sm

 To prove Lemma \ref{lemma}, we first show that its left side  is holomorphic in $z$ and thus holomorphic in  $y$ by symmetry under swapping $z$ and $y$. The $\partial_{\bar y}$ derivative of the left side of (\ref{lemma}) may be evaluated using the identities on the Arakelov Green function $\cG$, given in (\ref{AraG}) for genus 2.
 The $\delta(y,z)$-functions cancel between the first two terms, and $\Phi$ has been defined so as to cancel also the remaining terms in the $\partial_{\bar y}$ derivative. As a result,  the left side of (\ref{lemma}) is a single-valued holomorphic $(1,0)$-form in $z$ and $y$ which takes the form $M_I {}^{JK} \om _J(z) \om _K(y)$  for some constant tensor $M_I{}^{JK}$.  To show that $M=0$, we integrate the left side of (\ref{lemma}) against $\oom_L(z)$, use the fact that the contributions from the first and third terms vanish, and that those of the second and fourth terms cancel using the definition of $\Phi$. 
 
 \sm
 
The proof of the identity \eqref{MGI} proceeds by a judicious use of the formula (\ref{lemma}). 
We begin by considering the alternative integral for $\cZ_2$ given on the first line of (\ref{nuint}), 
\bea
\cZ_2 = - 4 \int_{\Sigma^3} \nu(x,y) \nu (y,x)  \kappa(z)  \cG(x,z) \cG(y,z) 
\eea
and eliminate $\nu(x,y)$ defined by (\ref{nu0}) using the second equation in (\ref{AraG}). The $\delta$-function produces the term $- \cZ_1$ on the right side, so that we obtain,
\bea
\cZ_1 + \cZ_2 = { 2 i \over \pi} \int _{\Sigma^3} \nu(y,x) \p_x \pby \cG(x,y) \cG(x,z) \kappa(z) \cG(z,y)
\eea
Integrating by parts in $x$ and writing out $\kappa(z)$ explicitly, we obtain, 
\bea
\cZ_1 + \cZ_2 =  { Y^{IJ} \over 2\pi} \int _{\Sigma^3} \nu(y,x) \om_I(z) \oom_J(z)  \pby \cG(x,y) \p_x \cG(x,z)  \cG(z,y)
\eea
Next, we use formula (\ref{lemma})  to re-express the combination $\om _I (z) \, \p_x \GA (x,z) $, 
 \bea
 \label{Z1Z2}
\cZ_1 + \cZ_2 & = &  { Y^{IJ} \over 2\pi} \int _{\Sigma^3} \nu(y,x) \oom_J(z)  \pby \cG(x,y)  \cG(z,y)
\no \\ && \quad \times 
\Big [ - \om_I(x) \, \p_z \GA (z,x) 
+ \p_z \Phi _I{}^K (z) \,  \om_K(x) +  \p_x \Phi _I{}^K (x) \, \om _K (z) \Big ]
\eea 
To evaluate the contributions from the first two terms inside the brackets, we integrate by parts in both $\bar y$ and $z$ and combine various Abelian differentials into $\nu(x,z)$. For the first term we obtain,
\bea
- { Y^{IJ} \over 2\pi} \int _{\Sigma^3} \nu(y,x) \oom_J(z)   \cG(x,y)  \p_z \pby \cG(z,y)  \om_I(x) \, \GA (z,x) 
= - \half \cZ_4 + \half \cZ_2
 \eea
 where we have used the second equation in (\ref{AraG}) for the mixed double derivative on $\cG$, the expression for $\cZ_4$ in (\ref{defZs}) to evaluate the contribution from the $\delta$-function, and the alternative formula for $\cZ_2$ given in the second line of (\ref{nuint}). For the second term we obtain,
\bea
&&
-i Y^{IJ} \int _{\Sigma ^2} \nu(y,x) \oom_J(y) \om_K(x) \cG(x,y) \Phi _I{}^K (y)
\no \\ &&
+i Y^{IJ} \int _{\Sigma ^3} \nu(y,x) \oom_J(z) \om_K(x) \nu(z,y) \cG(x,y) \Phi _I{}^K (z)
\eea
Using the formula (\ref{PhiIJ}) for $\Phi$ makes all Abelian differentials explicit, and regrouping these into $\nu$ differentials we obtain for the  second term inside the brackets of (\ref{Z1Z2}),
\bea
- 2 \int _{\Sigma ^3} \nu(x,z) \nu(z,y) \nu(y,x) \cG(x,y) \cG(y,z)
= \half \cZ_2 + \f^2
\eea
where we have used the alternative integral for $\cZ_2$ in the second line of (\ref{nuint}) as well as the formula for $\f^2$ on the third line of (\ref{nuint}).  Finally, to evaluate the contribution from the third term inside the brackets of (\ref{Z1Z2}), we integrate by parts in $x$ and again use the second equation in (\ref{AraG}) to obtain,
\bea
&&
i Y^{IJ} \int _{\Sigma ^2} \nu(x,x) \oom_J(z) \om_K(x) \cG(x,z) \Phi _I{}^K (x)
\no \\ &&
-i Y^{IJ} \int _{\Sigma ^3} \nu(y,x) \oom_J(z) \om_K(z) \nu(z,y) \cG(z,y) \Phi _I{}^K (x)
\eea
Using the formula (\ref{PhiIJ}) for $\Phi$ makes all Abelian differentials explicit, and regrouping these into $\nu$ differentials we obtain for  the third term inside the brackets of (\ref{Z1Z2}),
\bea
&&
4 \int _{\Sigma ^3} \kappa(x) \nu(z,y) \nu(y,z) \cG(x,y) \cG(x,z)
\no \\ &&
- 2 \int _{\Sigma ^4} \nu(x,y) \nu(y,x) \nu(z,w) \nu(w,z) \cG(x,w) \cG(y,z)
\no \\ && \hskip 0.3in
= - \cZ_2 - \cZ_3
\eea
where we have used the second line of (\ref{nuint}) to evaluate the first integral, and (\ref{Delnu}) twice to transform the second integral into the expression for $-\cZ_3$ with $\cZ_3$ given in (\ref{defZs}). Assembling all contributions proves formula \eqref{MGI}.

\sm

By evaluating the non-separating asymptotics of the function $\cZ_4$-function independently and using the asymptotics  obtained for $\f, \cZ_1, \cZ_2, \cZ_3$ in \cite{DHoker:2018mys}, we have shown in appendix \ref{sec:F2F4} that \eqref{MGI} implies a highly non-trivial identity between the genus-one elliptic modular graph functions $F_2$ and $F_4$.

\newpage

\section{Overall normalization of the genus-two amplitude}
\label{app:convert}

The five-point genus-two amplitude including its overall coefficient was determined in \cite{Gomez:2015uha} at the leading order $D^2\cR^5$ using the non-minimal pure spinor formalism. The normalization of the amplitude followed from a first principles calculation using the integrals over pure spinor space derived in \cite{Gomez:2009qd} together with the conventions for genus-two measures of \cite{Gomez:2010ad}. Consequently, the building blocks $T^m_{1,2,3|4,5}$ and $T_{12,3|4,5}$ (collectively denoted by $T^{\rm NMPS}$) featured in \cite{Gomez:2015uha} also depend on zero-modes of
the non-minimal pure spinor $\bar\lambda_\alpha$ whose integration gives rise to various combinatorial factors.

\sm 

In contrast, in this work we use the building blocks $T^m_{1,2,3|4,5}$ and $T_{12,3|4,5}$ (collectively denoted by $T^{\rm MPS}$) defined with the minimal pure spinor formalism in \cite{Mafra:2015mja}; which do not depend on the zero-modes of $\bar\lambda_\alpha$. Despite their different definitions, one can verify that BRST-invariant quantities written in either setup yield the same results with differing normalizations. The component expansion for bosonic external states of the building blocks used in this work can be downloaded from \cite{PSSsite}.  We will now show that their relative normalization is such that $T^{\rm NMPS} = 2^{10}\,3^3\,5\, T^{\rm MPS}$.

\sm

To show this we compare the component expansion of the BRST-invariant kinematic factor at order $D^2\cR^5$.
The pure spinor superspace representation $\cB_{(5)}|_{D^2\cR^5}$ obtained in this work \eqref{FT} coincides
with equation (5.44) from \cite{Gomez:2015uha},
\bea
\label{5.44}
\cK_5^{(2)} & = &
{\big|\langle T_{12,3|4,5}\rangle_0 \big|^2\over s_{12}} + {\big|\langle T_{12,4|3,5}\rangle_0 \big|^2\over s_{12}}
+ {\big|\langle T_{12,5|3,4}\rangle_0 \big|^2\over s_{12}} 
\no \\ &&
+\big|\langle T^m_{3,4,5|1,2}\rangle_0 \big|^2 +
(1,2|1,2,3,4,5)
\eea
up to an overall coefficient\footnote{We note the different
convention for Mandelstam invariants, where $s_{ij}^{\rm here} = (\ap/2)s_{ij}^{\rm there}$.},
\begin{equation}
\label{relate}
\cB_{(5)}|_{D^2\cR^5} = 2^6 \halfap{}\,\cK_5^{(2)}
\end{equation}
Straightforward calculations for 5 Type IIB gravitons show that
\bea
\label{5.46}
\cB_{(5)} \big|_{D^2 \cR^5} & = & - 2^{14} \invhalfap4 \cBtree{5}
\no \\
\cK_5^{(2)} \big|_{{\rm IIB}} & = & -2^{28}\,3^6\,5^2 \invhalfap5 \cBtree{5}
\eea
where the result in the non-minimal formalism is given in equation (5.46) of \cite{Gomez:2015uha}.
From \eqref{relate} and \eqref{5.46} it follows that $T^{\rm NMPS} = 2^{10}\,3^3\,5\, T^{\rm MPS}$.

\sm

It remains to explain the overall coefficient of the genus-two five-point amplitude in \eqref{twoloopfive}.
It matches
the normalization of the amplitude derived in equations (5.41) and (5.43) of \cite{Gomez:2015uha},
\begin{equation}
\label{5.41}
\cA_{(5)}^{\rm 2-loop} = (2\pi)^{10}\delta^{10}(k)\halfap5\! {\kappa^5 e^{2\l}\over 2^{45} \, 3^{6}\,5^2\,\pi^6}
\int_{{\cal M}_2} {| d^3\Omega |^2 \over (\det Y)^5}\int_{\Sigma^5}
\bigl|\langle \cK^{(2)}(z_1, \ldots,z_5)  \rangle_0 \bigr|^2 {\rm KN}_{(5)}
\end{equation}
where the integral over vertex points is given by, 
\begin{equation}
\label{5.43}
\int_{\Sigma_5} |\langle \cK^{(2)}(z_1, \ldots,z_5)\rangle_0 |^2  {\rm KN}_{(5)}= 2^6\pi\halfap{}(\det Y)^2\, \cK_5^{(2)} + \cO(\ap^2)
\end{equation}
Equation \eqref{5.43} is the origin of the different factor of $2^6(\ap/2)$ in \eqref{relate}, while the factor of $\pi$
is taken into account in the normalization of \eqref{twoloopfive} which contains $1/\pi^5$ instead of $1/\pi^6$ in 
\eqref{5.41}\footnote{In general, this difference is taken into account by the factor ${1\over \pi}$ in the definition of
\eqref{defB5}.}.

The precise normalization of the five-point
SYM tree amplitude used in section~\ref{sec:amps} follows from
the evaluation of
$A_{\rm YM}(1,2,3,4,5) = \langle E_{1234}V_5\rangle_0$ \cite{Mafra:2010jq}
with the measure normalized as $\langle (\lambda\gamma^m\theta)
(\lambda\gamma^n\theta)(\lambda\gamma^p\theta)(\theta\gamma_{mnp}\theta)\rangle_0 = 1$. The
five-point tree amplitude available in \cite{PSSsite} is $2880$ times bigger.

\newpage


\begin{thebibliography}{10}
\itemsep=0.01in

\bibitem{D'Hoker:1988ta}
E.~D'Hoker and D.~Phong, ``{The Geometry of String Perturbation Theory},''
\href{http://dx.doi.org/10.1103/RevModPhys.60.917}{{\em Rev.Mod.Phys.} {\bf 60}
  (1988)  917}.

\bibitem{D'Hoker:2002gw}
E.~D'Hoker and D.~Phong, ``{Lectures on two loop superstrings},'' {\em
  Conf.Proc.} {\bf C0208124} (2002)  85--123,
\href{http://arxiv.org/abs/hep-th/0211111}{{\tt  arXiv:hep-th/0211111}}.

\bibitem{Witten:2012ga}
E.~Witten, ``{Notes On Super Riemann Surfaces And Their Moduli},''
  \href{http://arxiv.org/abs/1209.2459}{{\tt  arXiv:1209.2459 [hep-th]}}.

\bibitem{Witten:2012bh}
E.~Witten, ``{Superstring Perturbation Theory Revisited},''
\href{http://arxiv.org/abs/1209.5461}{{\tt  arXiv:1209.5461 [hep-th]}}.

\bibitem{Green:1982sw}
M.~B. Green, J.~H. Schwarz, and L.~Brink, ``{N=4 Yang-Mills and N=8
  Supergravity as Limits of String Theories},''
\href{http://dx.doi.org/10.1016/0550-3213(82)90336-4}{{\em Nucl.Phys.} {\bf
  B198} (1982)  474--492}.

\bibitem{D'Hoker:2005jc}
E.~D'Hoker and D.~H. Phong, ``{Two-loop superstrings VI: Non-renormalization
  theorems and the 4-point function},''
  \href{http://dx.doi.org/10.1016/j.nuclphysb.2005.02.043}{{\em Nucl. Phys.}
  {\bf B715} (2005)  3--90},
\href{http://arxiv.org/abs/hep-th/0501197}{{\tt  arXiv:hep-th/0501197}}.

\bibitem{Berkovits:2005df}
N.~Berkovits, ``{Super-Poincare covariant two-loop superstring amplitudes},''
  \href{http://dx.doi.org/10.1088/1126-6708/2006/01/005}{{\em JHEP} {\bf 01}
  (2006)  005},
\href{http://arxiv.org/abs/hep-th/0503197}{{\tt  arXiv:hep-th/0503197}}.

\bibitem{Gomez:2015uha}
H.~Gomez, C.~R. Mafra, and O.~Schlotterer, ``{Two-loop superstring five-point
  amplitude and $S$-duality},''
  \href{http://dx.doi.org/10.1103/PhysRevD.93.045030}{{\em Phys. Rev.} {\bf
  D93} (2016) no.~4, 045030},
\href{http://arxiv.org/abs/1504.02759}{{\tt  arXiv:1504.02759 [hep-th]}}.

\bibitem{Gomez:2013sla}
H.~Gomez and C.~R. Mafra, ``{The closed-string 3-loop amplitude and
  S-duality},'' \href{http://dx.doi.org/10.1007/JHEP10(2013)217}{{\em JHEP}
  {\bf 1310} (2013)  217},
\href{http://arxiv.org/abs/1308.6567}{{\tt  arXiv:1308.6567 [hep-th]}}.

\bibitem{Berkovits:2005bt}
N.~Berkovits, ``{Pure spinor formalism as an N=2 topological string},''
  \href{http://dx.doi.org/10.1088/1126-6708/2005/10/089}{{\em JHEP} {\bf 10}
  (2005)  089}, \href{http://arxiv.org/abs/hep-th/0509120}{{\tt
  arXiv:hep-th/0509120}}.

\bibitem{DHoker:1989cxq}
E.~D'Hoker and D.~Phong, ``{Conformal Scalar Fields and Chiral Splitting on
  Superriemann Surfaces},'' \href{http://dx.doi.org/10.1007/BF01218413}{{\em
  Commun. Math. Phys.} {\bf 125} (1989)  469}.

\bibitem{DHoker:2020prr}
E.~D'Hoker, C.~R. Mafra, B.~Pioline, and O.~Schlotterer, ``{Two-loop
  superstring five-point amplitudes I: Construction via chiral splitting and
  pure spinors},'' \href{http://dx.doi.org/10.1007/JHEP08(2020)135}{{\em JHEP} {\bf 08}
  (2020)  135}, \href{http://arxiv.org/abs/2006.05270}{{\tt
  arXiv:2006.05270 [hep-th]}}.
  
\bibitem{Carrasco:2011mn}
J.~J. Carrasco and H.~Johansson, ``{Five-Point Amplitudes in N=4
  Super-Yang-Mills Theory and N=8 Supergravity},''
  \href{http://dx.doi.org/10.1103/PhysRevD.85.025006}{{\em Phys. Rev.} {\bf
  D85} (2012)  025006},
\href{http://arxiv.org/abs/1106.4711}{{\tt  arXiv:1106.4711 [hep-th]}}.

\bibitem{Mafra:2015mja}
C.~R. Mafra and O.~Schlotterer, ``{Two-loop five-point amplitudes of super
  Yang-Mills and supergravity in pure spinor superspace},''
  \href{http://dx.doi.org/10.1007/JHEP10(2015)124}{{\em JHEP} {\bf 10} (2015)
  124},
\href{http://arxiv.org/abs/1505.02746}{{\tt  arXiv:1505.02746 [hep-th]}}.

\bibitem{DMS3}
E.~D'Hoker and O.~Schlotterer, ``Two-loop superstring five-point
  amplitudes III, construction via the RNS formulation,'' to appear.

\bibitem{Green:1997tv}
M.~B. Green and M.~Gutperle, ``{Effects of D-instantons},''
  \href{http://dx.doi.org/10.1016/S0550-3213(97)00269-1}{{\em Nucl. Phys.} {\bf
  B498} (1997)  195--227},
\href{http://arxiv.org/abs/hep-th/9701093}{{\tt  arXiv:hep-th/9701093}}.

\bibitem{Green:1998by}
M.~B. Green and S.~Sethi, ``{Supersymmetry constraints on type IIB
  supergravity},'' \href{http://dx.doi.org/10.1103/PhysRevD.59.046006}{{\em
  Phys. Rev.} {\bf D59} (1999)  046006},
\href{http://arxiv.org/abs/hep-th/9808061}{{\tt  arXiv:hep-th/9808061}}.

\bibitem{Green:1999pu}
M.~B. Green, H.-h. Kwon, and P.~Vanhove, ``{Two loops in eleven dimensions},''
  \href{http://dx.doi.org/10.1103/PhysRevD.61.104010}{{\em Phys. Rev.} {\bf
  D61} (2000)  104010},
\href{http://arxiv.org/abs/hep-th/9910055}{{\tt  arXiv:hep-th/9910055}}.

\bibitem{Green:2005ba}
M.~B. Green and P.~Vanhove, ``{Duality and higher derivative terms in M
  theory},'' \href{http://dx.doi.org/10.1088/1126-6708/2006/01/093}{{\em JHEP}
  {\bf 0601} (2006)  093},
\href{http://arxiv.org/abs/hep-th/0510027}{{\tt  arXiv:hep-th/0510027}}.

\bibitem{Green:2014yxa}
M.~B. Green, S.~D. Miller, and P.~Vanhove, ``{$SL(2, \mathbb{Z})$-invariance
  and D-instanton contributions to the $D^6 R^4$ interaction},''
  \href{http://dx.doi.org/10.4310/CNTP.2015.v9.n2.a3}{{\em Commun. Num. Theor.
  Phys.} {\bf 09} (2015)  307--344},
\href{http://arxiv.org/abs/1404.2192}{{\tt  arXiv:1404.2192 [hep-th]}}.

\bibitem{Green:2010wi}
M.~B. Green, J.~G. Russo, and P.~Vanhove, ``{Automorphic properties of low
  energy string amplitudes in various dimensions},''
  \href{http://dx.doi.org/10.1103/PhysRevD.81.086008}{{\em Phys.Rev.} {\bf D81}
  (2010)  086008},
\href{http://arxiv.org/abs/1001.2535}{{\tt  arXiv:1001.2535 [hep-th]}}.

\bibitem{Green:2010kv}
M.~B. Green, S.~D. Miller, J.~G. Russo, and P.~Vanhove, ``{Eisenstein series
  for higher-rank groups and string theory amplitudes},'' {\em
  Commun.Num.Theor.Phys.} {\bf 4} (2010)  551--596,
\href{http://arxiv.org/abs/1004.0163}{{\tt  arXiv:1004.0163 [hep-th]}}.

\bibitem{Bossard:2020xod}
G.~Bossard, A.~Kleinschmidt, and B.~Pioline, ``{1/8-BPS Couplings and
  Exceptional Automorphic Functions},''
  \href{http://dx.doi.org/10.21468/SciPostPhys.8.4.054}{{\em SciPost Phys.}
  {\bf 8} (2020) no.~4, 054},
\href{http://arxiv.org/abs/2001.05562}{{\tt  arXiv:2001.05562 [hep-th]}}.

\bibitem{DHoker:1994gnm}
E.~D'Hoker and D.~H. Phong, ``{The Box graph in superstring theory},''
  \href{http://dx.doi.org/10.1016/0550-3213(94)00526-K}{{\em Nucl. Phys.} {\bf
  B440} (1995)  24--94},
\href{http://arxiv.org/abs/hep-th/9410152}{{\tt  arXiv:hep-th/9410152}}.

\bibitem{Green:1999pv}
M.~B. Green and P.~Vanhove, ``{The low energy expansion of the one-loop type II
  superstring amplitude},''
  \href{http://dx.doi.org/10.1103/PhysRevD.61.104011}{{\em Phys. Rev.} {\bf
  D61} (2000)  104011},
\href{http://arxiv.org/abs/hep-th/9910056}{{\tt  arXiv:hep-th/9910056}}.

\bibitem{Green:2008uj}
M.~B. Green, J.~G. Russo, and P.~Vanhove, ``{Low energy expansion of the
  four-particle genus-one amplitude in type II superstring theory},''
  \href{http://dx.doi.org/10.1088/1126-6708/2008/02/020}{{\em JHEP} {\bf 0802}
  (2008)  020},
\href{http://arxiv.org/abs/0801.0322}{{\tt  arXiv:0801.0322 [hep-th]}}.

\bibitem{D'Hoker:2015foa}
E.~D'Hoker, M.~B. Green, and P.~Vanhove, ``{On the modular structure of the
  genus-one Type II superstring low energy expansion},''
  \href{http://dx.doi.org/10.1007/JHEP08(2015)041}{{\em JHEP} {\bf 08} (2015)
  041},
\href{http://arxiv.org/abs/1502.06698}{{\tt  arXiv:1502.06698 [hep-th]}}.

\bibitem{DHoker:2005jhf}
E.~D'Hoker, M.~Gutperle, and D.~H. Phong, ``{Two-loop superstrings and
  S-duality},'' \href{http://dx.doi.org/10.1016/j.nuclphysb.2005.06.010}{{\em
  Nucl. Phys.} {\bf B722} (2005)  81--118},
\href{http://arxiv.org/abs/hep-th/0503180}{{\tt  arXiv:hep-th/0503180}}.

\bibitem{D'Hoker:2013eea}
E.~{D'Hoker} and M.~B. {Green}, ``{Zhang-Kawazumi invariants and superstring
  amplitudes},'' \href{http://dx.doi.org/10.1016/j.jnt.2014.03.021}{{\em {J.
  Number Theory}} {\bf 144} (2014)  111--150},
  \href{http://arxiv.org/abs/1308.4597}{{\tt  arXiv:1308.4597 [hep-th]}}.

\bibitem{DHoker:2014oxd}
E.~D'Hoker, M.~B. Green, B.~Pioline, and R.~Russo, ``{Matching the $D^{6}R^{4}$
  interaction at two-loops},''
  \href{http://dx.doi.org/10.1007/JHEP01(2015)031}{{\em JHEP} {\bf 01} (2015)
  031},
\href{http://arxiv.org/abs/1405.6226}{{\tt  arXiv:1405.6226 [hep-th]}}.

\bibitem{DHoker:2017pvk}
E.~D'Hoker, M.~B. Green, and B.~Pioline, ``{Higher genus modular graph
  functions, string invariants, and their exact asymptotics},''
  \href{http://dx.doi.org/10.1007/s00220-018-3244-3}{{\em Comm. Math. Phys.}
  (2017)  },
\href{http://arxiv.org/abs/1712.06135}{{\tt  arXiv:1712.06135 [hep-th]}}.

\bibitem{DHoker:2018mys}
E.~D'Hoker, M.~B. Green, and B.~Pioline, ``{Asymptotics of the $D^8 {\cal R}^4$
  genus-two string invariant},'' {\em Comm. Num. Theo. Phys.} {\bf 13} (2018),
\href{http://arxiv.org/abs/1806.02691}{{\tt  arXiv:1806.02691 [hep-th]}}.

\bibitem{DHoker:2015wxz}
E.~D'Hoker, M.~B. Green, {\"O}.~G{\"u}rdogan, and P.~Vanhove, ``{Modular Graph
  Functions},'' \href{http://dx.doi.org/10.4310/CNTP.2017.v11.n1.a4}{{\em
  Commun. Num. Theor. Phys.} {\bf 11} (2017)  165--218},
\href{http://arxiv.org/abs/1512.06779}{{\tt  arXiv:1512.06779 [hep-th]}}.

\bibitem{Schlotterer:2012ny}
O.~Schlotterer and S.~Stieberger, ``{Motivic Multiple Zeta Values and
  Superstring Amplitudes},''
  \href{http://dx.doi.org/10.1088/1751-8113/46/47/475401}{{\em J. Phys.} {\bf
  A46} (2013)  475401},
\href{http://arxiv.org/abs/1205.1516}{{\tt  arXiv:1205.1516 [hep-th]}}.

\bibitem{Green:2013bza}
M.~B. Green, C.~R. Mafra, and O.~Schlotterer, ``{Multiparticle one-loop
  amplitudes and S-duality in closed superstring theory},''
  \href{http://dx.doi.org/10.1007/JHEP10(2013)188}{{\em JHEP} {\bf 10} (2013)
  188},
\href{http://arxiv.org/abs/1307.3534}{{\tt  arXiv:1307.3534 [hep-th]}}.

\bibitem{Basu:2016mmk}
A.~Basu, ``{Simplifying the one loop five graviton amplitude in type IIB string
  theory},'' \href{http://dx.doi.org/10.1142/S0217751X17500749}{{\em Int. J.
  Mod. Phys. A} {\bf 32} (2017) no.~14, 1750074},
  \href{http://arxiv.org/abs/1608.02056}{{\tt  arXiv:1608.02056 [hep-th]}}.

\bibitem{Gaberdiel:1998ui}
M.~R. Gaberdiel and M.~B. Green, ``{An SL(2, Z) anomaly in IIB supergravity and
  its F theory interpretation},''
  \href{http://dx.doi.org/10.1088/1126-6708/1998/11/026}{{\em JHEP} {\bf 11}
  (1998)  026}, \href{http://arxiv.org/abs/hep-th/9810153}{{\tt
  arXiv:hep-th/9810153}}.

\bibitem{Green:1999qt}
M.~B. Green, ``{Interconnections between type II superstrings, M theory and N=4
  supersymmetric Yang-Mills},''
  \href{http://dx.doi.org/10.1007/BFb0104240}{{\em Lect. Notes Phys.} {\bf 525}
  (1999)  22}, \href{http://arxiv.org/abs/hep-th/9903124}{{\tt
  arXiv:hep-th/9903124}}.

\bibitem{Boels:2012zr}
R.~H. Boels, ``{Maximal R-symmetry violating amplitudes in type IIB superstring
  theory},'' \href{http://dx.doi.org/10.1103/PhysRevLett.109.081602}{{\em Phys.
  Rev. Lett.} {\bf 109} (2012)  081602},
\href{http://arxiv.org/abs/1204.4208}{{\tt  arXiv:1204.4208 [hep-th]}}.

\bibitem{Green:2019rhz}
M.~B. Green and C.~Wen, ``{Modular Forms and $SL(2, {\mathbb Z})$-covariance of
  type IIB superstring theory},''
  \href{http://dx.doi.org/10.1007/JHEP06(2019)087}{{\em JHEP} {\bf 06} (2019)
  087}, \href{http://arxiv.org/abs/1904.13394}{{\tt  arXiv:1904.13394
  [hep-th]}}.

\bibitem{Basu:2018bde}
A.~Basu, ``{Eigenvalue equation for genus two modular graphs},''
  \href{http://dx.doi.org/10.1007/JHEP02(2019)046}{{\em JHEP} {\bf 02} (2019)
  046}, \href{http://arxiv.org/abs/1812.00389}{{\tt  arXiv:1812.00389
  [hep-th]}}.

\bibitem{DHoker:2015sve}
E.~D'Hoker, M.~B. Green, and P.~Vanhove, ``{Proof of a modular relation between
  1-, 2- and 3-loop Feynman diagrams on a torus},''
  \href{http://dx.doi.org/https://doi.org/10.1016/j.jnt.2017.07.022}{{\em
  Journal of Number Theory} (2017)  },
  \href{http://arxiv.org/abs/1509.00363}{{\tt  arXiv:1509.00363 [hep-th]}}.

\bibitem{DHoker:2016mwo}
E.~D'Hoker and M.~B. Green, ``{Identities between Modular Graph Forms},''
  \href{http://dx.doi.org/https://doi.org/10.1016/j.jnt.2017.11.015}{{\em
  Journal of Number Theory} {\bf 189} (2018)  25 -- 80},
  \href{http://arxiv.org/abs/1603.00839}{{\tt  arXiv:1603.00839 [hep-th]}}.

\bibitem{DHoker:2016quv}
E.~D'Hoker and J.~Kaidi, ``{Hierarchy of Modular Graph Identities},''
  \href{http://dx.doi.org/10.1007/JHEP11(2016)051}{{\em JHEP} {\bf 11} (2016)
  051},
\href{http://arxiv.org/abs/1608.04393}{{\tt  arXiv:1608.04393 [hep-th]}}.

\bibitem{Basu:2016kli}
A.~Basu, ``{Proving relations between modular graph functions},''
  \href{http://dx.doi.org/10.1088/0264-9381/33/23/235011}{{\em Class. Quant.
  Grav.} {\bf 33} (2016) no.~23, 235011},
\href{http://arxiv.org/abs/1606.07084}{{\tt  arXiv:1606.07084 [hep-th]}}.

\bibitem{Bern:1998ug}
Z.~Bern, L.~J. Dixon, D.~C. Dunbar, M.~Perelstein, and J.~S. Rozowsky, ``{On
  the relationship between Yang-Mills theory and gravity and its implication
  for ultraviolet divergences},''
  \href{http://dx.doi.org/10.1016/S0550-3213(98)00420-9}{{\em Nucl. Phys.} {\bf
  B530} (1998)  401--456},
\href{http://arxiv.org/abs/hep-th/9802162}{{\tt  arXiv:hep-th/9802162}}.

\bibitem{Richards:2008jg}
D.~M. Richards, ``{The One-Loop Five-Graviton Amplitude and the Effective
  Action},'' \href{http://dx.doi.org/10.1088/1126-6708/2008/10/042}{{\em JHEP}
  {\bf 10} (2008)  042}, \href{http://arxiv.org/abs/0807.2421}{{\tt
  arXiv:0807.2421 [hep-th]}}.

\bibitem{Green:2010sp}
M.~B. Green, J.~G. Russo, and P.~Vanhove, ``{String theory dualities and
  supergravity divergences},''
  \href{http://dx.doi.org/10.1007/JHEP06(2010)075}{{\em JHEP} {\bf 1006} (2010)
   075},
\href{http://arxiv.org/abs/1002.3805}{{\tt  arXiv:1002.3805 [hep-th]}}.

\bibitem{Pioline:2018pso}
B.~Pioline, ``{String theory integrands and supergravity divergences},''
  \href{http://dx.doi.org/10.1007/JHEP02(2019)148}{{\em JHEP} {\bf 02} (2019)
  148},
\href{http://arxiv.org/abs/1810.11343}{{\tt  arXiv:1810.11343 [hep-th]}}.

\bibitem{DHoker:2005dys}
E.~D'Hoker and D.~H. Phong, ``{Two-loop superstrings. V. Gauge slice
  independence of the N-point function},''
  \href{http://dx.doi.org/10.1016/j.nuclphysb.2005.02.042}{{\em Nucl. Phys.}
  {\bf B715} (2005)  91--119},
\href{http://arxiv.org/abs/hep-th/0501196}{{\tt  arXiv:hep-th/0501196}}.

\bibitem{DHoker:2001kkt}
E.~D'Hoker and D.~H. Phong, ``{Two loop superstrings. 1. Main formulas},''
  \href{http://dx.doi.org/10.1016/S0370-2693(02)01255-8}{{\em Phys. Lett.} {\bf
  B529} (2002)  241--255},
\href{http://arxiv.org/abs/hep-th/0110247}{{\tt  arXiv:hep-th/0110247}}.

\bibitem{DHoker:2001qqx}
E.~D'Hoker and D.~H. Phong, ``{Two loop superstrings. 2. The Chiral measure on
  moduli space},'' \href{http://dx.doi.org/10.1016/S0550-3213(02)00431-5}{{\em
  Nucl. Phys.} {\bf B636} (2002)  3--60},
\href{http://arxiv.org/abs/hep-th/0110283}{{\tt  arXiv:hep-th/0110283}}.

\bibitem{DHoker:2001foj}
E.~D'Hoker and D.~H. Phong, ``{Two loop superstrings. 3. Slice independence and
  absence of ambiguities},''
  \href{http://dx.doi.org/10.1016/S0550-3213(02)00432-7}{{\em Nucl. Phys.} {\bf
  B636} (2002)  61--79},
\href{http://arxiv.org/abs/hep-th/0111016}{{\tt  arXiv:hep-th/0111016}}.

\bibitem{DHoker:2001jaf}
E.~D'Hoker and D.~H. Phong, ``{Two loop superstrings 4: The Cosmological
  constant and modular forms},''
  \href{http://dx.doi.org/10.1016/S0550-3213(02)00516-3}{{\em Nucl. Phys.} {\bf
  B639} (2002)  129--181},
\href{http://arxiv.org/abs/hep-th/0111040}{{\tt  arXiv:hep-th/0111040}}.

\bibitem{Witten:2013tpa}
E.~Witten, ``{Notes On Holomorphic String And Superstring Theory Measures Of
  Low Genus},'' \href{http://arxiv.org/abs/1306.3621}{{\tt  arXiv:1306.3621
  [hep-th]}}.

\bibitem{Berkovits:2005ng}
N.~Berkovits and C.~R. Mafra, ``{Equivalence of two-loop superstring amplitudes
  in the pure spinor and RNS formalisms},''
  \href{http://dx.doi.org/10.1103/PhysRevLett.96.011602}{{\em Phys. Rev. Lett.}
  {\bf 96} (2006)  011602},
\href{http://arxiv.org/abs/hep-th/0509234}{{\tt  arXiv:hep-th/0509234}}.

\bibitem{Gomez:2010ad}
H.~Gomez and C.~R. Mafra, ``{The Overall Coefficient of the Two-loop
  Superstring Amplitude Using Pure Spinors},''
  \href{http://dx.doi.org/10.1007/JHEP05(2010)017}{{\em JHEP} {\bf 05} (2010)
  017},
\href{http://arxiv.org/abs/1003.0678}{{\tt  arXiv:1003.0678 [hep-th]}}.

\bibitem{Berkovits:2000fe}
N.~Berkovits, ``{Super Poincare covariant quantization of the superstring},''
  \href{http://dx.doi.org/10.1088/1126-6708/2000/04/018}{{\em JHEP} {\bf 04}
  (2000)  018}, \href{http://arxiv.org/abs/hep-th/0001035}{{\tt
  arXiv:hep-th/0001035}}.

\bibitem{Berkovits:2006ik}
N.~Berkovits, ``{Explaining Pure Spinor Superspace},''
  \href{http://arxiv.org/abs/hep-th/0612021}{{\tt  arXiv:hep-th/0612021}}.

\bibitem{Mafra:2014gsa}
C.~R. Mafra and O.~Schlotterer, ``{Cohomology foundations of one-loop
  amplitudes in pure spinor superspace},''
  \href{http://arxiv.org/abs/1408.3605}{{\tt  arXiv:1408.3605 [hep-th]}}.

\bibitem{Mafra:2014oia}
C.~R. Mafra and O.~Schlotterer, ``{Multiparticle SYM equations of motion and
  pure spinor BRST blocks},''
  \href{http://dx.doi.org/10.1007/JHEP07(2014)153}{{\em JHEP} {\bf 07} (2014)
  153}, \href{http://arxiv.org/abs/1404.4986}{{\tt  arXiv:1404.4986 [hep-th]}}.

\bibitem{Mafra:2016nwr}
C.~R. Mafra and O.~Schlotterer, ``{One-loop superstring six-point amplitudes
  and anomalies in pure spinor superspace},''
  \href{http://dx.doi.org/10.1007/JHEP04(2016)148}{{\em JHEP} {\bf 04} (2016)
  148}, \href{http://arxiv.org/abs/1603.04790}{{\tt  arXiv:1603.04790
  [hep-th]}}.

\bibitem{PSSsite}
C.~R. Mafra and O.~Schlotterer,
{\tt http://www.southampton.ac.uk/\~{}crm1n16/pss.html}.

\bibitem{Mafra:2010pn}
C.~R. Mafra, ``{PSS: A FORM Program to Evaluate Pure Spinor Superspace
  Expressions},'' \href{http://arxiv.org/abs/1007.4999}{{\tt  arXiv:1007.4999
  [hep-th]}}.

\bibitem{Siegel}
C.~L. Siegel, {\em {Topics in complex function theory. Vol. I,II,III}}.
\newblock Wiley Interscience, 1971.

\bibitem{AlvarezGaume:1986es}
L.~Alvarez-Gaume, G.~W. Moore, and C.~Vafa, ``{Theta Functions, Modular
  Invariance and Strings},'' \href{http://dx.doi.org/10.1007/BF01210925}{{\em
  Commun. Math. Phys.} {\bf 106} (1986)  1--40}.

\bibitem{Pioline:2015qha}
B.~Pioline, ``{A Theta lift representation for the Kawazumi-Zhang and Faltings
  invariants of genus-two Riemann surfaces},''
  \href{http://dx.doi.org/10.1016/j.jnt.2015.12.021}{{\em J. Number Theor.}
  {\bf 163} (2016)  520--541},
\href{http://arxiv.org/abs/1504.04182}{{\tt  arXiv:1504.04182 [hep-th]}}.

\bibitem{Gerken:2018jrq}
J.~E. Gerken, A.~Kleinschmidt, and O.~Schlotterer, ``{Heterotic-string
  amplitudes at one loop: modular graph forms and relations to open strings},''
  \href{http://dx.doi.org/10.1007/JHEP01(2019)052}{{\em JHEP} {\bf 01} (2019)
  052}, \href{http://arxiv.org/abs/1811.02548}{{\tt  arXiv:1811.02548
  [hep-th]}}.

\bibitem{Gerken:2019cxz}
J.~E. Gerken, A.~Kleinschmidt, and O.~Schlotterer, ``{All-order differential
  equations for one-loop closed-string integrals and modular graph forms},''
  \href{http://dx.doi.org/10.1007/JHEP01(2020)064}{{\em JHEP} {\bf 01} (2020)
  064}, \href{http://arxiv.org/abs/1911.03476}{{\tt  arXiv:1911.03476
  [hep-th]}}.

\bibitem{Gerken:2020yii}
J.~E. Gerken, A.~Kleinschmidt, and O.~Schlotterer, ``{Generating series of all
  modular graph forms from iterated Eisenstein integrals},''
  \href{http://dx.doi.org/10.1007/JHEP07(2020)190}{{\em JHEP} {\bf 07} (2020)
  190}, \href{http://arxiv.org/abs/2004.05156}{{\tt  arXiv:2004.05156
  [hep-th]}}.
  
  
\bibitem{DHoker:2020uid}
E.~D'Hoker and O.~Schlotterer,
``Identities among higher genus modular graph tensors,''
\href{http://arxiv.org/abs/2010.00924}{{\tt  arXiv:2010.00924 [hep-th]}}.

\bibitem{Basu:2020pey}
A.~Basu,
``Poisson equations for elliptic modular graph functions,''
\href{http://arxiv.org/abs/2009.02221}{{\tt  arXiv:2009.02221
  [hep-th]}}.


\bibitem{Gomez:2009qd}
H.~Gomez, ``{One-loop Superstring Amplitude From Integrals on Pure Spinors
  Space},'' \href{http://dx.doi.org/10.1088/1126-6708/2009/12/034}{{\em JHEP}
  {\bf 12} (2009)  034}, \href{http://arxiv.org/abs/0910.3405}{{\tt
  arXiv:0910.3405 [hep-th]}}.

\bibitem{Mafra:2008ar}
C.~R. Mafra, ``{Pure Spinor Superspace Identities for Massless Four-point
  Kinematic Factors},''
  \href{http://dx.doi.org/10.1088/1126-6708/2008/04/093}{{\em JHEP} {\bf 04}
  (2008)  093}, \href{http://arxiv.org/abs/0801.0580}{{\tt  arXiv:0801.0580
  [hep-th]}}.

\bibitem{Vafa:1995fj}
C.~Vafa and E.~Witten, ``{A One loop test of string duality},''
  \href{http://dx.doi.org/10.1016/0550-3213(95)00280-6}{{\em Nucl. Phys. B}
  {\bf 447} (1995)  261--270}, \href{http://arxiv.org/abs/hep-th/9505053}{{\tt
  arXiv:hep-th/9505053}}.

\bibitem{Bern:2008qj}
Z.~Bern, J.~Carrasco, and H.~Johansson, ``{New Relations for Gauge-Theory
  Amplitudes},'' \href{http://dx.doi.org/10.1103/PhysRevD.78.085011}{{\em Phys.
  Rev. D} {\bf 78} (2008)  085011}, \href{http://arxiv.org/abs/0805.3993}{{\tt
  arXiv:0805.3993 [hep-ph]}}.

\bibitem{BjerrumBohr:2010hn}
N.~Bjerrum-Bohr, P.~H. Damgaard, T.~Sondergaard, and P.~Vanhove, ``{The
  Momentum Kernel of Gauge and Gravity Theories},''
  \href{http://dx.doi.org/10.1007/JHEP01(2011)001}{{\em JHEP} {\bf 01} (2011)
  001}, \href{http://arxiv.org/abs/1010.3933}{{\tt  arXiv:1010.3933 [hep-th]}}.

\bibitem{Mafra:2011nv}
C.~R. Mafra, O.~Schlotterer, and S.~Stieberger, ``{Complete N-Point Superstring
  Disk Amplitude I. Pure Spinor Computation},''
  \href{http://dx.doi.org/10.1016/j.nuclphysb.2013.04.023}{{\em Nucl. Phys. B}
  {\bf 873} (2013)  419--460}, \href{http://arxiv.org/abs/1106.2645}{{\tt
  arXiv:1106.2645 [hep-th]}}.

\bibitem{Berends:1987me}
F.~A. Berends and W.~Giele, ``{Recursive Calculations for Processes with n
  Gluons},'' \href{http://dx.doi.org/10.1016/0550-3213(88)90442-7}{{\em Nucl.
  Phys. B} {\bf 306} (1988)  759--808}.
  
\bibitem{Mafra:2010jq}
C.~R. Mafra, O.~Schlotterer, S.~Stieberger, and D.~Tsimpis, ``{A recursive
  method for SYM n-point tree amplitudes},''
  \href{http://dx.doi.org/10.1103/PhysRevD.83.126012}{{\em Phys. Rev. D} {\bf
  83} (2011)  126012}, \href{http://arxiv.org/abs/1012.3981}{{\tt
  arXiv:1012.3981 [hep-th]}}.  

\bibitem{Mafra:2015vca}
C.~R. Mafra and O.~Schlotterer, ``{Berends-Giele recursions and the BCJ duality
  in superspace and components},''
  \href{http://dx.doi.org/10.1007/JHEP03(2016)097}{{\em JHEP} {\bf 03} (2016)
  097}, \href{http://arxiv.org/abs/1510.08846}{{\tt  arXiv:1510.08846
  [hep-th]}}.

\bibitem{Boels:2013jua}
R.~H. Boels, ``{On the field theory expansion of superstring five point
  amplitudes},'' \href{http://dx.doi.org/10.1016/j.nuclphysb.2013.08.009}{{\em
  Nucl. Phys. B} {\bf 876} (2013)  215--233},
  \href{http://arxiv.org/abs/1304.7918}{{\tt  arXiv:1304.7918 [hep-th]}}.

\bibitem{mzvwebsite}
J.~Broedel, O.~Schlotterer, and S.~Stieberger,
\newblock {\tt https://wwwth.mpp.mpg.de/members/stieberg/mzv/index.html}.


\bibitem{Kawai:1985xq}
H.~Kawai, D.~Lewellen, and S.~Tye, ``{A Relation Between Tree Amplitudes of
  Closed and Open Strings},''
  \href{http://dx.doi.org/10.1016/0550-3213(86)90362-7}{{\em Nucl. Phys. B}
  {\bf 269} (1986)  1--23}.

\bibitem{Zerbini:2015rss}
F.~Zerbini, ``{Single-valued multiple zeta values in genus 1 superstring
  amplitudes},'' \href{http://dx.doi.org/10.4310/CNTP.2016.v10.n4.a2}{{\em
  Commun. Num. Theor. Phys.} {\bf 10} (2016)  703--737},
\href{http://arxiv.org/abs/1512.05689}{{\tt  arXiv:1512.05689 [hep-th]}}.

\bibitem{DHoker:2019xef}
E.~D'Hoker and M.~Green, ``{Absence of irreducible multiple zeta-values in
  melon modular graph functions},''
  \href{http://dx.doi.org/10.4310/CNTP.2020.v14.n2.a2}{{\em Commun. Num. Theor.
  Phys.} {\bf 14} (2020) no.~2, 315--324},
  \href{http://arxiv.org/abs/1904.06603}{{\tt  arXiv:1904.06603 [hep-th]}}.

\bibitem{Zagier:2019eus}
D.~Zagier and F.~Zerbini, ``{Genus-zero and genus-one string amplitudes and
  special multiple zeta values},''
  \href{http://dx.doi.org/10.4310/CNTP.2020.v14.n2.a4}{{\em Commun. Num. Theor.
  Phys.} {\bf 14} (2020) no.~2, 413--452},
  \href{http://arxiv.org/abs/1906.12339}{{\tt  arXiv:1906.12339 [math.NT]}}.

\bibitem{Vanhove:2020qtt}
P.~Vanhove and F.~Zerbini, ``{Building blocks of closed and open string
  amplitudes},'' in {\em {MathemAmplitudes 2019: Intersection Theory and
  Feynman Integrals}},
\newblock \href{http://arxiv.org/abs/2007.08981}{{\tt  arXiv:2007.08981
  [hep-th]}}.

\bibitem{Green:1997me}
M.~B. Green, M.~Gutperle, and H.-h. Kwon, ``{$\lambda^{16}$ and related terms
  in M-theory on $T^2$},''
  \href{http://dx.doi.org/10.1016/S0370-2693(97)01551-7}{{\em Phys. Lett.} {\bf
  B421} (1998)  149--161},
\href{http://arxiv.org/abs/hep-th/9710151}{{\tt  arXiv:hep-th/9710151}}.

\bibitem{fay73}
J.~D. Fay, ``Theta functions on {Riemann} surfaces,'' {\em Lecture Notes in
  Math.} {\bf 352} (1973).

\bibitem{Schiffer}
M.~Schiffer and D.~C. Spencer, {\em Functionals of finite Riemann surfaces}.
\newblock Courier Corporation, 2014.

\bibitem{DKS}
E.~D'Hoker, A.~Kleinschmidt, and O.~Schlotterer, ``Elliptic modular graph forms I: Identities and generating series,'' accepted for publication in {\em JHEP}, \href{http://arxiv.org/abs/2012.09198}{{\tt  arXiv:2012.09198
  [hep-th]}}.

\bibitem{Gerken:2018zcy}
J.~E. Gerken and J.~Kaidi, ``{Holomorphic subgraph reduction of higher-point
  modular graph forms},'' \href{http://dx.doi.org/10.1007/JHEP01(2019)131}{{\em
  JHEP} {\bf 01} (2019)  131}, \href{http://arxiv.org/abs/1809.05122}{{\tt
  arXiv:1809.05122 [hep-th]}}.

\bibitem{Gerken:2020aju}
J.~E. Gerken, ``{Basis Decompositions and a Mathematica Package for Modular
  Graph Forms},'' accepted for publication in {\em J.\ Phys.\ {\bf A}}, \href{http://arxiv.org/abs/2007.05476}{{\tt  arXiv:2007.05476
  [hep-th]}}.


\end{thebibliography}

\providecommand{\href}[2]{#2}\begingroup\raggedright\endgroup

\end{document}